%% file: ms_sy.tex
\documentclass{emulateapj}
\usepackage{apjfonts}
\usepackage{epsfig}
\usepackage{psfig}
\usepackage{graphicx}
\usepackage{natbib}

\slugcomment{Accepted for publication in {\it The Astrophysical Journal, Issue 730--2}}
\shorttitle{Seyferts on mas-scales}
\shortauthors{Lal et al.}

\begin{document}

\title{Seyfert galaxies: Nuclear Radio Structure and Unification}


\author{Dharam V. Lal$^{1,2,3}$, Prajval Shastri$^{1}$, Denise C. Gabuzda$^{4}$}
\affil{
$^1$Indian Institute of Astrophysics, Koramangala, Bangalore 560034, India}
\affil{
$^2$Joint Astronomy Programme, Department of Physics, Indian Institute of
Science, Bangalore 560012, India}
\affil{
$^3$Harvard-Smithsonian Center for Astrophysics, 60 Garden Street,
Cambridge, MA 02138, USA}
\affil{
$^4$Department of Physics, University College Cork, Cork, Republic of Ireland}

\begin{abstract}

A radio study of a carefully selected sample of 20~Seyfert galaxies
that are matched in orientation-independent parameters, which are
measures of intrinsic active galactic nuclei ({\small AGN}) power and
host galaxy properties
is presented to test the predictions of the unified scheme hypothesis.
Our sample sources have core flux densities greater than 8~mJy at 5~GHz
on arcsec-scales due to the feasibility requirements.
These simultaneous pc-scale and kpc-scale radio observations reveal
(i) that Seyfert~1 and Seyfert~2 galaxies have equal tendency
to show compact radio structures on mas-scales,
(ii) the distributions of pc-scale and kpc-scale radio luminosities
are similar for both Seyfert~1 and Seyfert~2 galaxies,
(iii) no evidence for relativistic beaming in Seyfert galaxies,
(iv) similar distributions of source spectral indices in spite of
the fact that Seyferts show nuclear radio flux density variations, and
(v) the distributions of projected linear size for Seyfert~1 and Seyfert~2
galaxies are not significantly different as would be expected in the
unified scheme.
The latter could be mainly due to a relatively large spread in the
intrinsic sizes.
We also find that a starburst alone cannot power these radio sources.
Finally, an analysis of the kpc-scale radio properties of the CfA
Seyfert galaxy sample shows results consistent with the predictions
of the unified scheme.
\end{abstract}

\keywords{Galaxies: Seyfert -- Radio continuum: galaxies}


\section{Introduction}
\label{intro}

Active galaxies are classified according to their appearance,
luminosity, and spectra into the following principal types:
Seyfert galaxies, radio galaxies, quasars, and {\small BL~L}acertae objects.
\citet{KW74} identified two types of Seyfert
galaxies on the basis of the widths of the nuclear emission lines.
While spectra of type~2 Seyfert galaxies have a
single set of relatively narrow emission lines
(whose width can be characterized in terms of full width
at half maximum, {\small FWHM}~$\approx$~300--1,000~km~s$^{-1}$),
the spectra of type~1 Seyfert galaxies have an additional much
broader component ({\small FWHM}~$\ge$~1,000~km~s$^{-1}$) of hydrogen
and helium lines. In the simplest cases, the
broad component is either absent (Seyfert~2) or strong and
dominant (Seyfert~1). With better data, it became clear that there
is a wide range in the relative strength of the broad and narrow
emission lines, and this led to refinement of the
Seyfert classification by introducing intermediate types
\citep{Osterbrock81,OP85}.
As the broad component
of H$\beta$ becomes weaker relative to the narrower component,
the Seyfert type changes from 1 to 1.2 to 1.5 to 1.8. For
Seyfert~1.8 galaxies, weak broad wings are just visible at the
base of H$\beta$, while in Seyfert~1.9 galaxies they are only
visible on the H$\alpha$ emission line at 6563~\AA. In practice,
these Seyfert sub-types have not been formally defined but
instead give an overall indication of the degree to which the
broad component is present.

The technique of spectropolarimetry yielded spectra
that showed broad lines in polarized light in Seyfert~2
galaxies. This polarized light was interpreted to
be the light that was initially
moving out of the nucleus in one direction which was then
reflected into our line of sight.
Such a technique could detect broad line regions ({\small BLR}s) in a
Seyfert~2 galaxy, {\it e.g.} NGC~1068 \citep{MG90}.
This led to the unified scheme model \citep{A93,LE10},
which is the key idea being used to organize and
make sense of our large and growing observational information
about Seyfert galaxies, {\it i.e.}
Seyfert~2 galaxies are intrinsically Seyfert~1 galaxies whose
continuum and broad-line emission is attenuated in the
direction of the observer.

Several investigations in the literature have yielded
results consistent with the predictions of this scheme, {\it e.g.},
the featureless continuum is stronger in Seyfert~1 than
in Seyfert~2 galaxies \citep{Lawrence87,MHetal94}.
\citet{Kinneyetal91} showed similar
ultraviolet slopes for Seyfert~2 and Seyfert~1 galaxies.
The active galactic nucleus (AGN) of Seyfert~2 galaxy is clearly seen in H$\alpha$,
but is barely detected in the ultraviolet images \citep{Colinaetal97}.
The ionizing radiation is roughly collimated before emerging
into the narrow line region perhaps due to an obscuring torus
\citep{Whittleetal88} and is sometimes cone-shaped
\citep{Pogge89,Evansetal91A,Evansetal91B,Evansetal93,Evansetal94},
which appear smaller in Seyfert~1 than Seyfert~2 galaxies
\citep{Krissetal94,MWT96,Colinaetal97,Heckmanetal97,Munozetal07,GDetal98}.
Furthermore, for a given far-infrared luminosity,
\citet{LE82} found a significant lack of soft X-ray emission in Seyfert~2
galaxies compared to Seyfert~1 galaxies \citep[see also][]{MHetal94,Cappietal96}
and \citet{MHetal94} found similar distributions of the hard X-ray emission
for both kinds of Seyfert galaxies.
\citet{Maiolinoetal97} and \citet{Curran00} found no differences in
the mean ratio of {\small CO} and far-infrared luminosity between the two
Seyfert classes suggesting that both Seyfert types
have the same amount of molecular gas.
\citet{Morgantietal99} found that
Seyfert~2 galaxies tend to have a larger projected radio linear size
than Seyfert~1 galaxies, whereas there is no statistically
significant difference in radio power between Seyfert~1 and
Seyfert~2 galaxies.
Recently, \citet{Gallimoreetal10}
found that Seyfert 1's show silicate emission on average and Seyfert 2's
show silicate absorption, which is broadly compatible with the obscuring torus
interpretation.
While the unified scheme is simple and attractive, there are
some observational results that are inconsistent with it,
such as,
the presence of relatively young ($\sim$1 Gyr) stellar populations
in Seyfert 2 galaxies \citep{Schmittetal99,GDetal01,Raimannetal03};
\citet{MGT98} found that Seyfert~2 galaxies, on average, tend to
have later morphological types than the Seyfert~1 galaxies;
\citet{DHetal99} confirmed that Seyfert~2 galaxies have an excess
of nearby companions over Seyfert~1 galaxies;
the scattered {\small BLR} is not detected in many Seyfert~2 galaxies
\citep{Tran01,Tran03};
a lack of X-ray absorption in several Seyfert 2 galaxies \citep{PB02};
Seyfert~2s having a higher propensity for nuclear starbursts \citep{Buchanan06};
and \citet{Royetal94} found a lower detection rate of compact radio cores
in Seyfert~1 than Seyfert~2 galaxies.

It was Roy et~al.'s (1994) observational result
which prompted us to commence this study.
In the unified scheme, since the torus is expected to be
transparent to emission at radio wavelengths, the compact
features should be similarly visible in Seyfert~1
and Seyfert~2 galaxies. Further, the inconsistency
with the unification scheme cannot be eased by invoking
relativistic beaming,
because then, the face-on {\small AGN}s, {\it viz.}, Seyfert~1 galaxies,
would be more likely to show compact structures. We aimed to
rigorously test the predictions of the unified scheme by
investigating the compact radio morphology of Seyfert galaxies.
To achieve this goal, we constructed a sample of Seyfert~1 and Seyfert~2
galaxies that are matched in orientation-independent parameters,
which are measures of intrinsic AGN power and host galaxy properties.

In this paper we first describe the construction of the sample
(Section~\ref{sample}) and use the radio maps for 15 objects presented in our
earlier paper \citep{LSG04} along with previously published data
for the remaining five objects to interpret our results and their
implications on the unification scheme hypothesis (Section~\ref{royetal}).
In Section~\ref{discuss} we also interpret the results of arcsec-scale
radio observations of
\citet{Kukulaetal95} for the CfA Seyfert galaxy sample \citep{HB92}
and their implications on the unification scheme hypothesis.
Finally, we summarize our conclusions in Section~\ref{summary}.

Throughout the paper we use the terms ``pc-scale'' and ``kpc-scale''
interchangeably for ``mas-scale'' and ``arcsec-scale'', respectively.
We also use ``face-on'' and ``edge-on'' interchangeably for ``Seyfert~1''
and ``Seyfert~2'' galaxies, respectively.
We assume a cosmology with $H_0$ =
75 km~s$^{-1}$~Mpc$^{-1}$ and $q_0$ = 0.
We define the spectral index
$\alpha$ in the sense that $S_\nu \propto \nu^{-\alpha}$, where $S_\nu$ and $\nu$
are flux density and frequency, respectively.
Since, we are dealing with small number statistics,
we use Mann-Whitney U test\footnote{Mann-Whitney U test is a non-parametric
statistical hypothesis test for small sample sizes, $\lesssim$20 \citep{SC81}.
It analyzes the degree of separation (or the amount of overlap)
between the two groups.
For example, the null hypothesis assumes that the two Seyfert sub-sample
types are homogeneous and come from the same population
(significance level, say $\lesssim$ 0.05).
The test involves the calculation of a statistic, called U, whose
distribution under the null hypothesis is known.
Significance is verified by using the computed test statistic (e.g., U)
and comparing this statistic (probability value) with the
Null Hypothesis value ($\lesssim$ 0.05).
If the former exceeds the latter, there is certainly sufficient
evidence to accept the Null Hypothesis.
For large samples, U is approximately normally distributed.}
\citep{SC81} to test the null hypothesis.

\section{Sample}
\label{sample}

The differences observed in samples of Seyfert galaxies can be
explained by the selection techniques used for assembling them.
For example, in the Markarian Survey, spectroscopic investigations have
shown that $\sim$ 10\% of all discovered 1500 galaxies with a strong
ultraviolet continuum are Seyfert galaxies \citep{Markarian67,Markarianetal86}.
Therefore, such a sample suffers from deficiency of Seyfert~2 galaxies,
which is most probably a result of the survey selection effect. This is
due to the fact that Seyfert~2 galaxies do not have excessive ultraviolet
continua due to obscuration and thus could easily elude the ultraviolet
search method \citep{MW84}.
The CfA Seyfert galaxy sample is
the first of the optically selected complete samples with
spectroscopic identifications and is due to \citet{HB92}.
The sample has an equal number of Seyfert~1 and Seyfert~2 galaxies (25 \& 23
respectively). The unresolved optical nucleus of
a Seyfert grows fainter with the square of distance, whilst the
surface brightness of its host galaxy remains constant over a constant
aperture. Therefore, in this sample the ratio of the two components,
the host galaxy surface brightness to the active nucleus surface brightness,
is highly variable.
\cite{HoU01} have discussed that the optical and ultraviolet selected samples
are likely to have inherent biases against the obscured  sources.
Similarly, the {\small IRAS} survey would most probably
detect reddened Seyfert~1 galaxies, but it may not be easy to
isolate them from the much more luminous starburst galaxies
\citep{Heckman90A,Heckman90B,HoU01,Buchanan06}.
In other words, the Seyfert samples based on the {\small IRAS} survey would be
contaminated due to the presence of luminous starburst galaxies.
Soft X-ray surveys may contain a larger fraction of soft X-ray
Seyfert~1 galaxies, since Seyfert~2 galaxies are weak soft X-ray sources
\citep{Veron86,LE10}.
%
It therefore seems that most of the Seyfert galaxy samples (optical, infrared,
and/or X-ray) have their biases and hence do not provide a good
platform to test the unification scheme hypothesis.
A Seyfert sample selected based on the orientation-independent
parameters, which are measures of {\small AGN} power and host galaxy
properties, would provide a good platform to test the predictions
of the unification scheme hypothesis.

\subsection{\textit{Bona fide} Seyfert galaxies}
\label{bona_fide}

The similarities between the nuclei of Seyfert galaxies
and QSOs have often been pointed out (Seyfert and starburst galaxies:
\citet{DD88}; low-ionization nuclear emission line regions ({\small LINER}s)
and radio-quiet quasars: \citet{HoU01}, \citet{Falckeetal00}, \citet{Hoetal97};
Low-Luminosity {\small AGN}: \citet{Nagaretal00}; etc.),
and numerous efforts have been made to
demonstrate a continuity between these objects. Any Seyfert sample
is rarely ever free from starburst galaxies, {\small LINER}s,
radio-quiet quasars, or radio-loud objects
\citep{Ketal03A,Ketal03B,Krongoldetal02,Levensonetal01,SB01,HuntMalkan99}.
Thus, we require the Seyfert galaxies that we select satisfy the
following definition:

\begin{enumerate}

\item [${\bullet}$] Its host is a spiral galaxy \citep{Weedman77}
of Hubble type S0 or later ({\it i.e.}~S0, Sa, Sab, Sb, Sbc,
and Sc) \citep{Sandage75}.
Radio-loud {\small AGN}s tend to reside in elliptical host galaxies
\citep{UP95}, and radio-quiet {\small AGN}s inhabit mostly spiral galaxies.
Thus we avoid any confusion due to the dichotomy of host-galaxy type
which may be linked with the radio-loud$/$radio-quiet dichotomy.

\item [${\bullet}$] Objects have low optical luminosity,
M$_B$~$\ge$~$-$23, in order to avoid radio-quiet quasars \citep{SG83}.

\item [${\bullet}$] It is a radio-quiet object, {\it i.e.} the
ratio of 5~GHz to $B-$Band flux density is less than 10
\citep{Kellermannetal89}.

\item [${\bullet}$] The nuclear line width of permitted line, 
H$\beta$$_{FWHM}$ (or H$\alpha$$_{FWHM}$), is more than
1,000 km~s$^{-1}$ \citep{KW74} for Seyfert~1 galaxies, and line intensity
ratio of [O~{\small III}]~$\lambda$5007 to H$\beta$ is greater than three
for Seyfert~2 galaxies \citep{DD88}.
LINERs (or H\,II region) galaxies occasionally show nuclear H$\beta$$_{FWHM}$
(or H$\alpha$$_{FWHM}$) line width more than 1,000 km~s$^{-1}$
\citep[][p.~24]{HoU01,Hoetal97,Peterson97}, but such galaxies never show line intensity
ratio of [O~{\small III}]~$\lambda$5007 to H$\beta$ greater than three
\citep[][p.~318]{HFS96,Krolik99}.
Or in other words, LINERs have a characteristically lower ionization state
than Seyfert nuclei.
Thus, we attempt to avoid any likely contamination due to the presence of
{\small LINER}s and H~{\small II} region galaxies in our Seyfert sample.

\end{enumerate}

\subsection{Criteria for the feasibility of our experiment}
\label{do_ability}

The sensitivity of an array is defined by the System Equivalent
Flux Density (${\small SEFD}$) \citep{Wrobel95,Wrobel00}.
The root-mean-square ({\small RMS}) thermal noise $\Delta$S in the visibility amplitude
of a single polarization baseline between antennas $i$ and $j$ is
$$
{\Delta}S~=~\frac{1}{{\eta}_S}~\times~\frac{\sqrt{SEFD_i~\times~SEFD_j}}{\sqrt{2\times{\Delta}{\nu}\times{\tau}}}~{\rm Jy}
$$
\citep{Wrobel95,Wrobel00},
where ${\eta}_S~\le~1$ accounts for the very-long-baseline-interferometry
({\small VLBI}) system inefficiency;
$\tau$ is the integration time (in seconds) for an individual scan,
which should be less than or equal to the coherence time,
and $\Delta\nu$ is the bandwidth (Hz).
Next,
the {\small RMS} thermal noise ${\Delta}I_m$, expected in a
single polarization image, assuming natural weighting \citep{Wrobel95} is
$$
{\Delta}I_m~=~\frac{1}{{\eta}_S}~\times~\frac{SEFD}{\sqrt{N\times(N-1)\times{\Delta}{\nu}\times{t_{int}}}}~{\rm Jy~beam}^{-1},
$$
where $N$ is the number of
antennas used and $t_{int}$ is the total integration time on
source.
For example,
the {\small RMS} noise level $\sigma$ on a baseline between two
very-long-baseline-array ({\small VLBA})
antennas for a data rate of 128~Mbits~s$^{-1}$, 2~min scan integration time,
and 6~cm observing wavelength is 4.7~mJy \citep{Wrobel00}.
A signal of 6$\sigma$ (=~28.2~mJy) is required to ensure reliable detection
of the correlated signal;
{\it i.e.}, the minimum detectable correlated flux density on each baseline.
We further adopted a scan integration time of 8.8~min,
which does not exceed the expected coherence time at 6~cm.
A single tape pass lasts 44 min;
with a scan duration of 8.8~min, exactly five scans will fit in each tape pass.
Increasing the scan integration time will decrease the detectable
flux density on all baselines; also adding the sensitive Effelsberg and
phased-very-large-array (phased-{\small VLA}) to the array will provide additionally reduced
detectable flux densities on baselines involving these antennas.
Hence, we decided to use 11 US stations (10 {\small VLBA} antennas and
the phased-{\small VLA}) and three European-VLBI-Network stations.
Torun and Noto stations along with Effelsberg provided us with a closure
triangle in Europe and gave a range of baseline lengths for the
sensitive baselines involving Effelsberg antenna.
Similarly {\small VLBA} stations spread all over the United States
provided us with a range of baseline lengths for sensitive baselines
involving phased-{\small VLA}.
Table \ref{correlated} gives 6$\sigma$ correlated flux densities,
minimum reliably detectable flux densities on various baselines
for a data rate of 128 Mbits~s$^{-1}$, a scan integration time of 8.8~min, and
an observing frequency of 5~GHz.
Finally, if we wish to have at least $\sim$~4~mJy of correlated flux density,
corresponding to the 6$\sigma$ detection limits for baselines between
one of the smaller antennas in the array and either the phased-{\small VLA}
or Effelsberg, the flux density in compact arcsec-scale structure must be
$\sim$~8~mJy, assuming that 50\% of the flux density seen with arcsec-scale
resolution would be detected with the above baselines.

\input{tab1.tex}

Therefore, the following constraint was enforced for the
feasibility requirements: The source must have been  
observed with at least arcsec-scale resolution,
at wavelengths of 6~cm ({\it i.e.}, {\small VLA}~$A$ or $B$~array
observations at these wavelengths), and it must have a detected
compact component with flux density greater than 8~mJy.
For objects that do not have $\lambda_{\rm 6~cm}$ measurements,
we used $\lambda_{\rm 3.6~cm}$ {\small VLA}~$A$ array flux
densities and assumed a flat spectral index between
these two wavelengths for these objects.
This constraint provides us with a reasonable sample size 
with minimum correlated flux density that is detectable on 
baselines involving Effelsberg antenna or phased-{\small VLA}.

\subsection{Criterion to minimize obscuration of optical properties}
\label{min_obs}

Next, we restrict our list of objects to those with a ratio
of minor to major isophotal diameter axes of the host galaxies
greater than half.
We thereby exclude edge-on spiral hosts and hence try to minimize
obscuration of optical properties due to transmission through an
edge-on galactic disk. Figure \ref{oursample}A shows the distribution of
the ratio of minor to major isophotal diameter axes, b$/$a, for the
final list of objects for the two Seyfert sub-classes.
The isophotal diameter ratios are gleaned from
\citet{deVetal91} and \citet{Lipovetskyetal88} catalogs.
We note that \citet{Pringleetal99,Schmittetal01} and \citet{NW99} have
shown that there is no correlation between the host galaxy rotation
axis and the direction of the radio jet.

\subsection{Criteria based on orientation-independent parameters}
\label{criteria_or_para}

We discussed in Section \ref{intro} that Roy et~al.'s (1994)
result is inconsistent with the unified scheme hypothesis
and the inconsistency is only made worse by 
invoking relativistic beaming. To rigorously
test the predictions of the unified scheme, the purportedly
face-on and edge-on Seyfert galaxies being compared should be
{\it intrinsically similar} within the framework of the
scheme. They should therefore be selected so that they
are matched in parameters that are orientation independent.
We attempt to do this with orientation-independent parameters
that are measures of intrinsic {\small AGN} power and host
galaxy properties.  Such a selection would enable us
to test the predictions of the unified scheme hypothesis
rigorously.
Therefore,
from the short list of objects that met the criteria
given in Sections \ref{bona_fide}, \ref{do_ability}, and \ref{min_obs},
we chose 10 Seyfert~1 and 10 Seyfert~2 galaxies,
i.e., two matched samples of Seyfert~1 and Seyfert~2 galaxies,
such that these two matched samples had similar distributions of
the following orientation-independent parameters.

\subsubsection{Heliocentric redshift}

In order to compare
the Seyfert~1 and Seyfert~2 galaxies 
from the same volume of space, we chose them to have similar
distribution of redshift. Figure~\ref{oursample}B shows the
distribution of redshift, $z$, for the two Seyfert sub-classes.

\subsubsection{Luminosity of the [O~{\small III}]~$\lambda$5007
emission line: Measure of intrinsic AGN power}

\begin{figure*}[ht]
\begin{tabular}{lll}
\includegraphics[width=5.78cm]{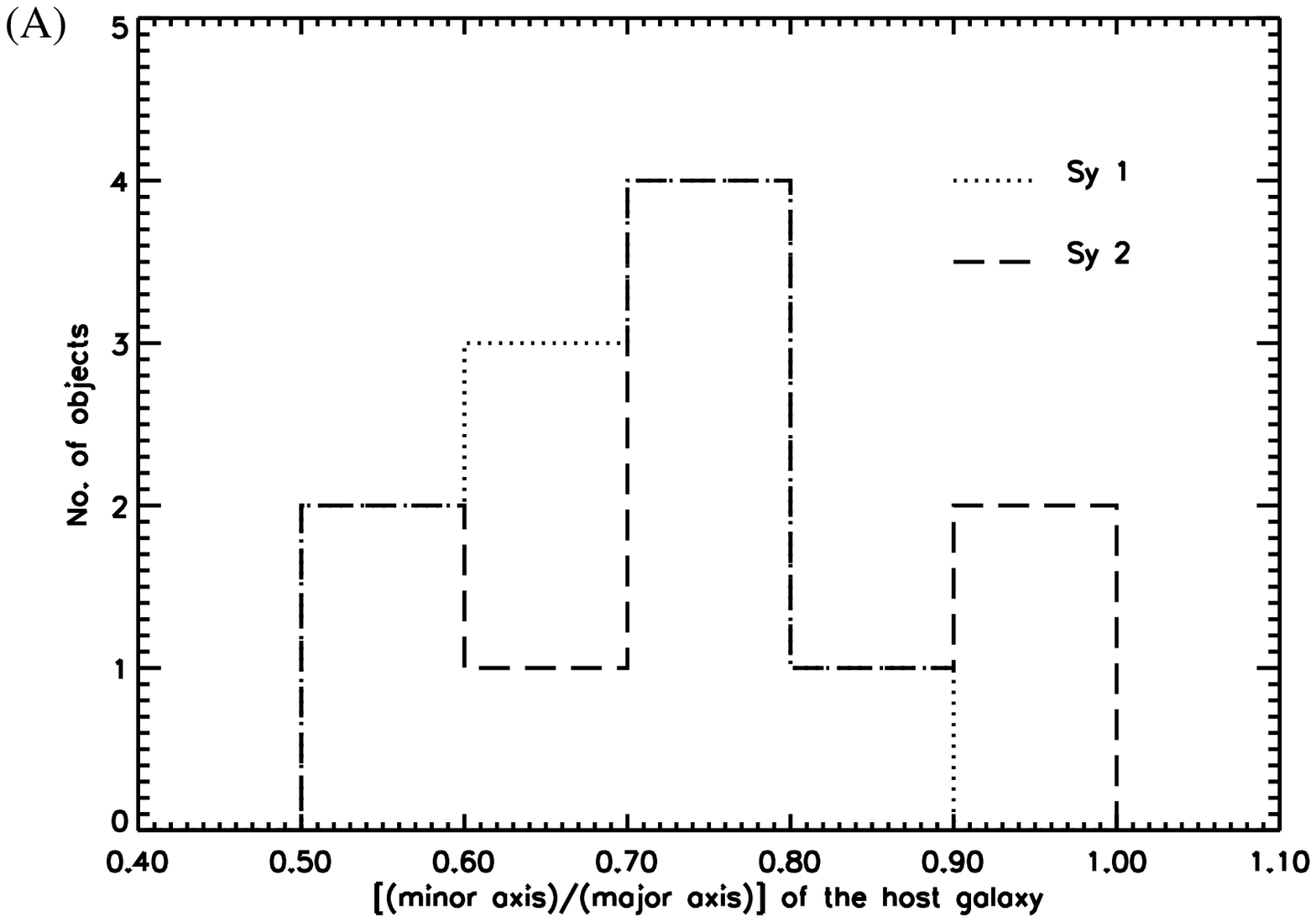} &
\includegraphics[width=5.78cm]{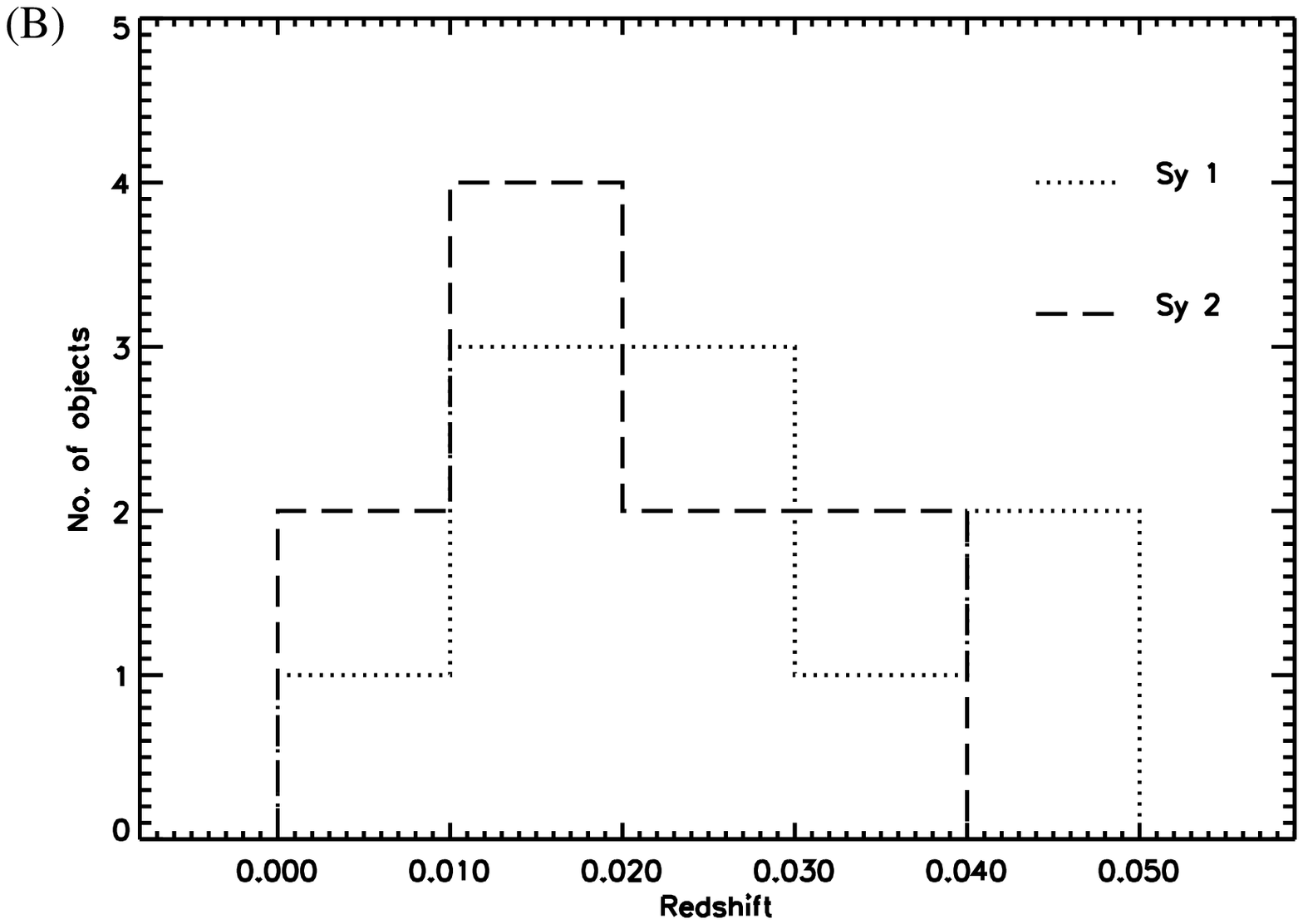} &
\includegraphics[width=5.78cm]{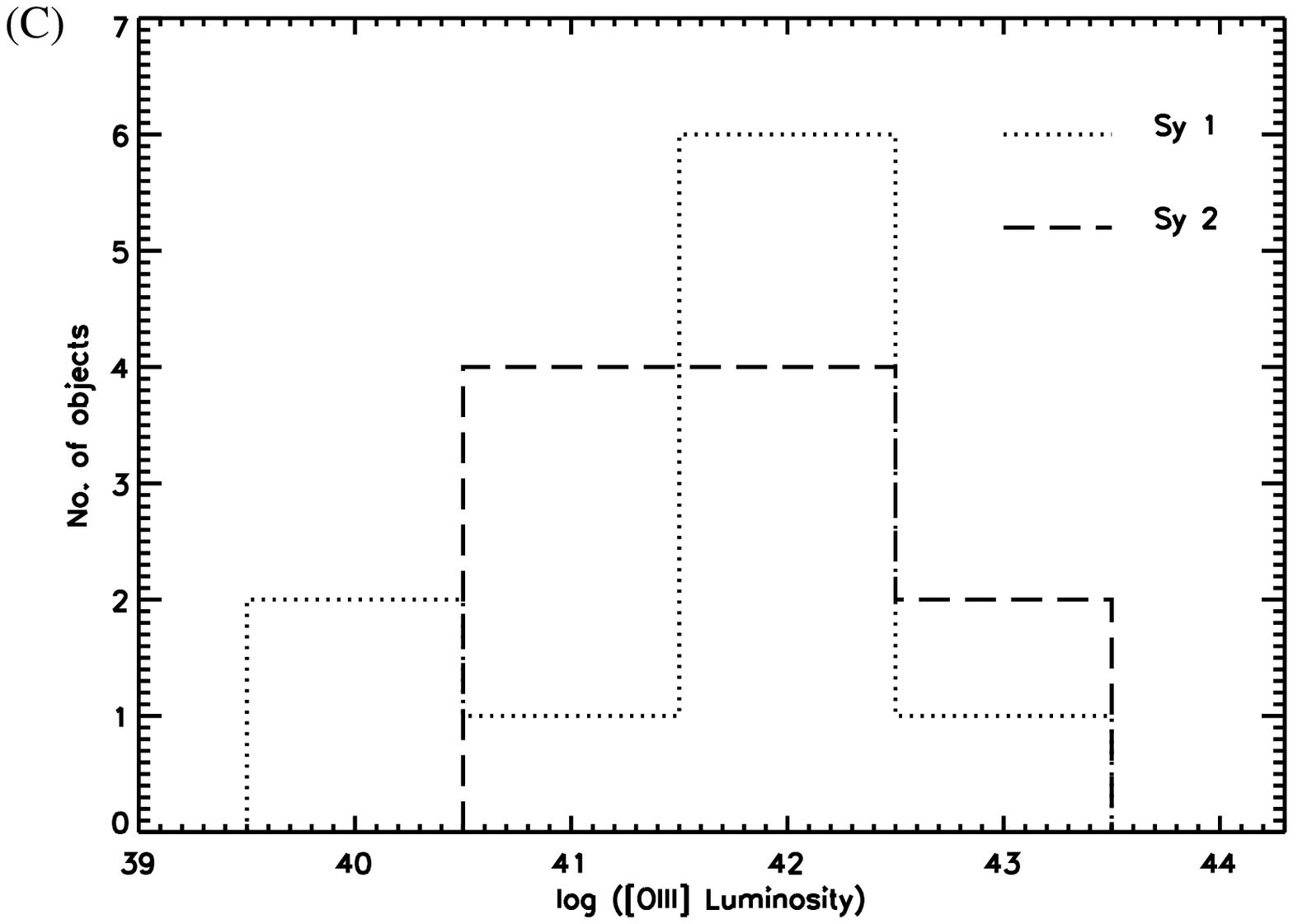} \\
\includegraphics[width=5.78cm]{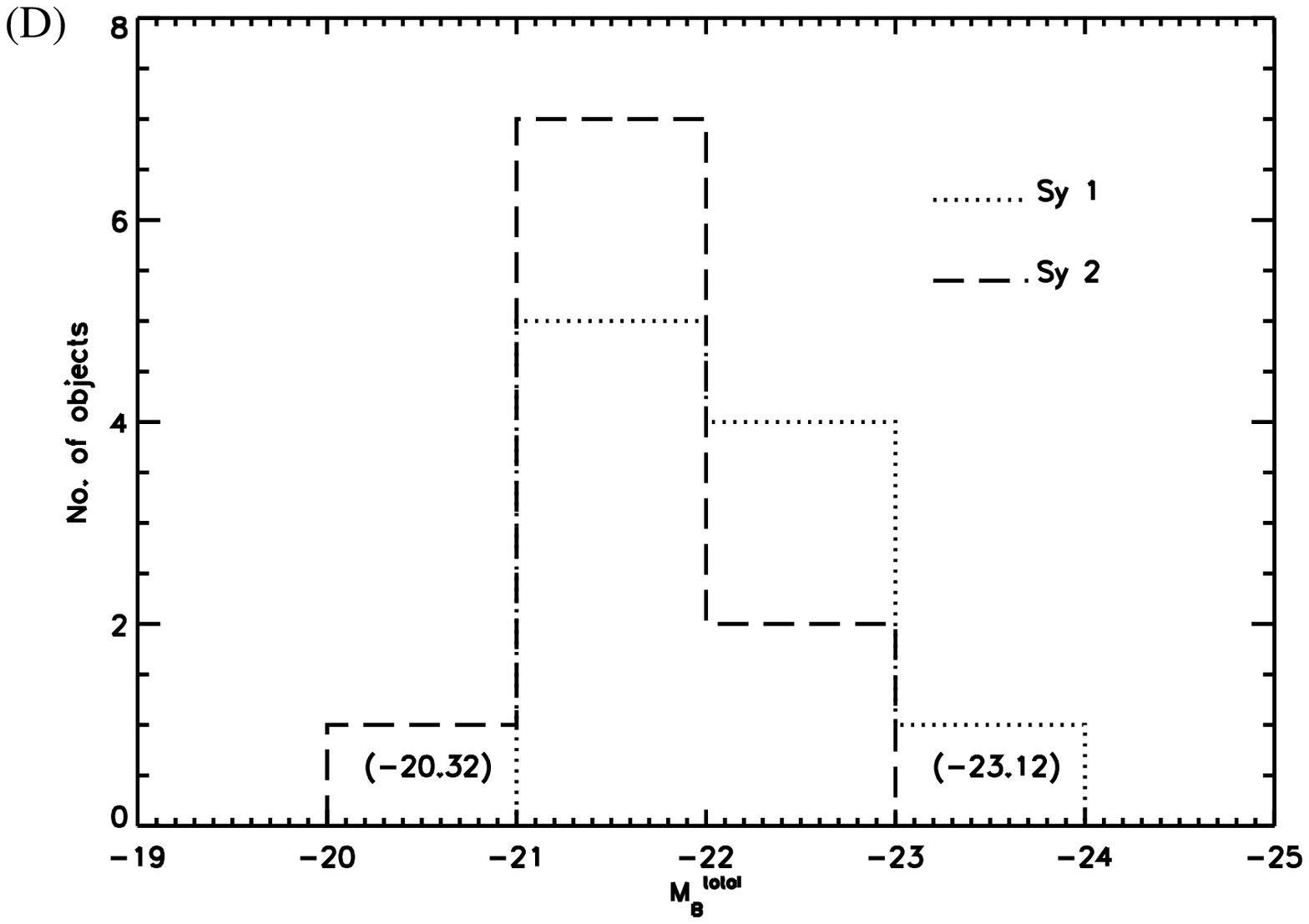} &
\includegraphics[width=5.78cm]{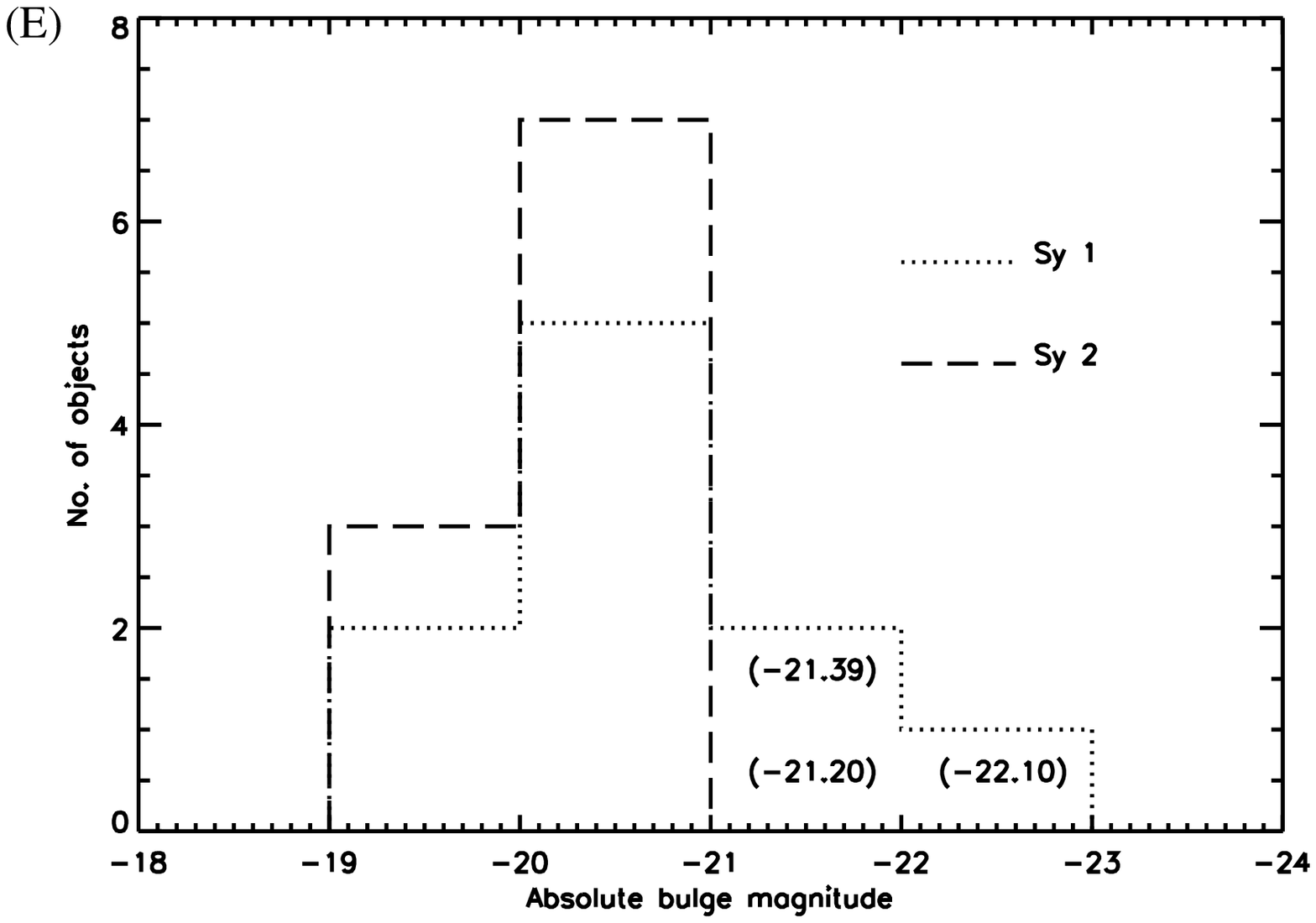} &
\includegraphics[width=5.78cm]{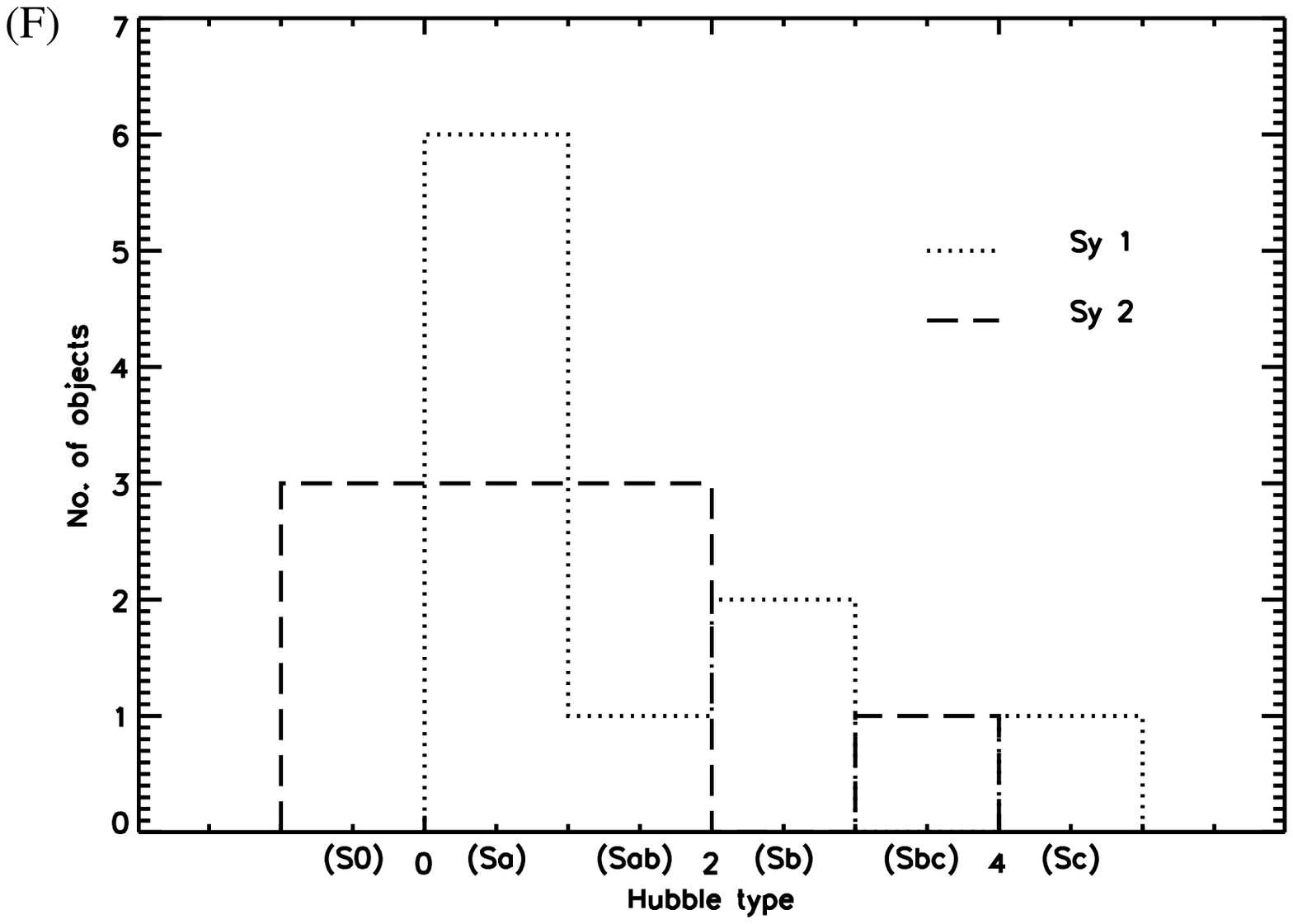} \\
\includegraphics[width=5.78cm]{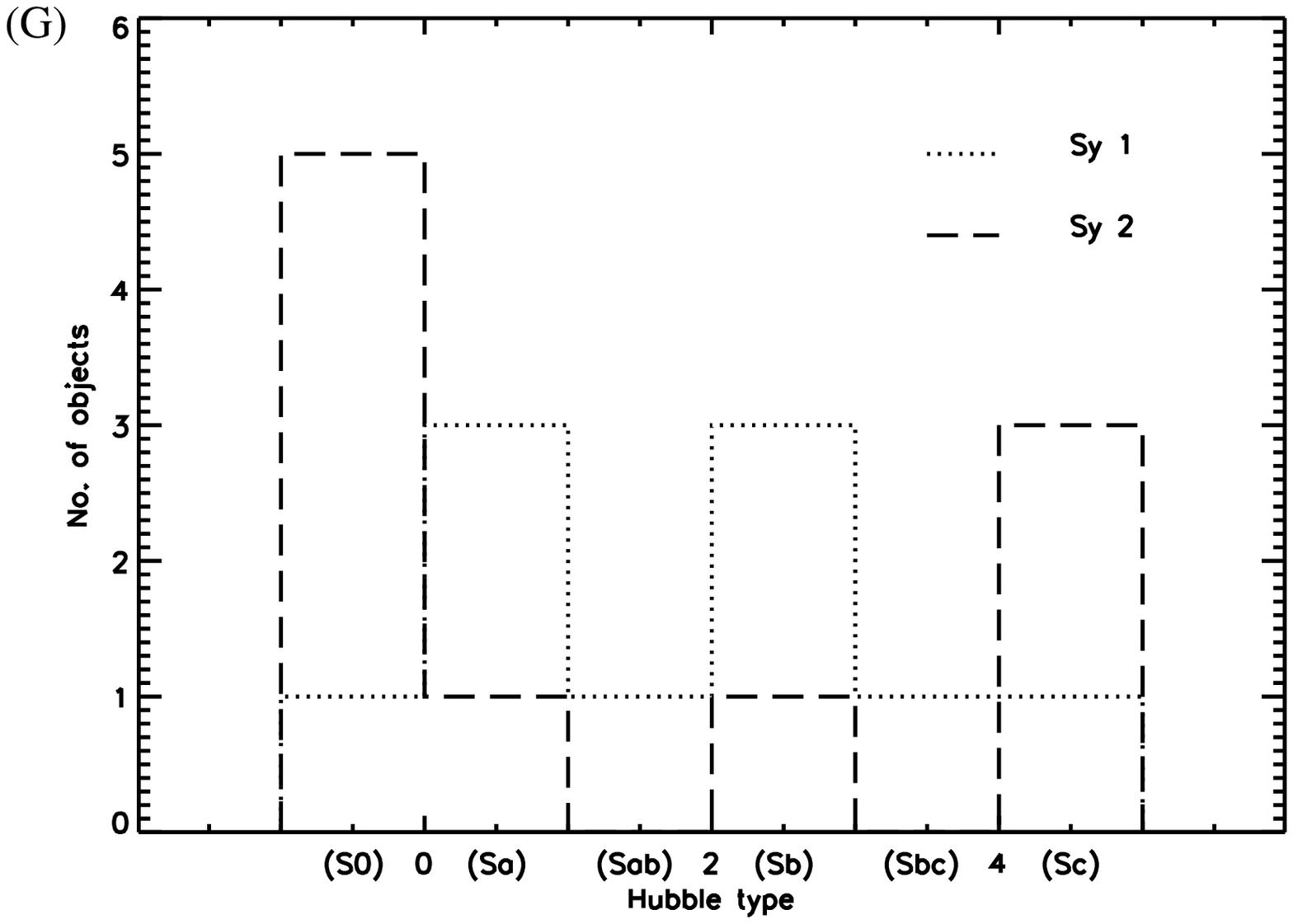} & &
\end{tabular}
\caption{{\small Histograms showing distributions of
ratio of the minor to the major isophotal diameter axes, b$/$a (A),
redshift, $z$ (B),
[O~{\small III}]~$\lambda$5007 luminosity (C),
total stellar absolute magnitude of the host galaxy, $M_B^{\rm total}$ (D;
values in the bin indicate $M_B^{\rm total}$ for the sources),
absolute magnitude of the bulge ($B$~band), $M_B^{\rm bulge}$ (E;
values in the bin indicate $M_B^{\rm bulge}$ for the sources),
the Hubble type, T \citep[F; mainly gleaned from RC3 catalog,
or][or Lipovetsky et~al. 1988]{Whittle92A}, and
the Hubble type, T (G; mainly gleaned from Malkan, Gorjian, \& Tam 1998,
or RC3 catalog, or Whittle 1992A, Lipovetsky et~al. 1988)
for the final list of objects for the two Seyfert sub-classes.}}
\label{oursample}
\end{figure*}

It is well known that narrow-line luminosities, {\it e.g.}, the luminosity
of the [O~{\small III}]~$\lambda$5007 emission line correlate strongly with
nuclear ionizing luminosity \citep{NW95,Whittle92B,Whittle92C,Yee80,Shuder81}.
Furthermore, although spatially
it could be distributed anisotropically \citep{Pogge89,Evansetal91A,
Evansetal91B} its luminosity is clearly orientation independent. We
therefore use the [O~{\small III}]~$\lambda$5007 luminosity
as a measure of the intrinsic {\small AGN} power \citep{NW95},
and we chose only those Seyfert~1 and Seyfert~2 galaxies that
had similar distribution of [O~{\small III}]~$\lambda$5007 luminosity.
Figure~\ref{oursample}C shows the distribution of [O~{\small III}]~$\lambda$5007
luminosity for the two Seyfert sub-classes.

\subsubsection{Stellar luminosity of the host galaxy}

The host galaxy properties of Seyfert galaxies ought to
be orientation independent within the unified scheme.
Hence, we chose stellar luminosity of the host galaxy (total luminosity
of the host corrected for the nuclear non-stellar and emission
line flux, redshift (K) correction, the internal absorption,
and the Galactic absorption)
as one of the parameters.
Most of our sample objects have stellar absolute magnitude of the
host galaxy, {\it i.e.}, M$_B^{\rm total}$ tabulated in \citet{Whittle92A}.
To determine corrections to total apparent magnitude,
B$_T$ (for Mrk~1218, NGC~2639, and
Mrk~231), for which M$_B^{\rm total}$ is not available in
\citet{Whittle92A}, we stick to the methodology of
\citet{Whittle92A}. Four factors, together
called ($\Delta$m)$^{\rm correction}$,
contribute to B$_T$, the first is the nuclear non-stellar and 
emission line flux, $\Delta$m$_A$, the second is the redshift (K)
correction, $\Delta$m$_K$, the third is the correction for internal
absorption, $\Delta$m$_i$, and the fourth is the
correction due to galactic absorption, $\Delta$m$_G$.
Thus
$$
B_T^c = B_T +  ({\Delta}m)^{\rm correction},
$$
where
$$
({\Delta}m)^{\rm correction} = {\Delta}m_A + {\Delta}m_K + {\Delta}m_i + {\Delta}m_G,
$$
which typically is $\lesssim$0.4 \citep{Whittle92A}.
The third reference catalogue of bright galaxies
\citep[RC3,][]{deVetal91} catalog gives B$_T^0$, {\it i.e.} the B$_T$
corrected for redshift (K) 
correction, the correction for internal absorption, and the
correction due to galactic absorption, {\it i.e.}
$$
B_T^0 = B_T + {\Delta}m_K + {\Delta}m_i + {\Delta}m_G.
$$ 
We then determine B$_T^c$, using
$$
B_T^c = B_T^0 + {\Delta}m_A.
$$
{$\Delta$}m$_A$, correction due to the non-stellar line 
and continuum emission is derived from the two line fluxes
F$_{5007}$ and F$_{H{\beta}}$, 
following the procedure described in \citet{Whittle92A}.
If only one of F$_{5007}$ or F$_{H{\beta}}$ is available,
as is the case for NGC~2639, 
the other is estimated using F$_{5007}/$F$_{H{\beta}}$~=~0.25 for
a Seyfert~1.0. 
As a function of redshift, $z$, the effective continuum flux
density, F$_c$,
$$
F_{c} = F_{cF} + F_{cH} + F_{cC},
$$
in ergs~s$^{-1}$~cm$^{-2}$~\AA$^{-1}$~ in the $B$~band due
to forbidden and Balmer
emission lines is approximated by
$$
F_{cF}~\simeq~(0.62 - 3.5z)~F_{5007}/980~~~~~~~~~(z \le 0.02),
$$
$$
F_{cF}~\simeq~0.55~F_{5007}/980~~~~~~~~~~~~~~(0.02 < z \le 0.07),
$$
$$
F_{cH}~\simeq~1.42~F_{H{\beta}}/980~~~~~~~~~~~~~~~~(z \le 0.03),
$$
$$
F_{cH}~\simeq~(1.51 - 3.14z)~F_{H{\beta}}/980~~~~~~(0.03 < z \le 0.07).
$$
The non-stellar continuum luminosity is derived from ${H{\beta}}$,
assuming a power law with spectral index $\alpha$ 
(F$_{\nu}$~$\propto$~${\nu}^{\alpha}$). The effective
continuum flux density in the $B$~Band is given by
$$
F_{cC}~\simeq~1.10^{({\alpha} + 2)}~(1 + z)^{({\alpha} + 2)}~F_{H{\beta}}/100.
$$ 
We adopt $\alpha$~=~$-$1.0 for the non-stellar continuum since
this corresponds to the canonical nuclear colors of Seyfert~1 galaxies
($\alpha$~=~$-$1.0 is equivalent to $U$$-$$B$~=~$-$0.75,
$B$$-$$V$~=~$+$0.41). The total non-stellar contribution
gives
$$
{\Delta}m_A = 2.5~log(\frac{7.2 \times 10^{-9}}{F_{c}}),
$$
which is subtracted from B$_T^0$ to give B$_T^c$
for Mrk~1218, NGC~2639, and Mrk~231.
M$_B^{\rm total}$ of the host galaxy is then determined using
$$
M_B^{\rm total}~=~B_T^c~-~5~log_{10}\left({\frac{r}{10~(pc)}}\right),
$$
where $r$ is the distance to the object in parsec.
Figure \ref{oursample}D shows the distribution of the total
stellar absolute magnitude, $M_B^{\rm total}$, for the two Seyfert sub-classes.

\subsubsection{Absolute bulge luminosity of the host galaxy}

\citet{Whittle92A} has argued that the nuclear stellar velocity
dispersion is a measure of the depth of the gravitational
potential within a scale of $\sim$~3~kpc.
Further, dispersion velocity of stars correlates with the
absolute bulge magnitude \citep{Whittle92A,NW95}.
We took this depth of gravitational potential, {\it i.e.} the absolute
magnitude of the bulge M$_B^{\rm bulge}$, to be an indicator of
intrinsic {\small AGN} power.  Once again,
the determination of M$_B^{\rm bulge}$ was performed in the following manner:
M$_B^{\rm bulge}$ is available for most of the
objects in \citet{Whittle92A}. For Mrk~1218, NGC~2639, Mrk~231,
and Mrk~477, we use the formulation adopted by \citet{Whittle92A}, {\it i.e.}
$$
M_B^{\rm bulge} = M_B^{\rm total} - ({\Delta}m)_{\rm disk},
$$
where
$$
({\Delta}m)_{\rm disk} = 0.324{\tau} - 0.054{\tau}^2 + 0.0047{\tau}^3,
$$
where $\tau$ = T~$+$~5, and T is the Hubble type \citep{Sandage75}.
Figure~\ref{oursample}E shows the distribution of the absolute bulge luminosity,
$M_B^{\rm bulge}$, of the host galaxy for the two Seyfert sub-classes.
We therefore tried to ensure that our two sub-samples did not
differ significantly in the distribution of this parameter.

\subsubsection{Hubble type of the host galaxy}

The Hubble type mainly depends on the size of the nuclear bulge
relative to the flattened disk \citep{Sandage75}.
\citet{MGT98} have argued that
the Seyfert~1 galaxies are of earlier Hubble type than
Seyfert~2 galaxies. We therefore have considered the
distribution of Hubble type of the host galaxy for
our sample of Seyfert~1 and Seyfert~2 galaxies.
We use the Hubble type given in the RC3 catalog
\citep{deVetal91} for our sample
sources, and when not available, we use values from \citet{Whittle92A}
or \citet{Lipovetskyetal88}. The Hubble type of one of the objects,
Mrk~1218, which was unavailable in \citet{deVetal91}, \citet{Whittle92A},
and \citet{Lipovetskyetal88}, is taken from \citet{MGT98}.
The distributions are shown in Figure \ref{oursample}F and are statistically
not significantly different for the two Seyfert sub-classes. 
We have thus controlled for the Hubble type in our sample.
Since, the morphological class given by \citet{MGT98} are based
on WFPC2, \textit{$Hubble~Space~Telescope$} images, we also in
Figure~\ref{oursample}G show
the distribution of the Hubble type for the two Seyfert sub-samples
where the Hubble types are preferentially gleaned from \citet{MGT98},
6 out of 10 Seyfert~1 and 7 out of 10 Seyfert~2 galaxies,
and then from RC3 catalog or \citet{Whittle92A} or \citet{Lipovetskyetal88}.

Thus, of the 126 Seyfert galaxies that had {\small VLA} data
in the literature, 54 met our feasibility criterion.
Twenty nine of these 54 had all the required parameters
in the literature. From these 29 we could pick 20 Seyferts
that met our selection criteria and were matched in the
orientation-independent parameters.

\input{tab2.tex}

\subsection{Our Seyfert sample}

Our aim was to study the pc-scale radio morphology of 
Seyfert galaxies so as to test the predictions of the unified
scheme hypothesis.
By matching Seyfert~1 and Seyfert~2 galaxies in the above
parameters, particularly [O~{\small III}]~$\lambda$5007
luminosity (an indicator
of intrinsic {\small AGN} power) and stellar luminosity of the host
galaxy, we ensured that the samples of Seyfert~1
and Seyfert~2 galaxies were intrinsically similar within the
framework of unified scheme.
Table~\ref{all_para} lists our sample of Seyfert 
galaxies giving all the orientation-independent parameters that were used to
construct it, {\it viz.}, 
ratio of minor to major axis of the host galaxy, the radio
flux density of the compact component, cosmological redshift, 
[O~{\small III}]~$\lambda$5007 emission line width,
the ratio of emission line intensities of fluxes in
[O~{\small III}]~$\lambda$5007 to H$\beta$,
luminosity of the [O~{\small III}]~$\lambda$5007 emission line,
stellar luminosity of the host galaxy,
bulge absolute luminosity of the host galaxy, and Hubble type for the two Seyfert
sub-classes.

The fact that we avoided edge-on host galaxies results in a
selection of samples of Seyfert~1 and Seyfert~2 galaxies
differing in one intrinsic respect.
Thus it follows that our sample has a paucity of
Seyfert~1 galaxies with their radio-jets in the plane of the host galaxy
and similarly there is a paucity of Seyfert~2 galaxies
with their jet-axis perpendicular to the plane of the host galaxy.
If the gaseous interstellar medium (ISM) of the host galaxy has an effect on the
propagation of the jet through it, then it implies that there
is a physical difference in this respect between our Seyfert~1
and Seyfert~2 galaxies.
However this is unlikely, since many studies indicate that the orientation
of the {\small AGN} axis (and therefore the radio-jet) relative to the
host galaxy axis is random \citep{Schmittetal01,Pringleetal99,NW99},
and radio-jets on all angular/linear scales are seen in all orientations.

In \citet{LSG04}, we presented the radio images that were obtained 
from our observations (project code GL022,
date of observations February 18, 1998) and described the properties
at these scales in the light of past observations.
The data from these observations along with published radio data for the
unobserved (well studied) sources, {\it viz.} Mrk~348, NGC~4151, Mrk~231
and Mrk~926 at both pc-scale and kpc-scales and NGC~5135 at kpc-scales
are used to discuss the impact on the unified scheme hypothesis.
Table \ref{vla_vlbi} lists the derived parameters (our and published radio
data) for our Seyfert galaxy sample; the sequence of sources is ordered in
right ascension.  

\input{tab3.tex}

\section{Interpretation}
\label {royetal}

{\bf The contradiction with the unification scheme and its resolution:}
\citet{Royetal94} observed
far-infrared selected, mid-infrared selected
and optically selected samples of Seyfert galaxies with a
275~km long, single-baseline, the Parkes-Tidbinbilla Interferometer
\citep[PTI,][]{Norrisetal88B,Norrisetal92B} at 1.7 and 2.3~GHz.
They reported that
compact radio structures are much more common in Seyfert~2 than
in Seyfert~1 galaxies in the far-infrared selected samples, as well as
in the combined mid-infrared and optically selected sample.
They deduced this result based on significantly different
detection rate of compact, high brightness temperature radio
structure.  Their surprising result that far-infrared selected Seyfert~1 
galaxies were detected less frequently
than were Seyfert~2 galaxies is inconsistent
with the standard unification scheme. The unified scheme would
predict an equal fraction of detections for Seyfert~1 and
Seyfert~2 galaxies. It is also the opposite sense of what is expected from
alternative models  in which Seyfert~1 galaxies are expected to
have more energetic cores than Seyfert~2 galaxies.
Also, even if the jets were relativistically beamed, we would 
expect Seyfert~1 galaxies to show systematically more prominent compact 
radio emission than Seyfert~2 galaxies, since they are the face-on 
objects, and it is their jets that would be pointed towards us
and therefore Doppler beamed.

However, \citet{Royetal94} invoked a model which attempts to
reconcile their result with the unification 
scheme. They considered the radio optical depth due to the free-free 
absorption of the {\small NLR} clouds which surround the radio emitting 
regions of the core and {\small NLR}, which is in line with the model
first proposed by \citet{Norrisetal92A}. There could be two 
distinct mechanisms; (i) obscuration by the {\small NLR} and (ii)
obscuration by individual {\small NLR} clouds that may contribute 
to the resulting radio appearance. These two mechanisms invoke
free-free absorption by the {\small NLR} clouds, which is highly 
dependent on the geometry of the {\small NLR}, the opening angle of the cone
(anisotropic escape of photons is in the form of a cone),
and the filling factor.
In contrast to Roy et~al.'s (1994) result,
all 19 of our 20 sample sources for which {\small VLBI} observations
are available have compact features.
We do not find any systematically different detection rate of
compact structures. We thus find that Seyfert~1 and Seyfert~2
galaxies have an equal tendency to show compact radio
structures, in contrast to the results of \citet{Royetal94}.
Although we chose sources with core flux density $>$~8~mJy for the
feasibility requirements, our result is thus consistent with
the prediction of the simple unified scheme for such a Seyfert sample.
Figure \ref{masradio}A shows the distribution of brightness temperatures,
T$_{\rm b}$, of the brightest component detected at 5~GHz on mas-scales and
Figure \ref{masradio}B shows the distribution of the ratio of flux density detected on
mas-scales, to that detected on arcsec-scales. 
Mann-Whitney~U test shows that the two distributions
(Figures~\ref{masradio}A and \ref{masradio}B) are same at a significance level of 0.10.
These distributions show that the brightness temperatures of the brightest
component detected on mas-scales and the fraction of total flux density
detected on mas-scales are not different for the two groups of Seyfert galaxies.
Note that our pc-scale and kpc-scale data are simultaneous, and hence, the
statistically similar ratio of flux density detected on mas-scales to that
detected on arcsec-scales for  the two Seyfert sub-classes is not affected
by possible radio variability in the compact radio flux densities in 15 out
of 20 cases.  This demonstrates that the fact that our detection
rate is higher than Roy et~al.'s (1994) is not just due to
the higher sensitivity of the interferometer we used. 

\subsection{Pc-scale radio luminosities}
\label{pc_us}

We derive the observed radio luminosity
from the observed flux density S$_{\nu}$
via
$$
 L_{\nu}~=~4{\pi}~S_{\nu}~(d)^2~(1 + z),
$$
$$
{\rm where}~~d = {\frac {cz~(1 + ({z}/{2}))}{H_0~(1 + z)}},~{\rm and~we~use}
$$
$$
 H_{0}~=~75~{\rm kms}^{-1}~{\rm Mpc}^{-1}~{\rm and}~~q_0~=~0.
$$
The values are given in Table \ref{vla_vlbi} and
Figure~\ref{masradio}C shows the distribution of the radio luminosity
detected on pc-scales for the two Seyfert sub-classes.
The distribution shows that the 
pc-scale radio luminosities of the Seyfert~1
and the Seyfert~2 galaxies are similar.
The Mann-Whitney~U test shows that
the distributions are statistically indistinguishable at a
significance level of 0.05.
Thus the distribution of pc-scale radio luminosities of 
Seyfert~1 and Seyfert~2 galaxies are consistent
with the unified scheme hypothesis.

\subsection{Kpc-scale radio luminosities}
\label{arcsec_vla}

\begin{figure*}[ht]
\begin{tabular}{lll}
\includegraphics[width=5.78cm]{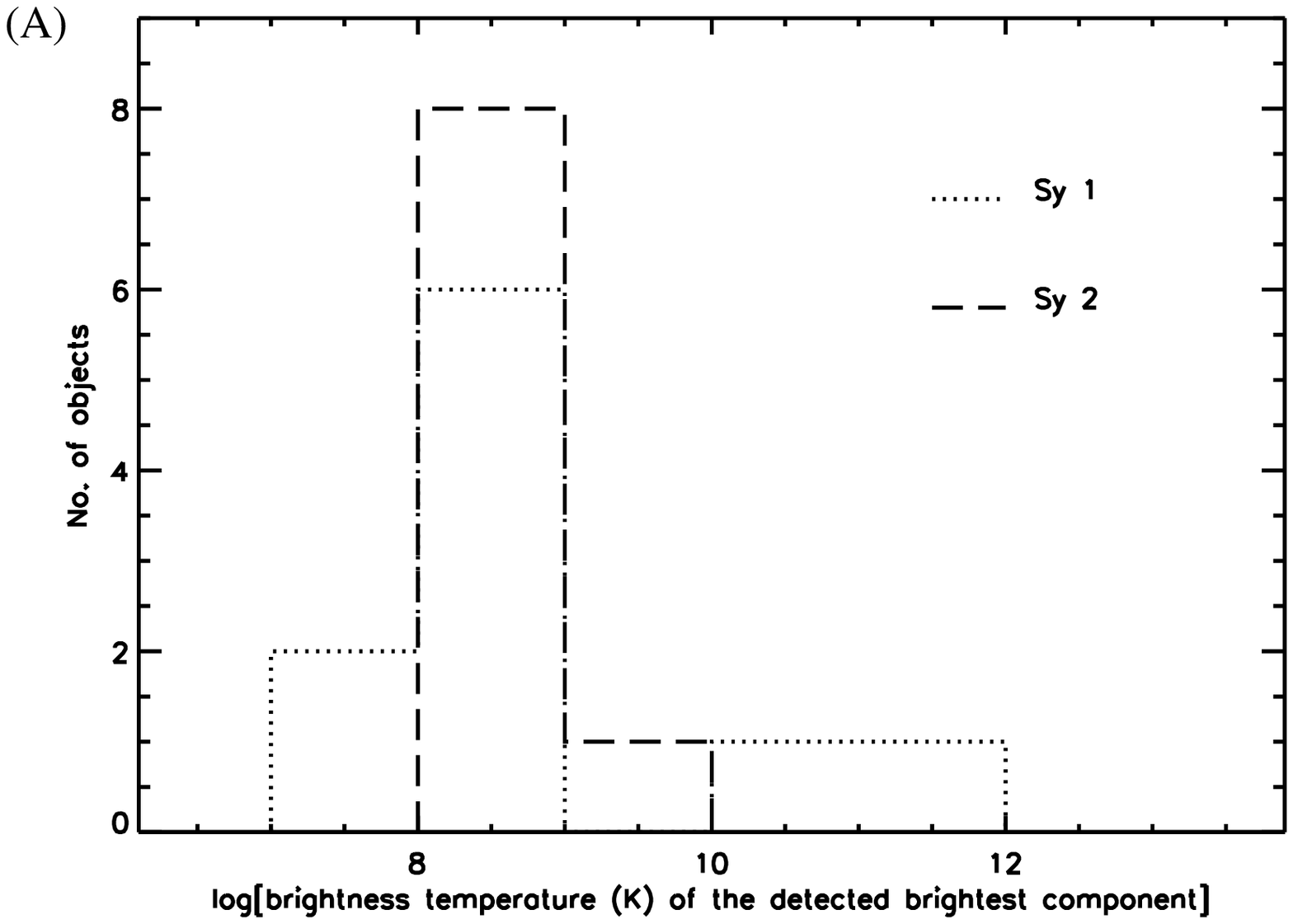} &
\includegraphics[width=5.78cm]{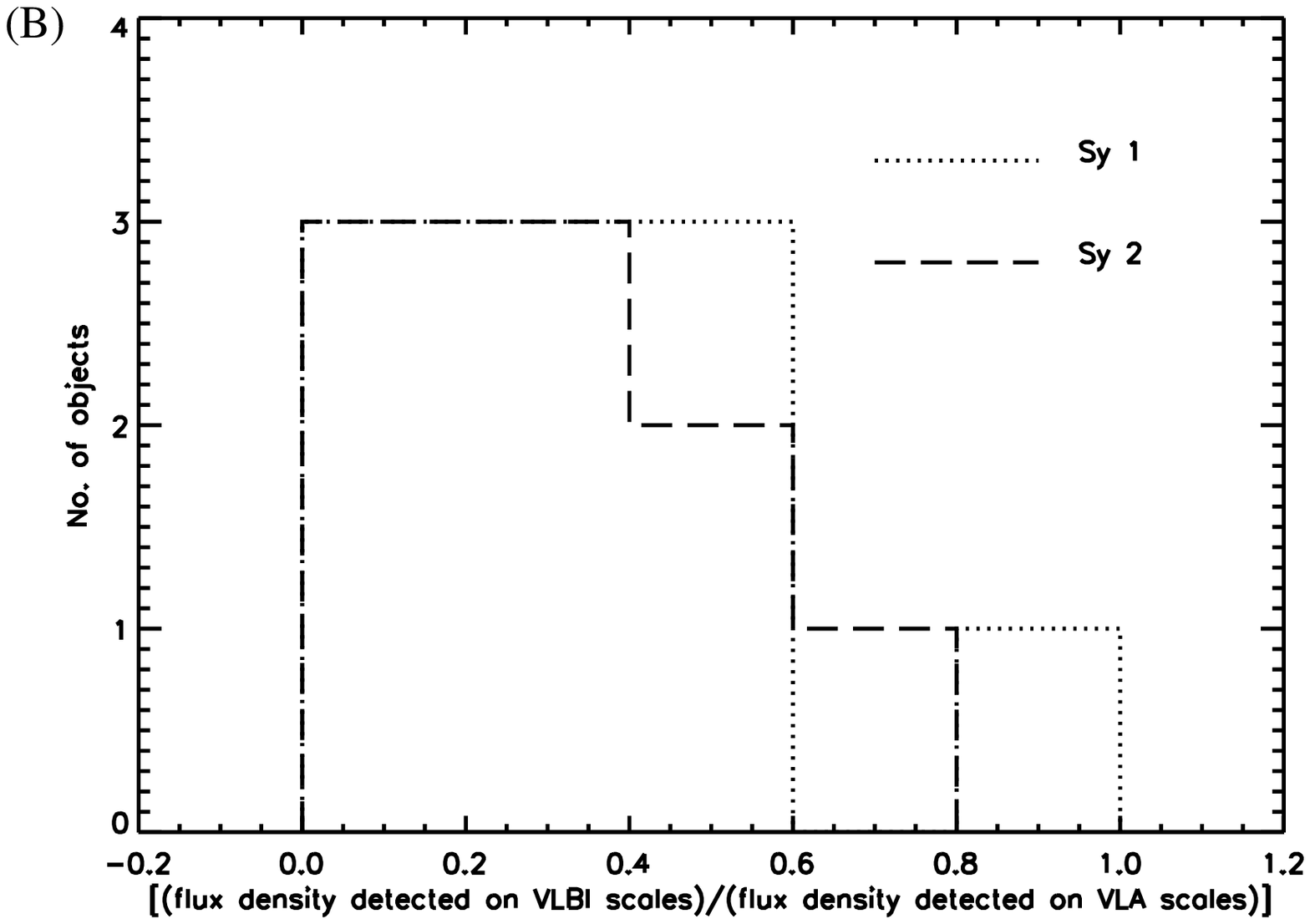} &
\includegraphics[width=5.78cm]{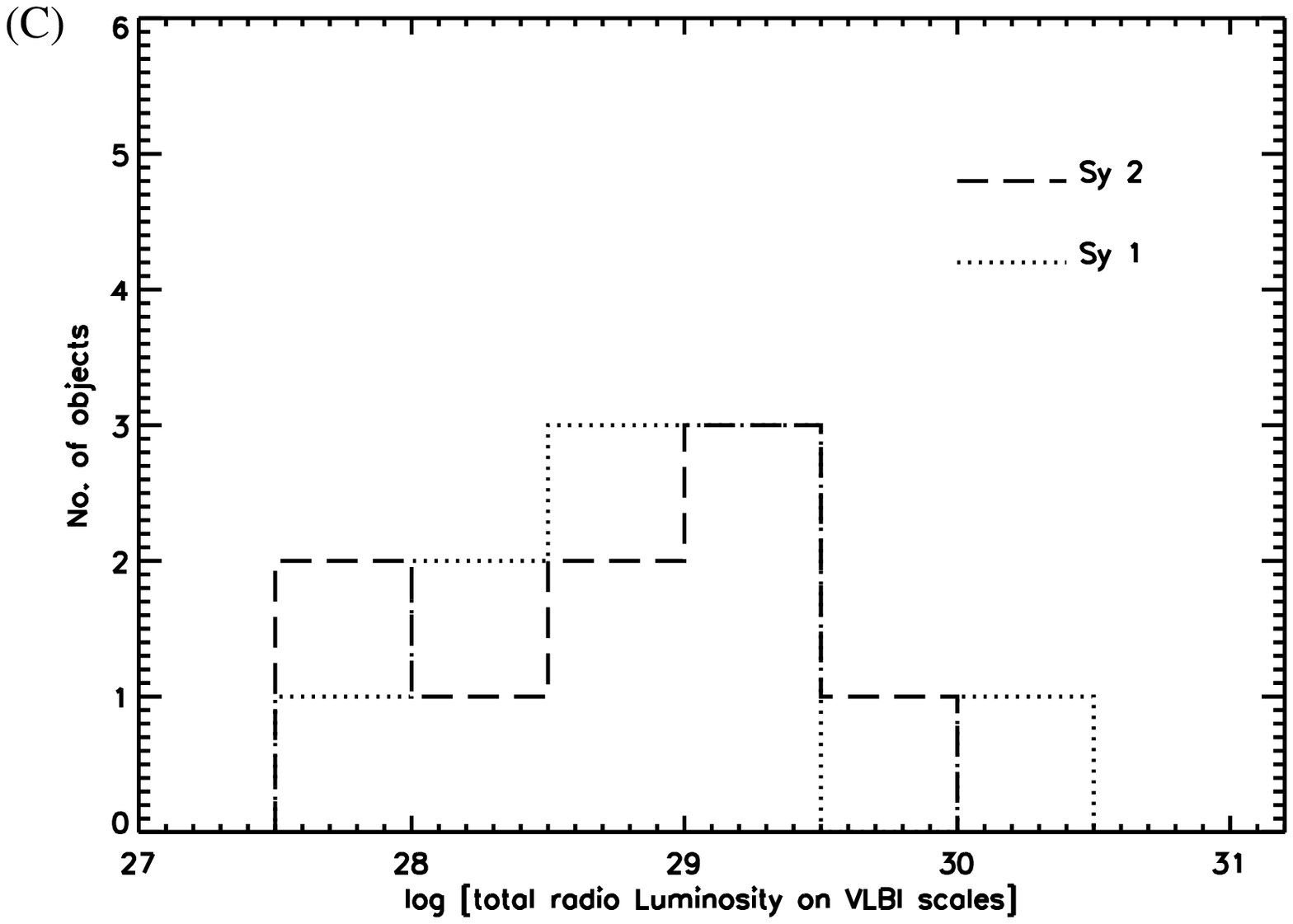} \\
\includegraphics[width=5.78cm]{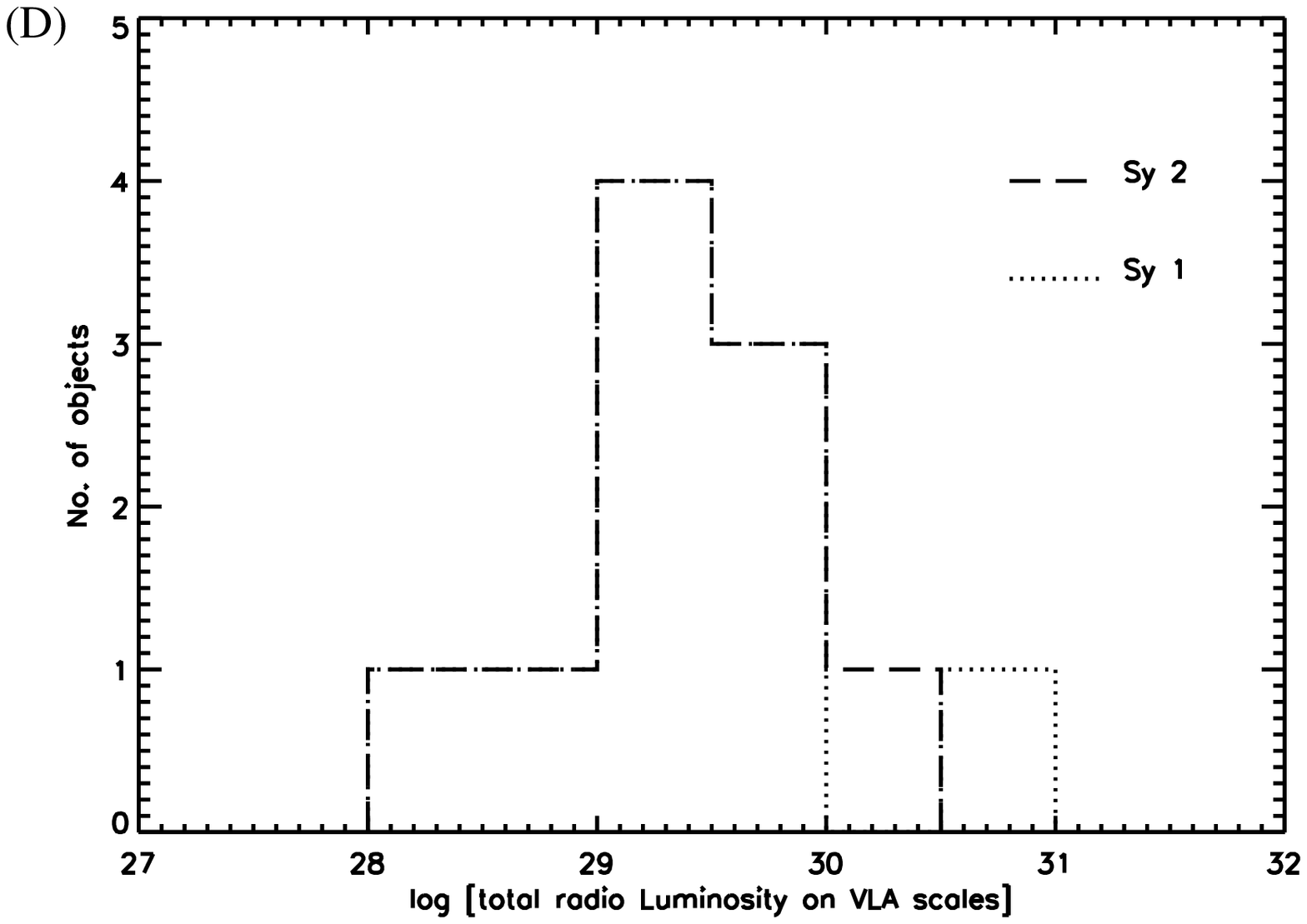} &
\includegraphics[width=5.78cm]{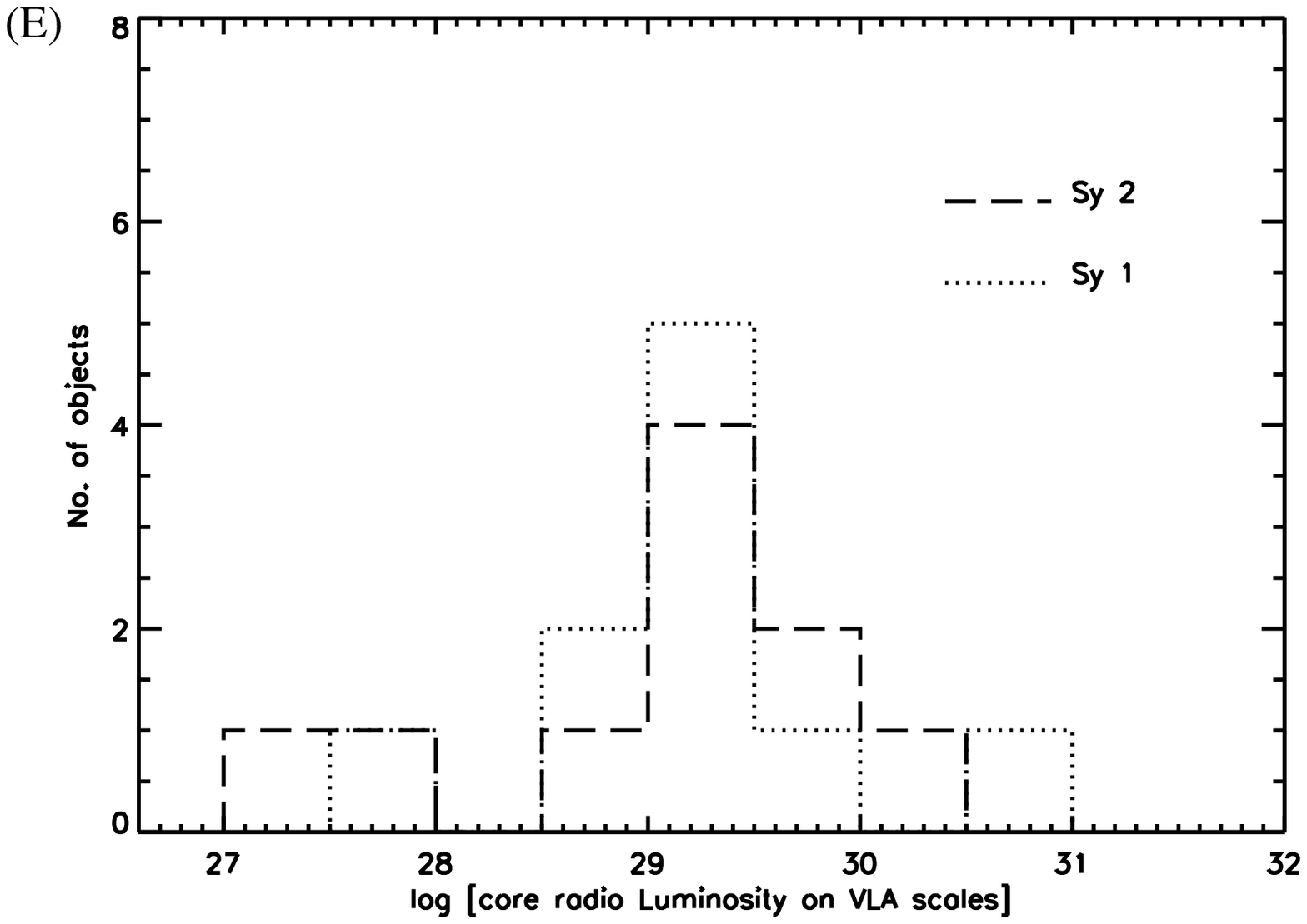} &
\includegraphics[width=5.78cm]{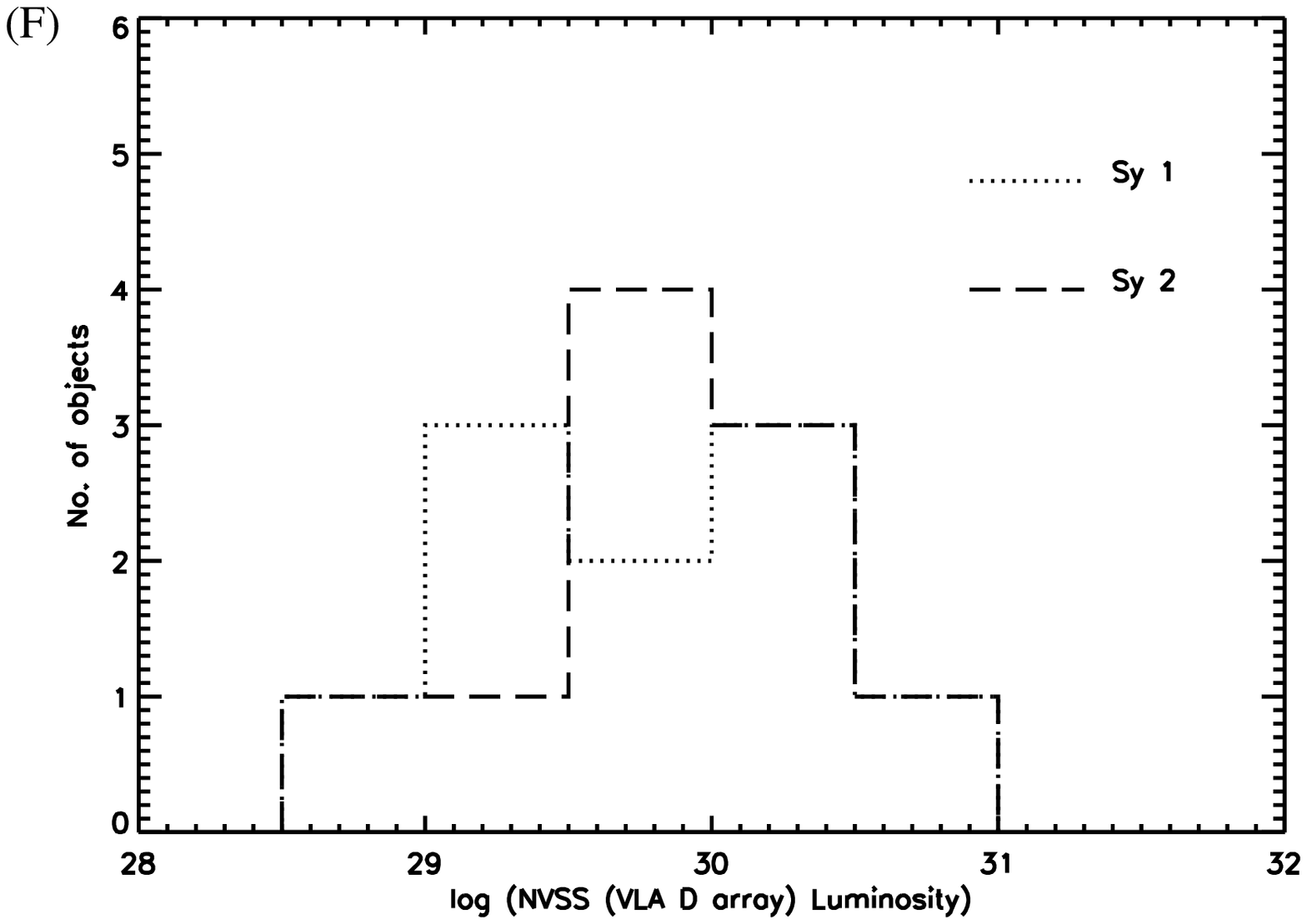} \\
\includegraphics[width=5.78cm]{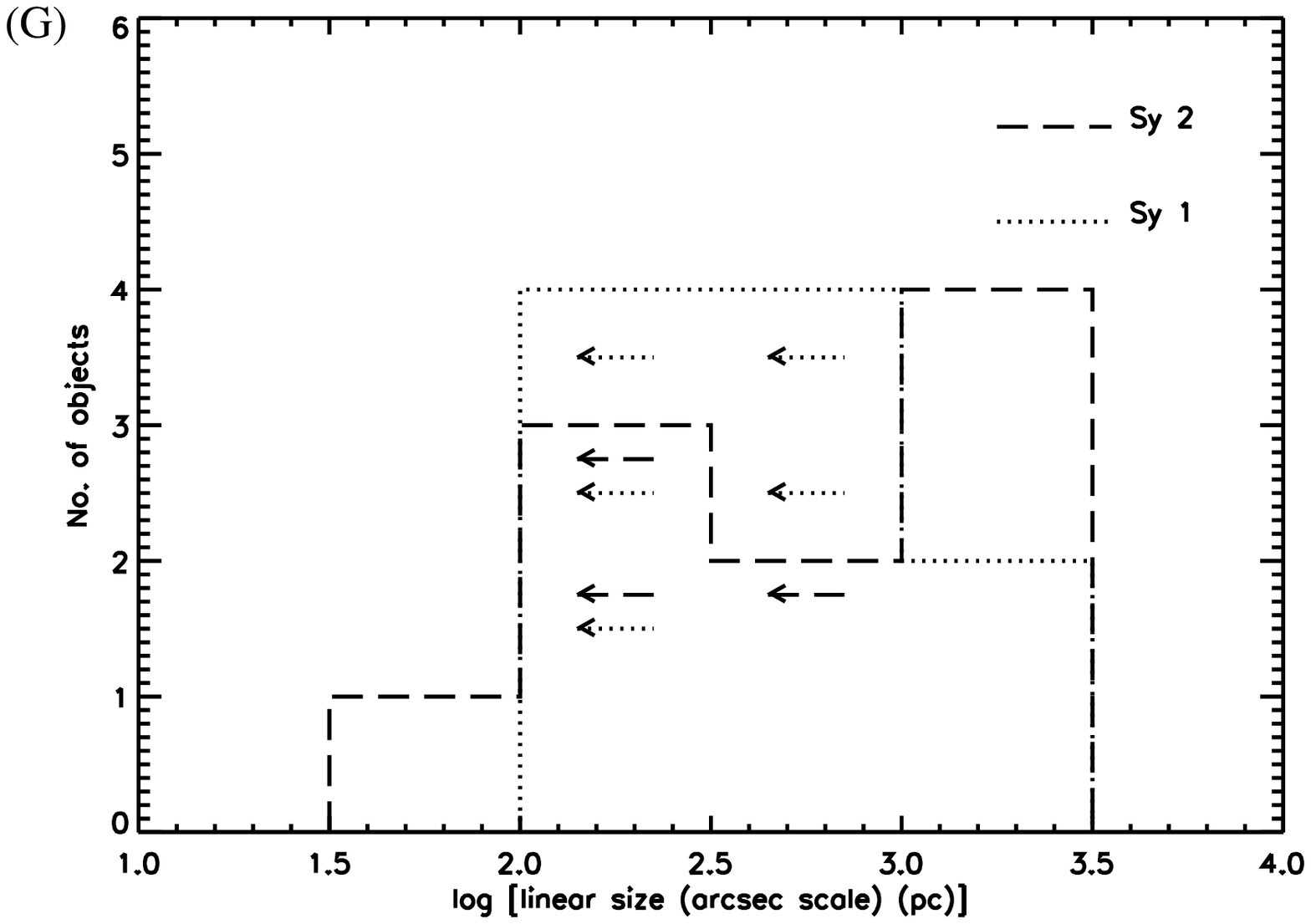} &
\includegraphics[width=5.78cm]{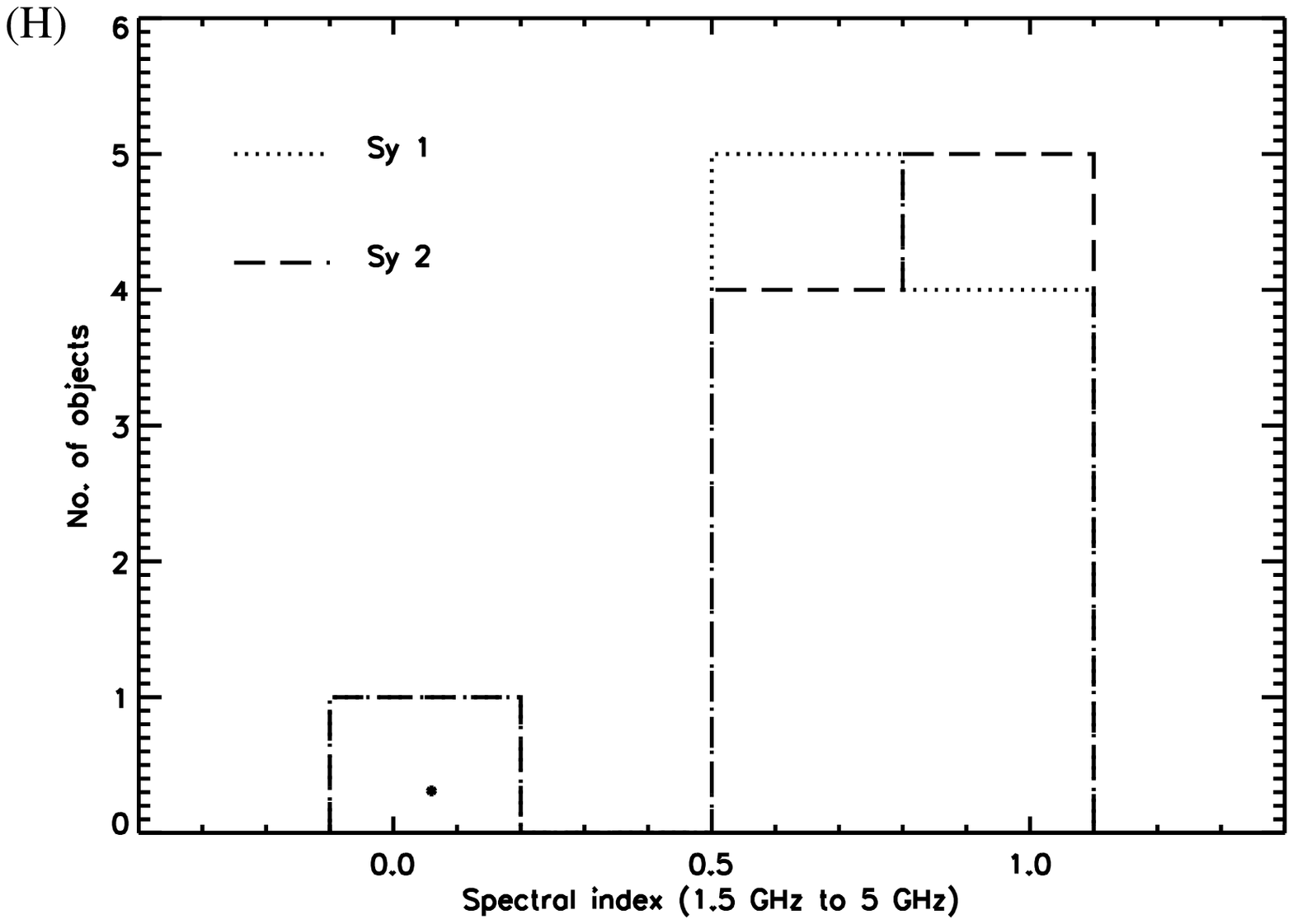} &
\includegraphics[width=5.78cm]{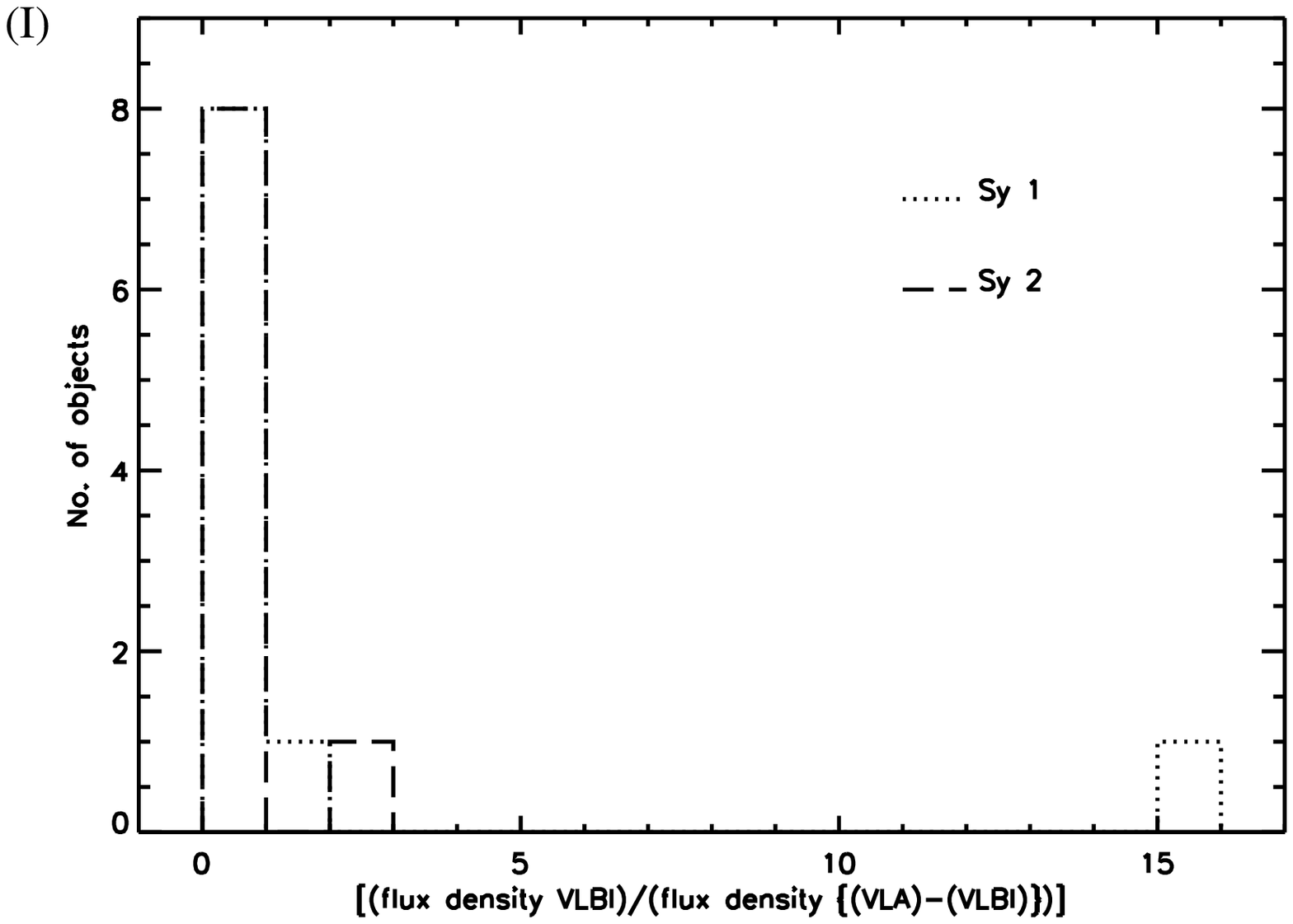}
\end{tabular}
\caption{{\small Histograms showing distributions of brightness temperature,
T$_{\rm b}$, of the brightest component detected on mas-scales (A),
ratio of the flux density detected on
mas-scales to the flux density detected on arcsec-scales (B),
total radio luminosity (ergs~s$^{-1}$Hz$^{-1}$) on mas-scales (C),
total radio luminosity (ergs~s$^{-1}$Hz$^{-1}$) on arcsec-scales (D),
core radio luminosity (ergs~s$^{-1}$Hz$^{-1}$) on arcsec-scales (E),
total radio luminosity (ergs~s$^{-1}$Hz$^{-1}$) on arcmin-scales at 1.4~GHz (F),
projected radio linear size (pc) on arcsec-scales;
arrows denote objects that are unresolved (G),
spectral indices between 1.5 and 5~GHz;
the spectral index between 1.5 and 8.4~GHz is
used for Mrk~348, denoted by asterisk (H), and
the fraction of the total radio flux
density detected on mas-scales as against the extended radio
emission, $R$ (I).}}
\label{masradio}
\end{figure*}

Figure \ref{masradio}D shows the distribution of the radio luminosity
detected on kpc-scales for the two Seyfert sub-classes.
The Mann-Whitney~U test shows that there is no statistically significant
difference in the distribution
at a significance level of 0.10.
The Mann-Whitney~U test also shows that there is no statistically
significant difference in the distribution of core (the component
of the source which is the closest to the optical nucleus within
errorbars) radio luminosity (Figure \ref{masradio}E)
on arcsec-scales for the two Seyfert sub-classes
at a significance level of 0.10.
Our results are thus consistent with the predictions of 
the unification scheme hypothesis and are also consistent
with the results of \citet{Nagaretal99}
but are inconsistent with \citet{Morgantietal99}, who
found that Seyfert~2 galaxies tend to be more luminous
than Seyfert~1 galaxies at marginal significance.
The {\small NRAO} {\small VLA} Sky Survey
\citep[{\small NVSS},][]{Condonetal98} radio 
observations are made at 1.4~GHz using {\small VLA}~$D$
configuration. All our sample objects have measurements
made using this instrument.
Figure~\ref{masradio}F shows that the two histograms of
Seyfert~1 and Seyfert~2 galaxies are statistically 
indistinguishable for total 1.4~GHz radio luminosity on arcmin-scales
at a significance level of 0.05 using Mann-Whitney~U test, which
is again consistent with the unified scheme hypothesis.

\subsection{Projected linear sizes}
\label{siz_eqn}

The linear sizes of Seyfert galaxies can be used to test the 
unification scheme because Seyfert~1 galaxies oriented at 
small angles to the line of sight should have systematically 
smaller linear sizes than Seyfert~2 galaxies in the plane of the sky.
For their distance limited sample of Seyfert galaxies, \citet{Morgantietal99}
found Seyfert~1 to be of systematically smaller sizes
than Seyfert~2 galaxies. This is consistent with the prediction
of the unified scheme.
We determine the projected linear size of the source using
$$
{\rm Linear~size}= 
({\rm angular~size})~\times~\left(\frac{(cz)(1+(z/2))}{H_{0}}\right);
$$
for~$q_0$~=~0.
Here, $c$ is the velocity of light, $H_0$ is the Hubble parameter and
$z$ is the redshift of the object.
We assume that the largest angular size of the source is the 
largest extent of the contour that is 5\% of the peak surface brightness
level, which is well above the noise level and use it to determine
the linear radio size of the source.
For Mrk 348 \citep{Ungeretal84}, NGC~4151, Mrk~231 \citep{Kukulaetal95},
NGC~5135 \citep{UW89}, and Mrk~926 \citep{UW84A}, we measure their
corresponding largest angular sizes directly from the published maps.
Table \ref{vla_vlbi} gives the projected
linear size of the source along with the reference, and
Figure \ref{masradio}G shows the distribution of projected 
linear size for our sample of Seyfert~1 and Seyfert~2 galaxies.
The Mann-Whitney~U test shows that the distributions are not 
significantly different at a significance level of 0.10.
It thus appears that the intrinsic variation in the projected
linear sizes is rather large and may swamp any systematic difference
between Seyfert~1 and Seyfert~2 galaxies due to projection.
Further, if the ISM of the host
galaxy affects the propagation of a Seyfert galaxy radio-jet, the
fact that our sample has a paucity of Seyfert~2 galaxies with
jet direction perpendicular to the plane of the host galaxy and
Seyfert~1 galaxies with their jets propagating in the plane
of the host galaxy disk, may also contribute to reduce the systematic
differences in projected linear sizes between Seyfert~1 and
Seyfert~2 galaxies.

\begin{figure}[ht]
\centerline{\includegraphics[width=8.0cm]{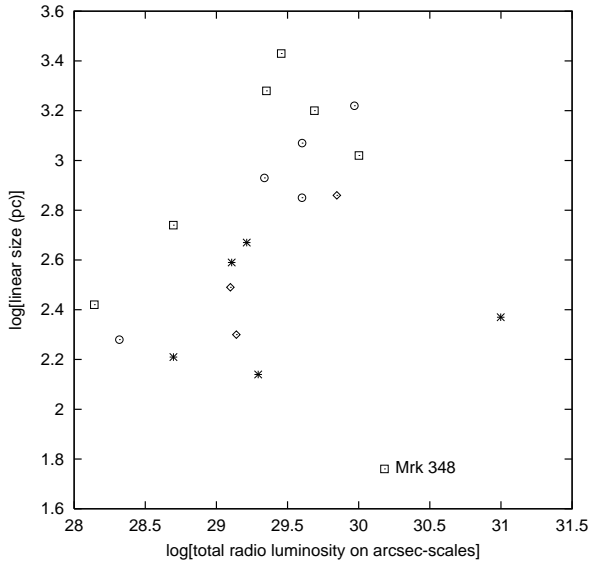}}
\caption{{\small Scatter plot of projected linear size (pc) {\it versus} radio
luminosity on kpc-scales (ergs~s$^{-1}$Hz$^{-1}$); symbols
$\circ$, {\tiny $\Box$}, $*$, and $\diamond$ denotes resolved Seyfert~1,
resolved Seyfert~2, unresolved Seyfert~1, and unresolved Seyfert~2
galaxies respectively.}}
\label{lum_line_size}
\end{figure}

We also show the scatter plot of the radio luminosity on 
kpc-scales {\it versus} the projected linear size in 
Figure \ref{lum_line_size}. Including 
the upper-limits to the linear size, the correlation is significant
(Spearman's rank correlation coefficient = 0.91) for the
Seyfert~1 and Seyfert~2 galaxies taken together. Thus, even though
we restricted the range of the intrinsic {\small AGN} power for our sample,
we still find a significant correlation of these two parameters, earlier
noted by \citet{UW89} and \citet{Morgantietal99}.
Mrk~348 is one of our outliers, and is noted by \citet{Morgantietal99};
a projected linear size of 5~kpc \citep{Baumetal93}
for this source puts it closer to the correlation.

\subsection{Source spectral indices}
\label{sourc_spec_index}

The unified scheme predicts that Seyfert~1 and~2 galaxies arise from the same
parent population of {\small AGN}, and the derived orientation-independent 
parameters should not show significantly different distributions. 
\citet{Morgantietal99} did not find significantly different
distribution of spectral index for Seyfert~1 and Seyfert~2 galaxies.
We use our observations along with measurements
at 2.0~cm, 3.6~cm, 6.0~cm and 20.0~cm, preferably {\small VLA}~$A$ or
B~configuration observations (when not available we use coarser
resolution measurements) to determine the spectral indices,
${\alpha}^{\rm 6~cm}_{\rm 20~cm}$, $\alpha^{\rm 3.6~cm}_{\rm 6~cm}$,
and $\alpha^{\rm 2~cm}_{\rm 3.6~cm}$, of the total
flux density emitted (core plus the extended radio emission) of
the source. Table \ref{yt_yc_spec} gives the measured flux densities
and the spectral indices for our Seyfert sample.
Mrk~348 and Mrk~231 show radio variability \citep{Lal02}.
\citet{Mundelletal09} using VLA at 8.4 GHz have 
shown that five sources from their sample of 12 optically selected,
early-type Seyfert galaxies show radio variability.
In this regard, we have attempted to use data that were
obtained as near in time as possible to ours, and with angular
resolution as close as possible to ours when calculating spectral indices
using our own data or data from the literature.
Mrk~231 between 1.4 and 5.0~GHz, Mrk~348 between 1.4 and
8.4~GHz, MCG~8-11-11 between 5.0 and 15~GHz, and NGC~5929
between 1.4 and 15~GHz show flat spectrum ($\alpha$ $\le$ 0.4)
radio cores.
Figure \ref{masradio}H shows the distribution of the source spectral index,
$\alpha_{\rm 20~cm}^{\rm 6~cm}$ between
1.5~GHz (or 1.4~GHz) and 5~GHz (except for Mrk~348, where it is 
plotted between 5.0~GHz and 8.4~GHz).
The Mann-Whitney~U test shows that the distributions
are statistically indistinguishable at a significance level of 0.02.
Our result at the stated significance level is consistent with the
prediction of the unified scheme, which does not have any preference for
either kind of Seyfert galaxies to show flat$/$steep source~spectrum.

\input{tab4.tex}

\subsection{Relativistic beaming in Seyfert galaxies?}
\label {beaming}

In radio loud objects when the emitting plasma has bulk 
relativistic motion relative to a fixed observer, its emission is 
Doppler enhanced or beamed in the forward direction (in the fixed 
frame), a direct consequence of the transformation of angles in special 
relativity.
An observer located in or near the path of this
plasma sees much more intense emission than if the same plasma 
were at rest. Strong relativistic beaming is thought to explain 
the superluminal motion and high luminosities that characterize blazars 
\citep{BR78}. If present in blazars, it may 
also be present in other {\small AGN}s where the radio-jet
is pointed close to the line of sight of the observer. 
\citet{BK79} formulated the
theory of bulk relativistic motion for a two-component
``jet version'' of the original model, and explored
some of the consequences. In this model, the radio emission
originates within a collimated supersonic jet that supplies
the extended radio lobes with mass, momentum and energy.
The required {\it in situ} acceleration of the emitting
particles is achieved by means of mildly relativistic
shocks propagating into the plasma and confined to a jet.
The ``fixed'' component observed in the {\small VLBI} observations is
identified in this model with the base of the jet.
One of the direct consequences of this bulk relativistic
motion is that when the motion is in directions close
to the line of sight, the observed radio flux density is
apparently enhanced due to Doppler effects. The observed
flux density $S_{\rm obs}$ of the jet at a frequency $\nu$ is
related to the emitted flux density $S_{\rm em}$ that would be observed in
the comoving frame at the same emitted frequency $\nu$ as 
$$
S_{\rm obs}~=~S_{\rm em}~D^3.
$$
$D$, the Doppler factor, is the ratio of observed to
emitted frequency and is given~by
$$
D~=~\gamma^{-1}(1-\frac{v}{c}~cos~\theta)^{-1},
$$
where $\theta$ is the angle of inclination of the direction
of bulk relativistic motion to the line of sight, $v$ is
velocity of the flow, and $\gamma$ is the Lorentz factor.
In this model, wherein the radio core component is
constituted of relativistically moving sub-components,
the observed flux density would be enhanced by Doppler
effects for directions close to the line of sight.

Seyfert galaxies show radio
emitting jet-like structures on small scales which appear to
be the low-power analogues of jets seen in radio powerful {\small AGN}s
\citep[][and references therein]{Nagaretal99}. If this {\small AGN}-linked
radio emission originates from plasma with sub-relativistic
bulk motion, the unified scheme would predict that Seyfert~1 
and Seyfert~2 galaxies should have similar radio morphologies 
on all scales and it would be independent of orientation of the 
Seyfert galaxy. But, if this radio emitting plasma has mildly 
relativistic bulk speeds \citep{Bicknelletal98,Ulvestadetal99B}, the 
unified scheme would predict that Seyfert~1 galaxies, in which 
the radio axis is oriented close to the observers line of sight, 
should show mild Doppler beaming of the radio cores, whereas 
Seyfert~2 galaxies should not exhibit such behavior.
So far, all the relativistically boosted jets with superluminal
motion have only been detected in radio-loud objects, except for
the Seyfert~1 galaxy, III~Zw~2 \citep{Brunthaleretal00}. III~Zw~2 is a
Seyfert~1 galaxy that conforms to our definition of Seyfert galaxies
and is the first detection of
superluminal motion in a Seyfert nucleus in a spiral galaxy.
Observations of the two Seyfert galaxies, Mrk~348 and 
Mrk~231 by \citet{Ulvestadetal99B}, showed sub-relativistic 
expansion in them.

In radio-loud {\small AGN}, the core flux density is Doppler 
enhanced due to relativistic
beaming, while the extended flux density (the flux density
of the lobes in radio-loud objects) is not enhanced.
Therefore, the ratio of the core and extended flux densities 
becomes a  beaming indicator \citep{KS82}.
We define an analogous parameter $R$ for Seyfert galaxies as
$$
 R = \frac{S_{\nu~{\rm compact}}}{S_{\nu~{\rm ext}}},
$$
where $S_{\nu~{\rm compact}}$ is the flux density that we measure on
pc-scales and
$S_{\nu~{\rm ext}}$ is defined as the difference of the total
flux density detected on kpc-scale and the flux density detected
on pc-scales. If the detected pc-scale emission is coming
primarily from ejected plasma close to the nucleus, then by
comparing the pc-scale
radio emission with the kpc-scale extended emission for the 
two classes of Seyfert galaxies, one could test the relativistic
beaming hypothesis.
Due to the availability of simultaneous arcsec-scale data ({\small VLA}) 
for our sample objects, we are able to test for the presence of beaming
without worrying about possible radio variability \citep{Mundelletal09}
in the compact radio flux densities in 15 out of 20 cases. For the 
rest of the sample objects (which are not observed by us) we 
use measurements on pc-scales and kpc-scales which are made at epochs 
as closely spaced in time as possible (see Table~\ref{vla_vlbi}).
Figure \ref{masradio}I shows the distribution of the ratio of the core and
extended flux densities, $R$ for the two Seyfert sub-classes.
Mann-Whitney~U test shows that the distributions are not 
significantly different at a level of 0.10.
In other words, the compact structures detected on pc-scales
are not boosted in Seyfert~ galaxies.

\subsection{Seyfert nuclei: Starburst or accretion-powered central engine ?}
\label{sb_bh}

A key question is, weather
a compact starburst alone power Seyfert galaxies, thus
not requiring the presence of a supermassive black hole in their nuclei.
Powerful circumnuclear starbursts have been unambiguously identified in
40 per~cent of nearby Seyfert 2 galaxies
\citep{cidF01,cidF04,GDetal98,GDetal01,Hetal2001}.
These starbursts were
originally detected by means of either ultraviolet or optical spectroscopy of the
central few 100~pc. Several spectroscopic works have detected features of young
and intermediate age stellar population
\citep{Heckmanetal97,GDetal98,GDetal01,SB00}, suggesting that these
populations are significant, if not dominant, in the nuclear region of many
Seyfert~2 galaxies.
Therefore, on a local scale, evidence is mounting that star formation and
nuclear activity are linked. Two possible evolutionary
progressions can be predicted: H\,II/Starburst
galaxies $\rightarrow$ Seyfert 2 \citep{SB01,Ketal03A,Ketal03B},
or a fuller scenario of H\,II/Starburst
galaxies $\rightarrow$ Seyfert 2 $\rightarrow$ Seyfert 1
\citep{HuntMalkan99,Levensonetal01,Krongoldetal02}.
These predict that galaxy interactions, leading to the concentration of a
large gas mass in the circumnuclear region of a
galaxy, trigger starburst emission. Then mergers
and bar-induced inflows can bring fuel to
a central black hole, stimulating AGN activity.
While emission from the hot stars in the torus might
account for the featureless continua in Seyfert~2 galaxies
\citep{FT95,GDetal98},
starbursts cannot produce the necessary collimation to
form radio jets. The existence of radio jets is, therefore,
often used as an indication of the presence of a black hole and
the accretion disk. Although some Seyfert galaxies are now known
to possess strikingly collimated jets \citep[][etc.]{Nagaretal99,
Kukulaetal95}, the resolution of the radio images
is often insufficient to demonstrate the high degree of
collimation seen in radio galaxies and radio-loud quasars.
For example, although Arp~220
shows a ``double-structure'' on kpc-scales, pc-scale observations
have revealed 13 faint sources in one of the components of the 
double, interpreted as radio supernovae \citep{SLL98B}.
\citet{SLL98B} show that these radio supernovae are
of the same class as radio supernova 1986J observed in the disk of
NGC~891 \citep{Rupenetal87}: Type II radio supernovae with luminosities
of order 10$^{28}$~ergs~s$^{-1}$Hz$^{-1}$ and exponential decay
time of 3~yr. The number of radio supernovae observed in Arp~220
would require a massive star formation rate of
70~M$_{\odot}$~yr$^{-1}$ \citep{Parraetal07}.

A question that can be addressed with the help of the pc-scale and kpc-scale
data for our Seyfert sample \citep{LSG04} is whether the cores of
Seyfert galaxies are primarily made up of radio supernovae.
To examine whether compact starbursts alone can
power Seyfert galaxies, we consider the following arguments.
\citet{FT95} and \citet{GDetal98} suggest that although emission from
hot stars in the torus might account for the featureless continua in Seyfert~2
galaxies, starbursts cannot produce the necessary collimation 
to form radio jets.
We note that elongated radio structures could be attributed
to star~formation along the galactic plane of a Seyfert host
galaxy if it is edge-on ({\it e.g.} M~82 a starburst galaxy,
\citet{Muxlowetal94}). In our sample, however this cannot explain
any of the elongated structures, since the sample was selected
so as to avoid edge-on host galaxies (ratio of minor to major
axis of the host galaxy is greater than 0.5 for all our Seyfert sources).
\citet{Ulvestadetal99B}, on pc-scales, confirmed that Mrk~231
and also Mrk~348 are jet-producing central engine systems.
Thus, clearly all objects with elongated or ``linear'' radio
structures, {\it viz.}, Mrk~348, MCG-8-11-11, NGC~2273, Mrk~78,
Mrk~1218, NGC~4151, NGC~5135, NGC~5929, NGC~7212,
Mrk~926, and Mrk~533 cannot be powered by a starburst alone.
We now try to examine whether radio supernovae or supernova
remnants in star forming regions can plausibly be retained as the
explanation for those objects in the sample that do not show 
linear structure, {\it viz.} Mrk~1 and NGC~7682 which are
essentially compact sources, and NGC~2639, Mrk~477, NGC~7469 and
Mrk~530 which are dominated by a compact source but have low
surface brightness extensions. Note that \citet{SLL98B}
have argued that Mrk~231 cannot be powered by  starburst alone.

Although an individual supernova can have brightness temperature 
higher than the brightness temperature of a radio-quiet quasar
\citep{Rupenetal87}, the most luminous
known radio supernova, 1986J \citep{Rupenetal87}, had a peak
luminosity of $\sim$~10$^{28}$ ergs~s$^{-1}$Hz$^{-1}$ 
at 5~GHz. Given that we have obtained pc-scale flux densities for
all the Seyfert galaxies, we find that 0.2 to 240
(median value of around 10) of such supernovae would be needed to
power them at 5~GHz (Table \ref{sn_rates}).
Since the typical lifetime of such a
supernova event is $\sim$~1~year, to sustain the observed radio
luminosities that we find, a supernova rate of
$\nu_{\rm SN}$~$\sim$~0.2 to 240 yr$^{-1}$ is required. Such rates are in line
with those required to power the luminous radio-quiet quasars in 
the starburst scenario \citep{Parraetal07,Terlevich90A,Terlevich90B}. 
However, since our observed luminosities are on pc-scales, and we
find evidence for one or several dominant compact components
in all our 19 observed Seyfert galaxies, these supernovae must be 
localized within a few cubic~parsec, corresponding to 
a density 10$^7$ times higher than that observed
in M~82 \citep{Muxlowetal94}, and higher than in the starburst model
of \citet{TB93} by a similar factor. 
Although the radio emission from
starburst region consists of synchrotron radiation from SNRs plus
thermal free-free emission from H~{\small II} regions, the 
brightness temperature of such a region cannot exceed 10$^5$~K at
$\nu~>$~1~GHz \citep{Condon92}.
Using the component size along with its flux density on mas-scales,
we find that the brightness temperature, T$_{\rm b}$ (Table \ref{sn_rates}),
is in the range
of $\sim$~0.4--7000~$\times$~10$^{8}$~K for our Seyfert sample.
We therefore conclude on the basis of the high brightness temperatures
($>~10^8$), small sizes ($<$~1~pc), and high supernova rates
to explain the detected compact components, 
that starbursts alone cannot explain the observed radio luminosities
in Seyfert galaxies. 
We can also rule out individual or a collection of extremely
bright radio supernovae as an explanation for the compact
emission from the Seyfert galaxies.

\input{tab5.tex}

\subsection{CfA Seyfert galaxy sample: Kpc-scale radio morphology}
\label{discuss}

The CfA Seyfert sample \citep{HB92} is drawn from
2399 galaxies in the CfA Redshift survey \citep{Davisetal83,
Huchraetal83} and consists of 48 objects (24 Seyfert~1.0
\& Seyfert~1.5, 4 Seyfert~1.8, 4~Seyfert~1.9, and 15~Seyfert~2.0)
chosen solely on the basis of strong emission lines in their
spectra. \citet{Kukulaetal95} made observations of the
optically selected complete spectroscopic sample
at 8.4~GHz~with the {\small VLA} in $A$ and $C$ configuration in 1991~June
and 1992~April respectively, and the observational results of the
sample along with individual source morphology, radio maps,
flux densities, etc. have been presented in their paper.
Note that the unresolved optical nucleus of
a Seyfert grows fainter with the square of distance, whilst the
surface brightness of its host galaxy remains constant over
a constant aperture.
In other words, in this sample, the ratio of the
two components (the host galaxy to the active nucleus surface brightness)
is highly variable.
Nevertheless, as it has a larger number of
objects than in our sample, we therefore use it to test the unified
scheme hypothesis. We use \citet{Kukulaetal95} data and plot the 
distribution for Seyfert~1 galaxies (Seyfert~1.0, 1.5 
and 1.8) and Seyfert~2 galaxies (Seyfert~1.9 and~2.0) to compare 
radio luminosities, projected linear sizes, and relativistic
beaming. Note that, as per our definition,
Seyfert~1 galaxy has Doppler widths of H$\beta$ (or H$\alpha$)
emission lines greater than 1000~km~s$^{-1}$,
whereas Seyfert~1.9 and 2.0 do not
(Seyfert~1.9 shows faint H$\alpha$ and not H$\beta$).

\subsubsection{Radio luminosity comparisons}

\begin{figure*}[ht]
\begin{tabular}{lll}
\includegraphics[width=5.78cm]{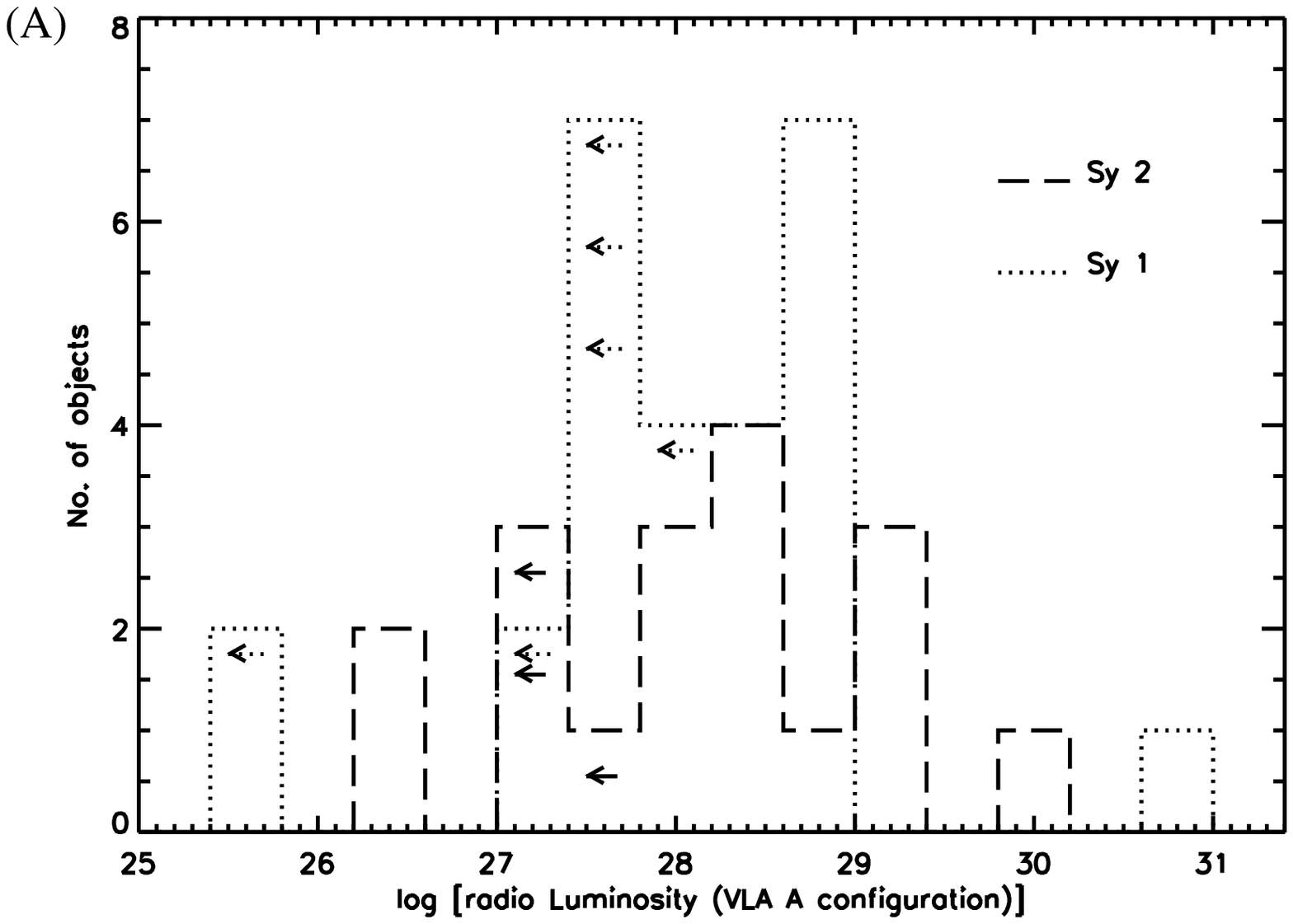} &
\includegraphics[width=5.78cm]{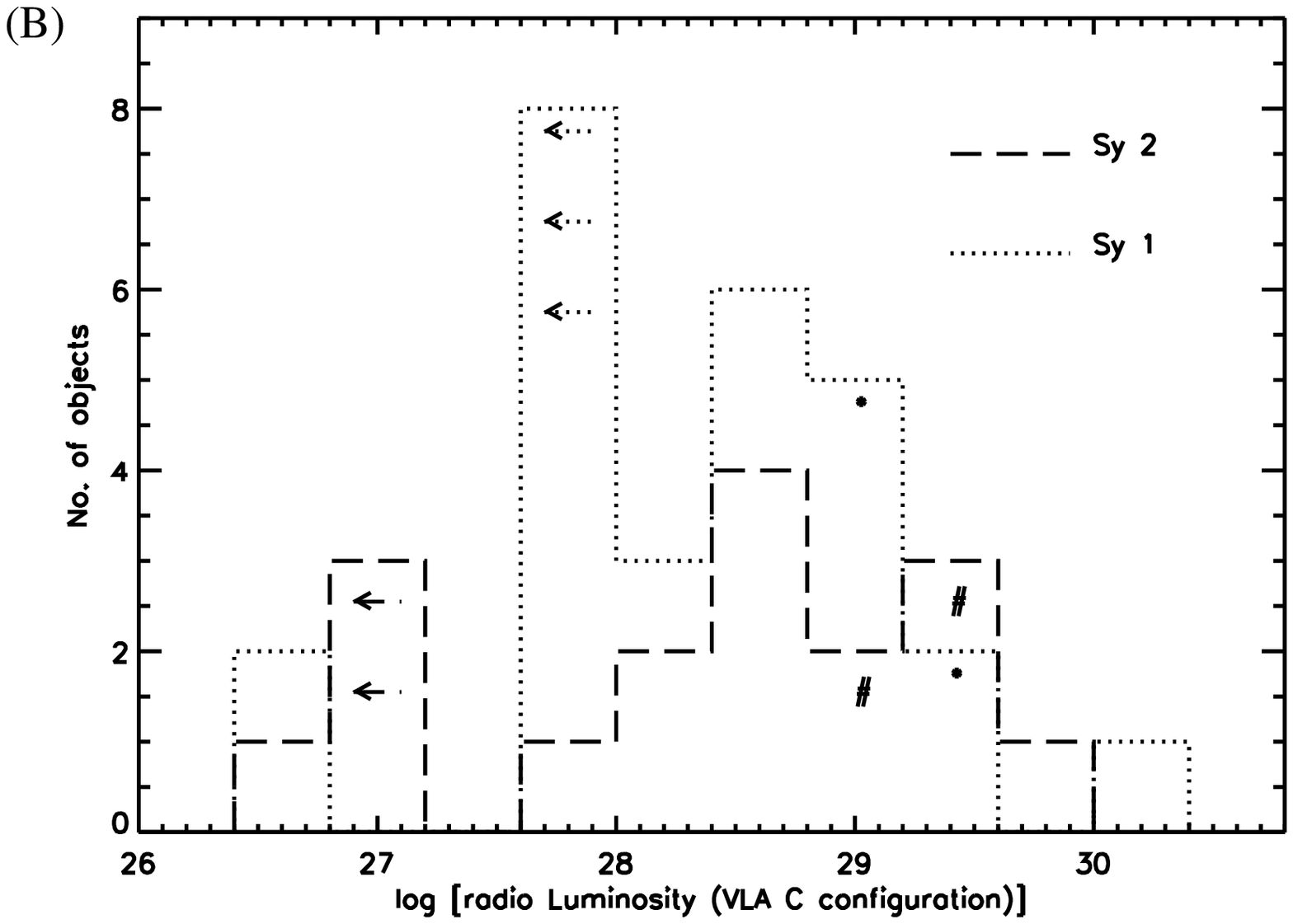} &
\includegraphics[width=5.78cm]{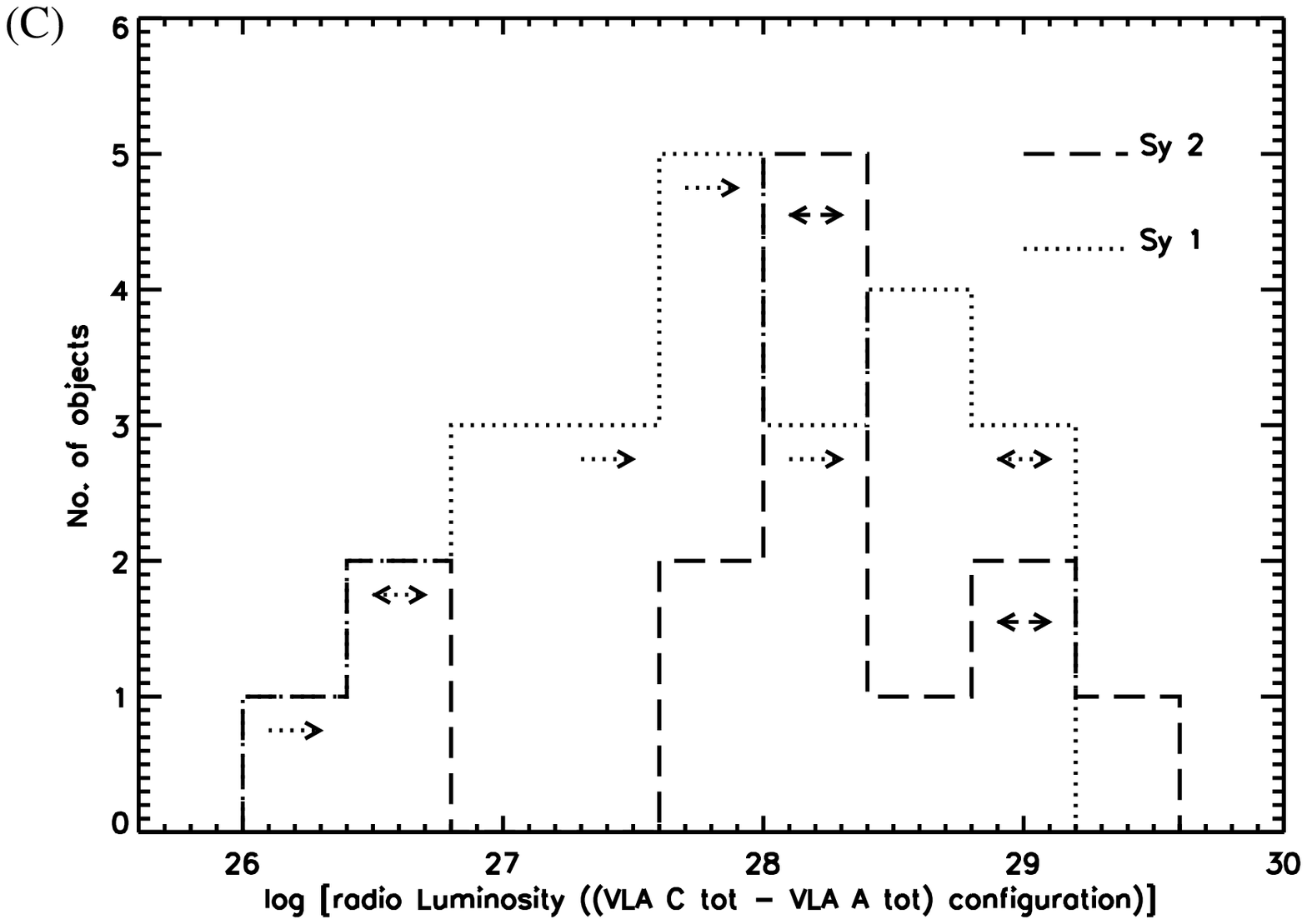} \\
\includegraphics[width=5.78cm]{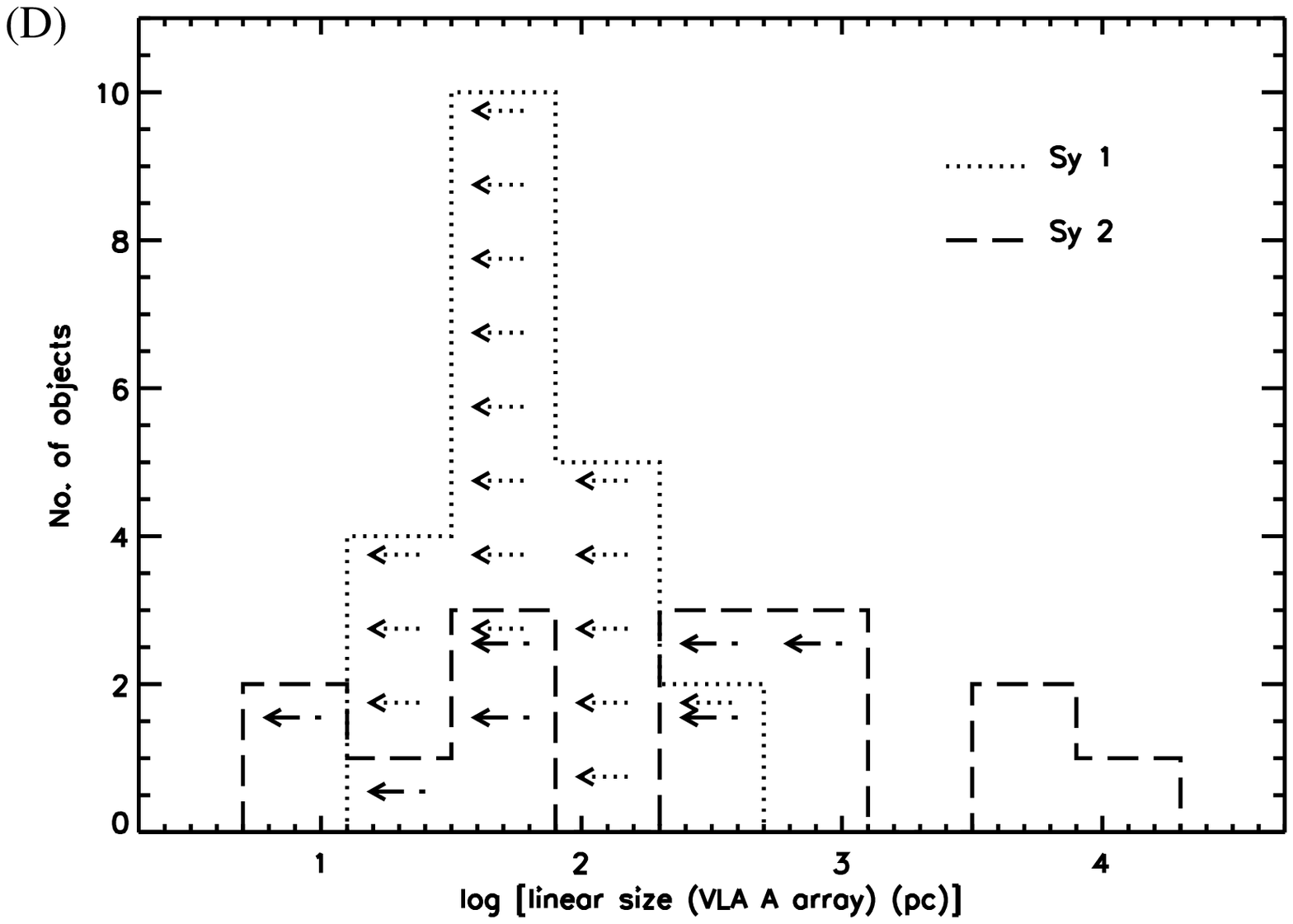} &
\includegraphics[width=5.78cm]{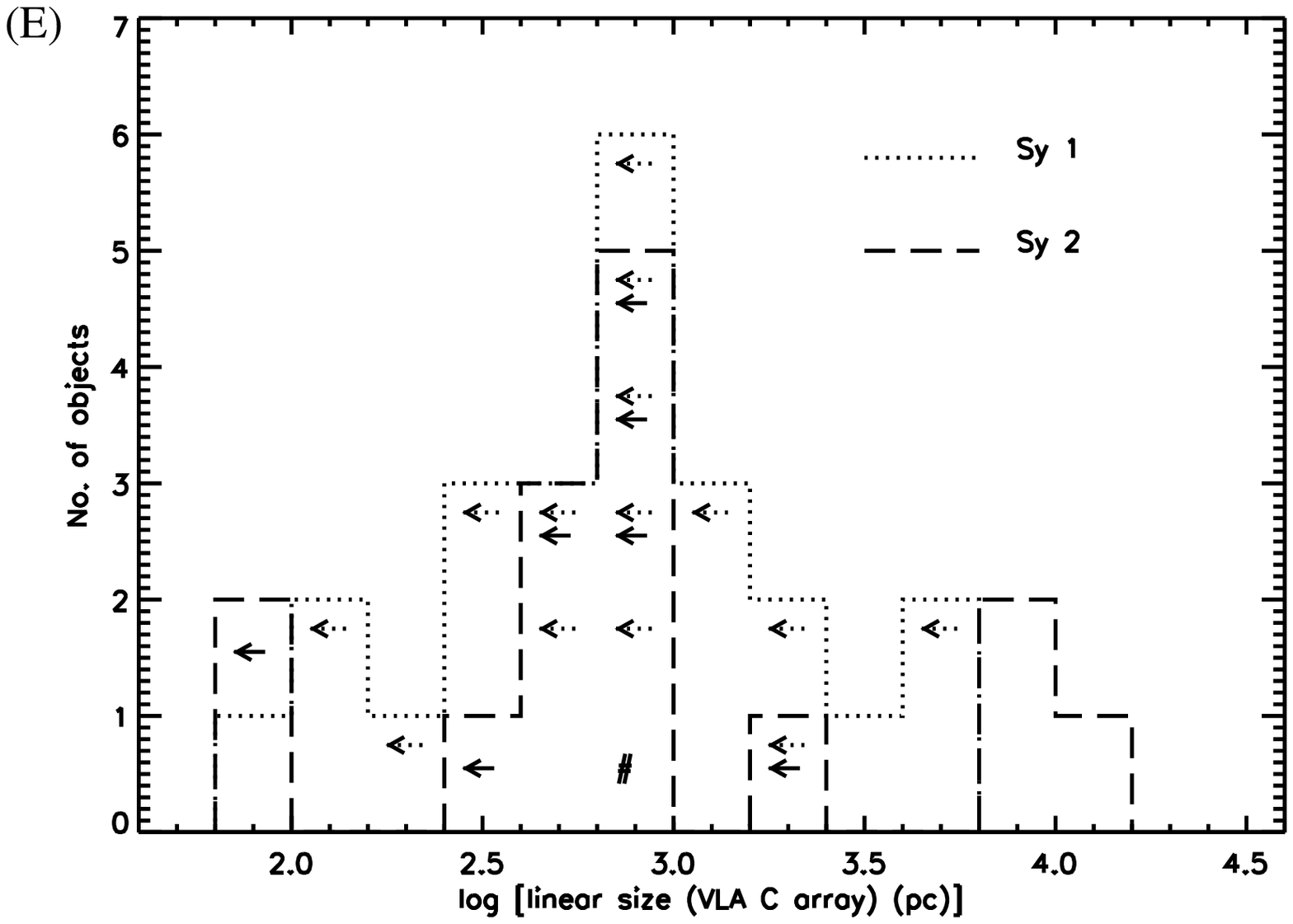} &
\includegraphics[width=5.78cm]{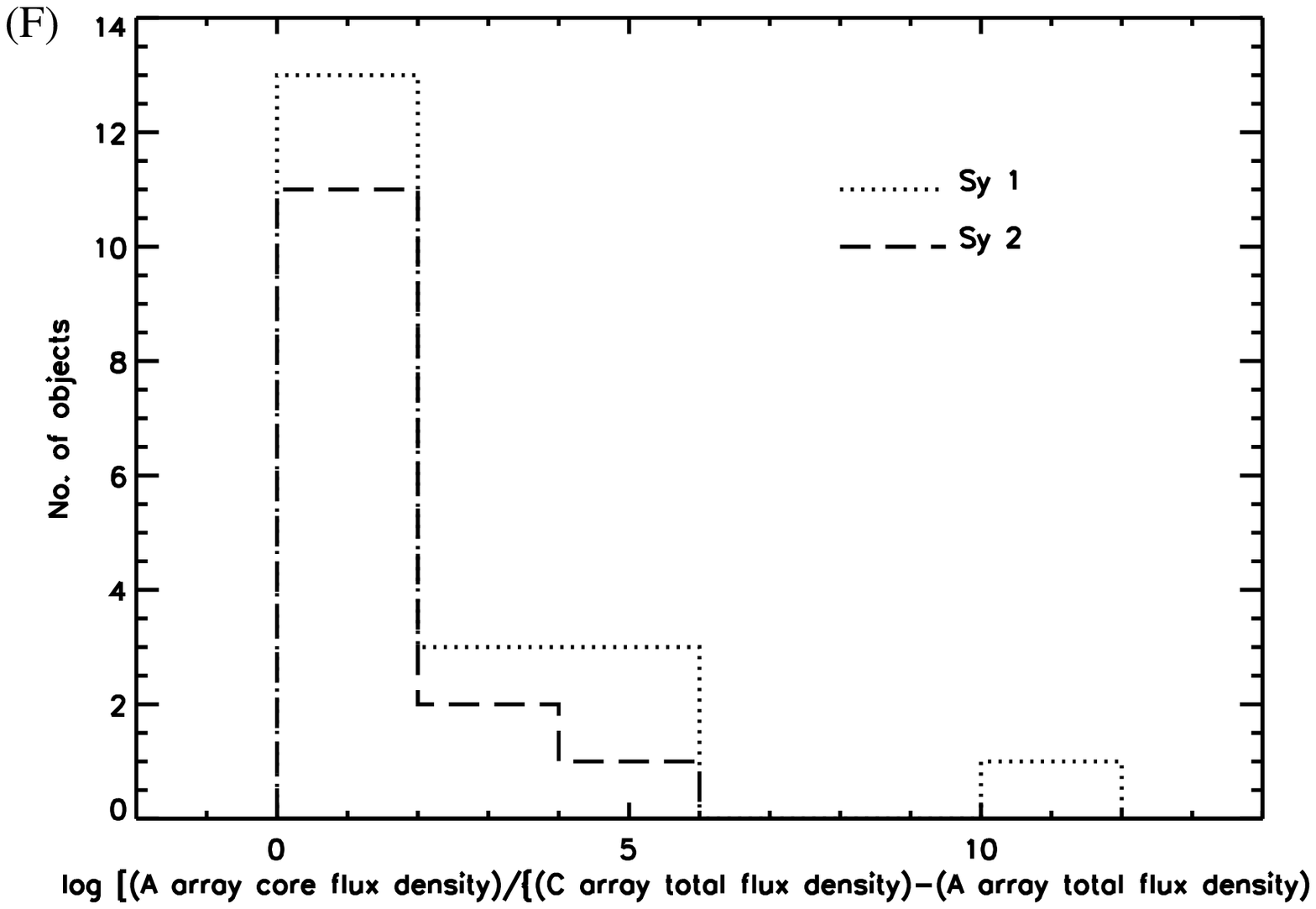}
\end{tabular}
\caption{{\small Histograms showing distributions of
total detected radio luminosity
(ergs~s$^{-1}$Hz$^{-1}$) for the CfA Seyfert sample at 8.4~GHz using
{\small VLA}~$A$ array; arrows denote undetected objects (A),
the radio luminosity (ergs~s$^{-1}$Hz$^{-1}$)
for the CfA Seyfert sample at 8.4~GHz using {\small VLA}~$C$ array;
`$*$' and `$\#$' denote an error of $\sim$~20\% in flux density
calibration for Seyfert~1 and~2 galaxies respectively;
arrows denote undetected objects (B),
the total extended radio luminosity
(ergs~s$^{-1}$Hz$^{-1}$) for the CfA Seyfert sample at 8.4~GHz;
single-headed arrow indicates non-detection on {\small VLA}~$A$ configuration
and double-headed arrow indicates non-detection on both the configurations (C),
projected linear size (pc) of the source
as measured by {\small VLA}~$A$ array for the CfA Seyfert galaxy sample
at 8.4~GHz; arrows indicate unresolved objects (D),
projected linear size (pc) of the source
as measured by {\small VLA}~$C$ array for the CfA Seyfert galaxy sample
at 8.4~GHz; `$\#$'
for an object in a bin denotes the {\small VLA}~$A$ array linear size for it,
and arrows indicate unresolved objects (E), and fraction of
the radio core flux density detected by {\small VLA}~$A$ array as against the
extended flux density, $R$ for the CfA Seyfert sample at 8.4~GHz (F).}}
\label{kukulares}
\end{figure*}

Almost all the objects in this well defined sample were detected
at radio wavelengths (39 of the 48 by {\small VLA}~$A$ array 
and 42 of the 48 by {\small VLA}~$C$ array),
so it is valid to test for the distribution of total
{\small VLA}~$A$ array and {\small VLA}~$C$ array radio luminosity.
Figures~\ref{kukulares}A and \ref{kukulares}B
show the distributions of the total detected
radio luminosities by {\small VLA}~$A$ array and by {\small VLA}~$C$ array
configurations, respectively.
It suggest that there is a decent
similarity in the two distribution for Seyfert~1 and 
Seyfert~2 galaxies.
We use the Mann-Whitney~U test to test the 
hypothesis that the two distributions of total radio 
luminosities detected using {\small VLA}~$A$ configuration and {\small VLA}~$C$ 
configuration are similar. The results of this test indicate
that the distribution of Seyfert~1 and Seyfert~2 galaxies
are similar at a significance level of 0.15 for the CfA Seyfert sample using 
both VLA array configurations at 8.4~GHz.
We also compare the distribution of the extended radio luminosity,
corresponding to the difference of the flux density detected by the
{\small VLA}~$C$ and {\small VLA}~$A$ array shown in
Figure \ref{kukulares}C.
The Mann-Whitney~U test gives a significance level
of 0.20 that the distribution of Seyfert 1 and
Seyfert 2 galaxies are same.  This significance level is not
high enough to demonstrate statistically that the distributions are
the same, but it does not contradict the possibility that they are
the same.
We thus conclude that the distributions of the radio luminosity of
Seyfert~1 and Seyfert~2 galaxies for the CfA Seyfert sample at 8.4~GHz
are probably similar at all scales
and is consistent with the predictions of the unified scheme.

\subsubsection{Projected linear sizes}

Table~\ref{kukula_sum} summarizes the classification of the radio
morphology for the CfA Seyfert galaxy sample, following the scheme of 
\citet{UW84B}. The numbers in the corresponding columns in the Table shows the
number of sources seen of the kinds mentioned in
\citet{UW84A,UW84B} and \citet{UHo01}
on each {\small VLA} configuration.
The Seyfert galaxy samples used for the comparison are
(i) Markarian, based on ultraviolet-excess selection criteria
containing 29 sources \citep{UW84A};
(ii) distance-limited, heterogeneous selection criteria containing 57 sources
\citep{UW84B,UW89}; and
(iii) Palomar, optical selection criteria containing 45 sources
\citep{HoU01, UHo01}.
As compared to \citet{Wilson91}, Kukula et~al.'s (1995) measurements show
a higher fraction of unresolved sources and a lower fraction of galaxies
with diffuse$/$linear radio emission. This result may be due to an
observational bias; these observations are not as sensitive to diffuse,
extended, steep spectrum emission as are the 20~cm {\small {\small VLA}}
observations of \citet{UW89}.

\input{tab6.tex}

We use angular sizes tabulated in Kukula et~al.'s (1995) paper 
along with our formulation in Section~\ref{siz_eqn} to determine the 
linear size and plot the distribution of linear sizes derived
from {\small VLA}~$A$ array shown in Figure~\ref{kukulares}D
and {\small VLA}~$C$ array shown in Figure~\ref{kukulares}E
measurements for the two classes of Seyfert galaxies
(arrows denote the upper limits to the sizes).
The Mann-Whitney~U tests show that
the two distributions are statistically distinguishable at a
significance level of 0.10 and 0.05 for the linear projected sizes
of the two classes of the Seyfert galaxies on the basis of
{\small VLA}~$C$ array and {\small VLA}~$A$ array measurements, respectively.
We thus believe that Seyfert~2 galaxies tend to show larger
projected linear sizes than Seyfert~1 galaxies when found
using {\small VLA}~$A$ configuration and {\small VLA}~$C$ configuration,
and this is consistent with the unified scheme.
This is also consistent with the result obtained by \citet{Morgantietal99}.

\subsubsection{Relativistic beaming}

To investigate relativistic beaming, we use the distribution of 
the ratio between the possibly beamed and extended radio flux densities. 
We assume that the emission associated with the difference of 
the emissions between {\small VLA}~$C$ and {\small VLA}~$A$
configurations would not be Doppler boosted and we call
this as an extended emission. The {\small VLA}~$A$ configuration
measurement probes structures on scales smaller than {\small VLA}~$C$
configuration and we use {\small VLA}~$A$ configuration emission as the
one which would suffer from Doppler boosting if beaming is present.
Figure \ref{kukulares}F shows the distribution of the ratio $R$ (defined
in Section~3.5), of the two, {\small VLA}~$A$ configuration emission to
the extended emission as an indicator of relativistic beaming.
The Mann-Whitney U test is not conclusive about whether the distributions
are the same or not at a significance level of 0.25.
Note that the two observations, {\small VLA}~$C$ and {\small VLA}~$A$
array for the CfA Seyfert sample, are not simultaneous as they were for
our Seyfert sample, and we have discussed earlier that Seyfert galaxies
show radio variability.
Furthermore, here the distribution of $R$ and its statistical significance
does not actually compare the flux density detected on pc-scales
against the extended emission and hence is not really a measure of
relativistic beaming. But nevertheless we use it as an indicator to
probe the boosting of core flux density in Seyfert galaxies.
Since it is equally likely that the distributions obtained 
here are either same or different, and considering also the results
based on our Seyfert sample, we conclude 
that Seyfert galaxies do not show relativistic bulk motion in their nuclei.

\section{Conclusions}
\label{summary}

Unification of Seyfert~1 and Seyfert~2 galaxies has been
attempted by various authors in the past. Their radio emission
has been extensively studied at $\sim$~arcsec-scales,
but at mas-scales a systematic study has not been done before.
We carefully selected and made a list of 20~Seyfert galaxies
that are matched in orientation-independent parameters which are
measures of intrinsic {\small AGN} power and host galaxy properties.
Additionally, the sample met the feasibility requirements,
i.e., each sample source had a detectable compact core component at
arcsec-scale resolution with flux density greater than 8~mJy at 5~GHz.
This sample was used to test the unification scheme hypothesis
rigorously by observing it at pc-scales.
Although these results are based on a small sample size,
these are valuable data for the faintest and least luminous radio cores
of AGN using Global-VLBI.

Using our measured radio flux densities of the pc-scale and kpc-scale
structures for Seyfert galaxies and their derived
detection rates, radio luminosities, spectral indices,
and projected linear sizes,
we find that: 
(i) A starburst alone cannot power these radio sources because, 
they have high brightness temperature, and the core radio luminosity
at 5~GHz is $\sim$ 10$^{28}$~ergs~s$^{-1}$Hz$^{-1}$ 
and arises from a region smaller than a few cubic~pc.
(ii) Our sample of Seyfert~1 and Seyfert~2 galaxies
have equal tendency to show compact radio structures,
in contrast to the results of \citet{Royetal94}, who concluded that
compact radio structures were much more common in Seyfert~2 galaxies
than in Seyfert~1 galaxies.
(iii) The distributions of pc-scale and kpc-scale radio luminosities
are similar for both Seyfert~1 and Seyfert~2 galaxies. This is
consistent with the prediction of the unified scheme hypothesis.
(iv) We do not find any
evidence for relativistic beaming in Seyfert galaxies. 
(v) Although nuclei of Seyfert galaxies show a variation in their
nuclear flux density \citep{Mundelletal09}, our sample of
Seyfert~1 and Seyfert~2 galaxies
show similar distributions of source spectral indices.
(vi) The unification scheme hypothesis predicts that Seyfert~1 galaxies
oriented at small angles to the line of sight should have 
systematically smaller projected linear size than Seyfert~2 galaxies.
The distributions for Seyfert~1 and Seyfert~2 galaxies
of our sample are not significantly different as would be
expected in the unified scheme. This could be mainly due to a relatively
large spread in the intrinsic sizes.

Additionally, from the radio observations of the CfA Seyfert galaxy sample
\citep{Kukulaetal95} and the kpc-scale data from them, the radio luminosities,
projected linear size, and the 
ratio of flux density detected on {\small VLA}~$A$ array configuration 
and the extended emission detected on {\small VLA}~$C$ array configuration
for the two classes of Seyfert galaxies
are consistent with the unification scheme hypothesis.

\acknowledgements

We thank the anonymous referee for her/his prompt review
of the manuscript and for useful and detailed comments that
lead to significant improvement of the paper.
DVL is grateful to L. Lanz for her grammatical corrections.
This project was done with financial support from the Indo-Russian
International Long Term Programme of the Department of Science and
Technology, Government of India and the Russian Academy of Sciences.
Financial support in the initial phase from the Indian National Science
Academy exchange programme is also acknowledged. DVL acknowledges support
from the Joint Institute for VLBI in Europe for a visit there.
DG acknowledges support from the European Commission under TMR contract
No. ECBFMGECT950012. The VLA and VLBA are operated
by the National Radio Astronomy Observatory.
The National Radio Astronomy Observatory is a facility of the National Science
Foundation operated under cooperative agreement by Associated Universities,
Inc.
The European VLBI Network is a joint facility of
European, Chinese and other radio astronomy institutes funded by their
national research councils.
This research has made use of NASA's
Astrophysics Data System bibliographic services, the NASA/IPAC Extragalactic
Database (NED) which is operated by the Jet Propulsion Laboratory,
California Institute of Technology, under contract with NASA, and the SIMBAD
database, operated by CDS, Strasbourg, France.



\end{document}

%% file: tab1.tex
\begin{deluxetable}{cc}
\tabletypesize{\scriptsize}
\tablewidth{0pc}
\tablecaption{Table showing 6$\sigma$ correlated
flux densities, minimum reliably detectable flux densities on various baselines
for a data rate of 128 Mbits~s$^{-1}$, a scan integration time of 8.8~min, and
an observing frequency of 5~GHz.}
\tablehead{
\multicolumn{1}{c}{Baseline}           & Correlated flux density}
\startdata
Phased-{VLA} and Effelsberg & $\sim$~0.7~mJy \\
Phased-{VLA} and {VLBA}-station       & $\sim$~2.6~mJy \\
Effelsberg and {VLBA}-station       & $\sim$~3.4~mJy \\
      {VLBA}-station and {VLBA}-station       & $\sim$~13.5~mJy
\enddata
\label{correlated}
\end{deluxetable}

%% file: tab2.tex
\begin{deluxetable*}{lcccrcccclr}
\tabletypesize{\scriptsize}
\tablecaption{
\label{all_para}
Table showing the list of Seyfert~1 and
Seyfert~2 galaxies that constitute our sample, with the
orientation-independent parameters that were used to constraint it.}
\tablewidth{0pt}
\tablehead{
                        &
                        &
\colhead{VLA $A$~array} &
                        &
                        &
                        &
                        &
                        &
                        &
                        &   \\
 \colhead{Object}       &
                        &
     \colhead{compact}  &
                        &
 \colhead{[O\,III]} &
                        &
\colhead{[O\,III]}  &
                        &
\colhead{}   &
                        &   \\
                        &
\colhead{(b/a)}     &
\colhead{component}     &
\colhead{Redshift}      &
\colhead{line width}    &
\colhead{$\frac{F_{5007}}{F_{H\beta}}$} &
\colhead{luminosity}    &
\colhead{$M_B^{\rm total}$} &
\colhead{$M_B^{\rm bulge}$}    &
\colhead{T}         &
\colhead{T}             \\
                        &
                        &
\colhead{S$_{\nu}$  (mJy)}  &
                        &
\colhead{(km~s$^{-1}$)} &
                        &
\colhead{(ergs~s$^{-1}$)}   &
                        &
\colhead{} &
                        &   \\
(1) & (2) & (3) & (4) & (5) & (6) & (7) & (8) & (9) & (10) & (11)
}
\startdata
Seyfert~1s &    &     &        &             &            &           &           &        &              & \\
           &    &     &        &             &            &           &           &        &              & \\ [-0.3cm]
MCG 8-11-11&0.71&32.5 & 0.025  & ~605.0$^3$  &~3.5$^1$    & 42.14$^3$ & $-$23.1$^3$ & $-$22.10 & {\tt 1}$^{9}$  & {\tt 0} \\
Mrk 1218   &0.58&23.0 & 0.029  & 1078.0$^6$  &~3.8$^3$    & 41.82$^1$ & $-$21.1$^7$ & $-$20.04 & {\tt 1}$^{10}$ & {\tt 1} \\
NGC 2639   &0.60&23.4 & 0.011  & ~400.0$^2$  & 0.3$^7$    & 39.88$^2$ & $-$21.4$^7$ & $-$20.33 & {\tt 1}$^{11}$ & {\tt 3} \\
NGC 4151   &0.71&34.0 & 0.003  & ~425.0$^3$  &~3.3$^1$    & 42.19$^3$ & $-$21.2$^3$ & $-$19.99 & {\tt 2}$^{11}$ & \\
Mrk 766    &0.87&14.5 & 0.013  & ~180.0$^3$  &~3.7$^3$    & 41.77$^3$ & $-$21.0$^3$ & $-$20.01 & {\tt 1}$^{11}$ & {\tt 3} \\
Mrk 231    &0.74&155.0& 0.042  &$>$~600.0$^5$&~0.4$^1$    & 40.27$^1$ & $-$22.3$^7$ & $-$19.75 & {\tt 5}$^{11}$ & \\
Ark 564    &0.65&~8.0 & 0.024  & ~240.0$^3$  &~1.0$^3$    & 41.72$^c$ & $-$21.7$^3$ & $-$20.11 & {\tt 3}$^{9}$  & \\
NGC 7469   &0.72&21.0 & 0.016  &~360.0$^3$   &~0.6$^3$    & 41.84$^3$ & $-$22.0$^3$ & $-$20.99 & {\tt 1}$^{11}$ & {\tt 4} \\
Mrk 926    &0.67&~9.0 & 0.047  & ~365.0$^3$  &~0.6$^1$    & 42.53$^3$ & $-$22.4$^3$ & $-$21.39 & {\tt 1}$^{9}$  & \\
Mrk 530    &0.66&10.0 & 0.030  & ~490.0$^3$  &~0.3$^1$    & 41.26$^3$ & $-$22.7$^3$ & $-$21.20 & {\tt 3}$^{11}$ & {\tt 1} \\
           &    &     &        &             &            &           &             &          &          & \\ [-0.2cm]
Seyfert~2s &    &     &        &             &            &           &             &          &          & \\
           &    &     &        &             &            &           &             &          &          & \\ [-0.3cm]
Mrk 348    &1.00&480.0& 0.015  & ~365.0$^3$  &10.5$^3$    & 41.69$^3$ & $-$21.1$^3$ & $-$20.27 & {\tt 0}$^{11}$ & {\tt 0} \\
Mrk 1      &0.62&26.0 & 0.016  & ~520.0$^3$  &11.1$^3$    & 41.85$^3$ & $-$20.3$^3$ & $-$19.46 & {\tt 0}$^{9}$  & {\tt 5} \\
NGC 2273   &0.76&~8.9 & 0.006  & ~110.0$^3$  &10.0$^3$    & 40.92$^3$ & $-$21.0$^3$ & $-$19.97 & {\tt 1}$^{11}$ & {\tt 3} \\
Mrk 78     &0.55&~8.0 & 0.037  & 1075.0$^3$  &13.2$^3$    & 42.62$^3$ & $-$22.0$^3$ & $-$20.78 & {\tt 1}$^{9}$  & \\
NGC 5135   &0.71&15.8 & 0.014  & ~165.0$^3$  &~5.1$^3$    & 41.28$^3$ & $-$22.1$^3$ & $-$20.91 & {\tt 2}$^{11}$ & {\tt 5} \\
Mrk 477    &0.78&18.3 & 0.038  & ~370.0$^3$  &~8.8$^3$    & 43.02$^3$ & $-$21.0$^3$ & $-$20.13 & {\tt 0}$^{12}$ & \\
NGC 5929   &0.93&13.5 & 0.009  & ~415.0$^3$  &~4.0$^3$    & 40.63$^3$ & $-$21.4$^3$ & $-$20.13 & {\tt 2}$^{11}$ & {\tt 0} \\
NGC 7212   &0.50&30.0 &0.027   & ~435.0$^3$  &10.8$^3$    & 42.34$^3$ & $-$21.2$^3$ & $-$20.22 & {\tt 1}$^{9}$  & \\
Mrk 533    &0.78&27.0 &0.029   & ~350.0$^3$  &12.9$^3$    & 42.26$^3$ & $-$22.7$^3$ & $-$20.68 & {\tt 4}$^{11}$ & {\tt 5} \\
NGC 7682   &0.89&13.5 &0.017   & ~255.0$^3$  &$>$~4.0$^3$ & 41.46$^3$ & $-$21.1$^3$ & $-$19.88 & {\tt 2}$^{11}$ & {\tt 0}
\enddata

\tablecomments{Col. (1), source name;
col. (2), ratio of minor to major isophotal axes of the host galaxies
(all entries are from RC3 catalog \citep{deVetal91} except for
Mrk~926 source, which is from \citet{Lipovetskyetal88});
col. (3), core flux densities on acrsec-scale resolution
(flux density of the detected compact core component at $\lambda_{\rm 6~cm}$ on
arcsec-scale resolution. NGC~7212: Correlated flux density detected at 13~cm by \citet{Royetal94} using
the PTI interferometer.
Mrk 533 and NGC~7682: The measurements are at $\lambda_{\rm 3.6~cm}$ on arcsec-scale resolution);
col. (4), cosmological redshift;
col. (5), [O\,III] $\lambda$5007 emission line widths;
col. (6), line intensity ratios of  [O\,III] $\lambda$5007 and H$\beta$;
col. (7), luminosities of the [O\,III] $\lambda$5007 emission line;
col. (8), stellar luminosities of the host galaxies;
col. (9), absolute bulge luminosities of the host galaxies;
col. (10), Hubble types of the host galaxies (see Section 2.4.5$-$the
morphological class (Hubble type) given in RC3 catalog \citep{deVetal91}
for our sample sources and when not available, we use values from, in order,
\citet{Whittle92A}, or \citet{Lipovetskyetal88}, or \citet{MGT98});
col. (11), same as col. (10), but the  morphhological class preferentially
gleaned from \citet{MGT98}, which are based on WFPC2, $HST$ images,
RC3 catalog, or \citet{Whittle92A}, or \citet{Lipovetskyetal88}.}
\tablerefs{
(1) \citet{DD88};
(2) \citet{Keel83};
(3) \citet{Whittle92A};
(4) see Section 2.4.4;
(5) \citet{KVS98};
(6) \citet{Goodrich89};
(7) see Section 2.4.3;
(8) see Section 2.4.5;
(9) \citet{Whittle92A};
(10) \citet{MGT98};
(11) \citet{deVetal91};
(12)  \citet{Lipovetskyetal88}.}
\end{deluxetable*}

%% file: tab3.tex
\begin{deluxetable*}{lccccrccccr}
\tablecolumns{11}
\tablewidth{0pc}
\tabletypesize{\scriptsize}

\tablecaption{
\label{vla_vlbi}
Table giving derived parameters for our Seyfert galaxy sample
(sources are ordered in increasing right ascension).
}
\tablehead{
\colhead{Object} & \colhead{Distance} & \colhead{S$_{\nu}^{\rm total}$} & \colhead{S$_{\nu}^{\rm total}$} & \colhead{S$_{\nu}^{\rm core}$} & \colhead{S$_{\nu}^{\rm total}$} & \colhead{L$_{(\rm NVSS)}^{\rm total}$} & \colhead{L$_{(\rm kpc-scale)}^{\rm total}$} & \colhead{L$_{(\rm kpc-scale)}^{\rm core}$} & \colhead{L$_{(\rm pc-scale)}^{\rm total}$} &\colhead{Size} \\
 &  & \colhead{($^{\rm arcmin-}_{\rm scale}$)} & \colhead{($^{\rm arcsec-}_{\rm scale})$} & \colhead{($^{\rm arcsec-}_{\rm scale})$} & \colhead{($^{\rm mas-}_{\rm scale})$} & \colhead{$_{(\rm 1.4~GHz)}$}  & \colhead{$_{(\rm 5~GHz)}$} & \colhead{$_{(\rm 5~GHz)}$} & \colhead{$_{(\rm 5~GHz)}$} & \colhead{$_{(\rm log(size))}$} \\
   & \colhead{(Mpc)} & \multicolumn{4}{c}{(mJy)} & \multicolumn{4}{c}{(ergs~s$^{-1}$Hz$^{-1}$)} & \colhead{(pc)} \\
   \multicolumn{1}{c}{(1)} & (2) & (3)     & (4)     & (5)     & (6)                    & (7)                    & (8)                    & (9)                   & (10)  & \multicolumn{1}{c}{(11)}
}
\startdata
Mrk 348     & 59.6&277.3&346.0$^1$&310.3$^1$&163.0$^2$& 1.2 $\times$ 10$^{30}$ & 1.5 $\times$ 10$^{30}$ & 1.3 $\times$ 10$^{30}$ & 7.0 $\times$ 10$^{29}$& 1.76$^{10}$  \\
Mrk 1       & 63.5& 70.8& 27.8  & 24.1  &  4.4        & 3.5 $\times$ 10$^{29}$ & 1.4 $\times$ 10$^{29}$ & 1.2 $\times$ 10$^{29}$ & 2.1 $\times$ 10$^{28}$&$<$~2.30$^{11}$  \\
MCG 8-11-11 & 98.8&228.3& 78.0  & 23.6  &  4.9        & 2.7 $\times$ 10$^{30}$ & 9.1 $\times$ 10$^{29}$ & 2.8 $\times$ 10$^{29}$ & 5.9 $\times$ 10$^{28}$& 3.22$^{11}$  \\
NGC 2273    & 23.9& 59.7& 22.0  & 12.1  &  6.3        & 4.1 $\times$ 10$^{28}$ & 1.4 $\times$ 10$^{28}$ & 8.4 $\times$ 10$^{27}$ & 4.4 $\times$ 10$^{27}$& 2.42$^{11}$ \\
Mrk 78      &145.4& 35.0& 13.1  &  8.9  &  8.5        & 9.2 $\times$ 10$^{29}$ & 2.8 $\times$ 10$^{29}$ & 2.3 $\times$ 10$^{29}$ & 2.1 $\times$ 10$^{29}$& 3.43$^{11}$  \\
Mrk 1218    &114.4& 63.6& 27.7  & 18.7  & 12.3        & 1.0 $\times$ 10$^{30}$ & 3.9 $\times$ 10$^{29}$ & 3.0 $\times$ 10$^{29}$ & 1.2 $\times$ 10$^{29}$& 2.85$^{11}$  \\
NGC 2639    & 43.8&105.5& 85.8  & 63.5  & 39.5        & 2.5 $\times$ 10$^{29}$ & 1.9 $\times$ 10$^{29}$ & 1.5 $\times$ 10$^{29}$ & 9.2 $\times$ 10$^{28}$& $<$~2.14$^{11}$ \\
NGC 4151    & 12.0&345.8&120.0$^3$&43.0$^3$&10.0$^4$  & 6.0 $\times$ 10$^{28}$ & 2.0 $\times$ 10$^{28}$ & 7.4 $\times$ 10$^{27}$ & 3.5 $\times$ 10$^{27}$& 2.57$^{14}$  \\
Mrk 766     & 51.7& 37.6& 17.9  & 13.8  &  3.6        & 1.2 $\times$ 10$^{29}$ & 4.9 $\times$ 10$^{28}$ & 4.5 $\times$ 10$^{28}$ & 1.2 $\times$ 10$^{28}$&$<$~2.21$^{11}$  \\
Mrk 231     &164.6&282.9&282.0$^5$&270.0$^5$&173.0$^5$& 9.6 $\times$ 10$^{30}$ & 5.9 $\times$ 10$^{30}$ & 5.5 $\times$ 10$^{30}$ & 2.3 $\times$ 10$^{30}$&$<$~2.37$^{14}$  \\
NGC 5135    & 55.6&190.0& 58.8$^6$&16.2$^6$ &         & 7.2 $\times$ 10$^{29}$ & 2.2 $\times$ 10$^{29}$ & 6.1 $\times$ 10$^{28}$ &                       & 3.28$^{12}$  \\
Mrk 477     &149.2& 60.7& 27.3  & 23.0  &  8.1        & 1.7 $\times$ 10$^{30}$ & 6.8 $\times$ 10$^{29}$ & 6.6 $\times$ 10$^{29}$ & 2.3 $\times$ 10$^{29}$&$<$~2.86$^{13}$  \\
NGC 5929    & 35.8&105.5& 34.4  &1.3$^7$&  6.2        & 1.6 $\times$ 10$^{29}$ & 4.9 $\times$ 10$^{28}$ & 2.0 $\times$ 10$^{27}$ & 9.8 $\times$ 10$^{27}$& 2.78$^{11}$  \\
NGC 7212    &106.6&108.9& 31.0  &19.6    &  7.0       & 1.5 $\times$ 10$^{30}$ & 4.8 $\times$ 10$^{29}$ & 2.7 $\times$ 10$^{29}$ & 9.2 $\times$ 10$^{28}$& 3.20$^{11}$  \\
Ark 564     & 94.9& 27.7& 11.4   &  8.6  &  3.2       & 3.1 $\times$ 10$^{29}$ & 1.3 $\times$ 10$^{29}$ & 9.5 $\times$ 10$^{28}$ & 3.6 $\times$ 10$^{28}$& $<$~2.59$^{11}$  \\
NGC 7469    & 63.5&167.7& 47.9  & 22.0  &  6.1        & 8.3 $\times$ 10$^{29}$ & 2.1 $\times$ 10$^{29}$ & 1.1 $\times$ 10$^{29}$ & 3.0 $\times$ 10$^{28}$& 2.93$^{11}$  \\
Mrk 926     &183.8& 33.0& 9.0$^8$&7.7$^8$&5.0$^9$     & 1.4 $\times$ 10$^{30}$ & 3.8 $\times$ 10$^{29}$ & 3.3 $\times$ 10$^{29}$ & 2.1 $\times$ 10$^{29}$& 3.07$^{15}$  \\
Mrk 530     &118.3& 23.5& 10.2  &  8.0  &  8.6        & 4.6 $\times$ 10$^{29}$ & 1.6 $\times$ 10$^{29}$ & 1.4 $\times$ 10$^{29}$ & 1.5 $\times$ 10$^{29}$&$<$~2.67$^{11}$  \\
Mrk 533     &114.4&206.4& 58.8  & 38.2   &16.9        & 3.3 $\times$ 10$^{30}$ & 9.7 $\times$ 10$^{29}$ & 6.2 $\times$ 10$^{29}$ & 2.7 $\times$ 10$^{29}$& 3.02$^{11}$  \\
NGC 7682    & 67.4& 58.6& 22.6  & 22.0  & 11.5        & 3.3 $\times$ 10$^{29}$ & 1.2 $\times$ 10$^{29}$ & 1.2 $\times$ 10$^{29}$ & 5.0 $\times$ 10$^{28}$& $<$~2.49$^{11}$
\enddata

\tablecomments{Col. (1), source name;
col. (2), distance to the source assuming a cosmology with $H_0 = 75~{\rm km~s}^{-1}~{\rm Mpc}^{-1}$ and $q_0 = 0$;
col. (3), total flux density on arcmin-scale at 1.4~GHz (NVSS: Condon et~al. 1998);
col. (4), total flux density on arcsec-scale at 5~GHz;
col. (5), core flux density on arcsec-scale at 5~GHz;
col. (6), core flux density on mas-scale at 5~GHz;
col. (7), radio power on arcmin-scale at 1.4~GHz (NVSS: Condon et~al. 1998);
col. (8), total radio power on arcsec-scale at 5~GHz;
col. (9), core radio power on arcsec-scale at 5~GHz;
col. (10), core radio power on mas-scale at 5~GHz;
col. (11), logarithm of the projected linear size on mas-scale at 5~GHz.}
\tablerefs{
(1) Nov 1996, VLA 8.4~GHz measurement \citep{Theanetal01};
(2) Apr 1995, VLBA 8.4~GHz measurement \citep{BL98};
(3) Mar 1980, VLA 5.0~GHz measurement \citep{Johnstonetal82};
(4) May$/$Jun 1996, VLBA 5.0~GHz measurement \citep{Ulvestadetal98};
(5) Dec 1996, VLA  5.0~GHz measurement \citep{Ulvestadetal99A};
(6) Feb 1985, VLA 5.0~GHz measurement \citep{UW89};
(7) Mar 1989, MERLIN measurements and ${\alpha}^{\rm 2~cm}_{\rm 18~cm}$~=~0.32; \citep{Suetal96};
(8) May 1982, VLA 5.0~GHz measurement \citep{UW84A,UW84B};
(9) Jul 1997, VLBA 8.4~GHz measurement \citep{Mundelletal00};
(10) \citep{Ungeretal84};
(11) Our \citep{LSG04} measurements;
(12) \citep{UW89};
(13) \citep{Pedlaretal93};
(14) \citep{Kukulaetal95};
(15) \citep{UW84A}.}
\end{deluxetable*}

%% file: tab4.tex
\begin{deluxetable*}{lrrrrrrrrrl}
\tablecolumns{11}
\tabletypesize{\scriptsize}

\tablecaption{
\label{yt_yc_spec}
Source total flux densities and spectral indices; we define the spectral index
$\alpha$ in the sense that $S_\nu \propto \nu^{-\alpha}$, where $S_\nu$ and $\nu$
are flux density and frequency, respectively.
}
\tablewidth{0pt}
\tablehead{
 & \multicolumn{4}{c}{{Total flux density}} & & \multicolumn{4}{c}{{Spectral index (total)}} & \\
\cline{2-5} \cline{7-10} \\
Object &$S_{\rm 1.5~GHz}$&$S_{\rm 5.0~GHz}$&$S_{\rm 8.4~GHz}$&$S_{\rm 15~GHz}$& &$\alpha_{\rm 20~cm}^{\rm 3.6~cm}$&$\alpha_{\rm 20~cm}^{\rm 6~cm}$&$\alpha_{\rm 6~cm}^{\rm 3.6~cm}$&$\alpha_{\rm 3.6~cm}^{\rm 2~cm}$ & References \\
 & \multicolumn{4}{c}{{(mJy)}}&  & \multicolumn{4}{c}{{}} &
}
\startdata
Mrk 348     &302.2    &       &238.0    &        && 0.14&     &     &       &1.5\,GHz$-$1, 8.4\,GHz$-$1  \\
Mrk 1       & 68.0    & 27.65 & 15.4    &        && 0.86& 0.75& 1.13&       &1.5\,GHz$-$2, 5.0\,GHz$-$3, 8.4\,GHz$-$4  \\
MCG 8-11-11 &180.0    & 75.76 & 38.4    &20.0    && 0.90& 0.72& 1.31& 1.13  &1.5\,GHz$-$5, 5.0\,GHz$-$3, 8.4\,GHz$-$4, 15\,GHz$-$~6 \\
NGC 2273    & 52.0    & 19.92 & 10.2    &        && 0.95& 0.80& 1.29&       &1.5\,GHz$-$1, 5.0\,GHz$-$3, 8.4\,GHz$-$1  \\
Mrk 78      & 31.0    & 10.46 &         &        &&     & 0.90&     &       &1.5\,GHz$-$7, 5.0\,GHz$-$3 \\
Mrk 1218    & 65.0    & 24.04 &         &        &&     & 0.83&     &       &1.5\,GHz$-$6, 5.0\,GHz$-$3 \\
NGC 2639    &104.0    & 54.50 &         &        &&     & 0.54&     &       &1.5\,GHz$-$8, 5.0\,GHz$-$8 \\
NGC 4151    &330.0    &120.00 & 72.32   &23.9    && 0.88& 0.84& 0.98& 1.91  &1.5\,GHz$-$9, 5.0\,GHz$-$9, 8.4\,GHz$-$10, 15\,GHz$-$11 \\
Mrk 766     & 39.3    & 15.16 &  8.68   &        && 0.88& 0.79& 1.09&       &1.5\,GHz$-$1, 5.0\,GHz$-$3, 8.4\,GHz$-$1  \\
Mrk 231     &280.0    &355.00 &         &$<$~284.0&&    &$-$0.20&   &       &All measurements are from 12  \\
NGC 5135    &163.2    & 58.80 &         &        &&     & 0.85&     &       &1.5\,GHz$-$8, 5.0\,GHz$-$8  \\
Mrk 477     & 60.7    & 24.39 &         &        &&     & 0.76&     &       &1.5\,GHz$-$NVSS, 5.0\,GHz$-$3 \\
NGC 5929    & 64.7    & 31.74 & 16.75   & 9.0    && 0.78& 0.59& 1.23& 1.07  &All measurements are from 13  \\
NGC 7212    &108.8    & 33.94 &$>$~9.0  &        &&     & 0.97&     &       &1.5\,GHz$-$NVSS, 5.0\,GHz$-$3, 8.4\,GHz$-$14   \\
Ark 564     & 27.7    & 11.30 &  7.0    &        &&     & 0.74& 0.92&       &1.5\,GHz$-$NVSS, 5.0\,GHz$-$3, 8.4\,GHz$-$4  \\
NGC 7469    &134.0    & 43.56 & 15.97   & 8.0    && 1.23& 0.93& 1.93& 1.19  &1.5\,GHz$-$5, 5.0\,GHz$-$3, 8.4\,GHz$-$10, 15\,GHz$-$15  \\
Mrk 926     & 33.0    &  9.00 &         &        &&     & 1.07&     &       &1.5\,GHz$-$NVSS, 5.0\,GHz$-$16  \\
Mrk 530     & 23.5    &  9.17 &  3.26   &        &&     & 0.78& 2.00&       &1.5\,GHz$-$NVSS, 5.0$\,GHz-$3, 8.4\,GHz$-$10  \\
Mrk 533     &160.0    & 60.37 & 39.77   &        && 0.81& 0.81& 0.80&       &1.5\,GHz$-$5, 5.0\,GHz$-$3, 8.4\,GHz$-$10  \\
NGC 7682    & 61.6    & 22.21 & 13.46   &        && 0.88& 0.84& 0.97&       &1.5\,GHz$-$17, 5.0$-$3, 8.4\,GHz$-$10
\enddata
\tablerefs{
(1) Nagar et~al. (1999);
(2) \citet{deBW76};
(3) Lal, Shastri \& Gabuzda (2004);
(4) \citet{Schmittetal01};
(5) \citet{Ungeretal86};
(6) \citet{UW86};
(7) \citet{UW84A};
(8) \citet{UW89};
(9) \citet{Johnstonetal82};
(10) \citet{Kukulaetal95};
(11) \citet{WU82};
(12) \citet{Ulvestadetal99A};
(13) \citet{Suetal96};
(14) \citet{FWS98};
(15) \citet{Wilsonetal91};
(16) \citet{UW84A};
(17) \citet{Edelson87}.
}
\end{deluxetable*}

%% file: tab5.tex
\begin{deluxetable}{lccrcrc}
\tabletypesize{\scriptsize}
\tablewidth{0pc}
\tablecaption{Table showing the
brightness temperature (T$_{\rm b}$) of compact bright component
and the supernova rates ($\nu_{\rm SN}$) necessary to reproduce
radio emission for our sample sources.}
\tablehead{ \colhead{Object}    & & & & \colhead{T$_{\rm b}$} & & \colhead{$\nu_{\rm SN}$}  \\
            & & & &  \colhead{(K)} & & \colhead{(yr$^{-1}$)}
}
\startdata
 Mrk 348$^{1,2}$&&&& 4.4 $\times$ $10^{9}$ &  &72 \\
 Mrk 1      & & &  & 2.2 $\times$ $10^{8}$ &  & 2 \\
 MGC 8-11-11& & &  & 4.4 $\times$ $10^{8}$ &  & 6 \\
 NGC 2273   & & &  & 2.6 $\times$ $10^{8}$ &  & 0.4\\
 Mrk 78     & & &  & 6.1 $\times$ $10^{8}$ &  & 21 \\
 Mrk 1218   & & &  & 2.8 $\times$ $10^{8}$ &  & 17 \\
 NGC 2639   & & &  & 7.0 $\times$ $10^{11}$&  & 9 \\
 NGC 4151$^3$&& &  & 3.7 $\times$ $10^{7}$ &  & 0.2 \\
 Mrk 766    & & &  & 2.8 $\times$ $10^{8}$ &  & 1 \\
 Mrk 231$^4$& & &  & 2.6 $\times$ $10^{10}$&  & 240 \\
 NGC 5135   & & &  & 2.3 $\times$ $10^{8}$ &  & 24 \\
 Mrk 477    & & &  & 2.3 $\times$ $10^{8}$ &  & 24 \\
 NGC 5929   & & &  & 5.3 $\times$ $10^{8}$ &  & 1 \\
 NGC 7212   & & &  & 3.0 $\times$ $10^{8}$ &  & 10 \\
 Ark 564    & & &  & 2.5 $\times$ $10^{8}$ &  & 4 \\
 NGC 7469   & & &  & 3.8 $\times$ $10^{8}$ &  & 3 \\
 Mrk 926$^2$& & &  & 9.7 $\times$ $10^{7}$ &  & 22 \\
 Mrk 530    & & &  & 5.9 $\times$ $10^{8}$ &  & 15 \\
 Mrk 533    & & &  & 4.3 $\times$ $10^{8}$ &  & 28 \\
 NGC 7682   & & &  & 2.9 $\times$ $10^{8}$ &  & 5
\enddata
\tablerefs{
(1) \citet{BL98};
(2) \citet{Mundelletal00};
(3) \citet{Ulvestadetal98};
(4) \citet{Ulvestadetal99A}.}
\label{sn_rates}
\end{deluxetable}

%% file: tab6.tex
\begin{deluxetable}{lcc}
\tabletypesize{\scriptsize}
\tablewidth{0pc}
\tablecaption{Table showing summary of the classification of the
radio morphologies, and the number of sources seen
on the two VLA configurations for the CfA Seyfert galaxy sample.}
\tablehead{
  Morphology$^{1,2,3}$       & {\small VLA}~$A$ array & {\small VLA}~$C$ array}
\startdata
  not detected        & 8           & 6           \\
  unresolved          &17           &20           \\
  slightly resolved   & 5           & 9           \\
  resolved            &14           & 8           \\
  Compact double      & 1           & 1           \\
  no data             & 3           & 4         
\enddata
\tablerefs{
(1) \citet{UW84A};
(2) \citet{UW84A};
(3) \citet{ UHo01}.}
\label{kukula_sum}
\end{deluxetable}

%% file: ms_sy.bbl
\begin{thebibliography}{}

\bibitem[Antonucci(1993)]{A93} Antonucci, R.R.J., 1993, ARA\&A, 31, 473. 
\bibitem[Barvainis \& Lonsdale(1998)]{BL98} Barvainis, R., \& Lonsdale, C.J. 1998, ApJ, 115, 885.
\bibitem[Baum et~al.(1993)]{Baumetal93} Baum, S.A., O'Dea, C.P., Dallacassa, D., de~Bruyn, A.G., \& Pedlar, A. 1993, ApJ, 419, 553.
\bibitem[Bicknell et~al.(1998)]{Bicknelletal98} Bicknell, G.V., Dopita, M.A., Tsvetanov, Z.I., \& Sutherland, R.S., 1998, ApJ, 495, 680.
\bibitem[Blandford \& Rees(1978)]{BR78} Blandford, R.D., \& Rees, M.J., 1978, PhyS, 17, 265.
\bibitem[Blandford \& K\"onigl(1979)]{BK79} Blandford, R.D., \& K\"onigl, A., 1979, ApJ, 232, 34.
\bibitem[Brunthaler et~al.(2000)]{Brunthaleretal00} Brunthaler, A., Falcke, H., Bower, G.C., Aller, M.F., Aller, H.D., Teraesranta, H., Lobanov, A.P., Krichbaum, T.P., \& Patnaik, A.R., 2000, A\&AL, 357, L45.
\bibitem[Buchanan et~al.(2006)]{Buchanan06} Buchanan, C.L., Gallimore, J.F., O'Dea, C.P., Baum, S.A.; Axon, D.J., Robinson, A., Elitzur, M., Elvis, M., 2006, AJ, 132, 401
\bibitem[de~Bruyn \& Wilson(1976)]{deBW76} de~Bruyn, A.G., \& Wilson, A.S. 1976, A\&A, 53, 93.
\bibitem[Cappi et~al.(1996)]{Cappietal96} Cappi, M., Mihara, T., Matsuoka, M., Brinkmann, W., Prieto, M.A., \& Palumbo, G.G.C., 1996, ApJ, 456, 141.
\bibitem[Cid~Fernandes \& Terlevich(1995)]{FT95} Cid Fernandes, \& Terlevich, R., 1995, MNRAS, 272, 423.
\bibitem[Cid~Fernandes et~al.(2001)]{cidF01} Cid Fernandes, Heckman T., Schmitt H., Gonz\'alez Delgado R. M., Storchi-Bergmann T., 2001, ApJ, 558, 81.
\bibitem[Cid~Fernandes et~al.(2004)]{cidF04} Cid Fernandes, Gu Q., Melnick J., Terlevich E., Terlevich R., Kunth D., Rodrigues Lacerda R., \& Joguet B., 2004, MNRAS, 355, 273.
\bibitem[Colina et~al.(1997)]{Colinaetal97} Colina, L., Vargas, M.L.G., Rosa M. Gonzalez, Mas-Hesse, J.M., Perez, E., Alberdi, A., \& Krabbe, A., 1997, ApJL, 488, L71.
\bibitem[Condon(1992)]{Condon92} Condon, J.J., 1992, ARA\&A, 30, 575.
\bibitem[Condon et~al.(1998)]{Condonetal98} Condon, J.J., Cotton, W.D., Greisen, E.W., Yin, Q.F., Perley, R.A., Taylor, G.B., \& Broderick, J.J., 1998, AJ, 115, 1693.
\bibitem[Curran(2000)]{Curran00} Curran, S.J., 2000, A\&AS, 144, 271.
\bibitem[Dahari \& De~Robertis(1988)]{DD88} Dahari, O., \& De~Robertis, M.M. 1988, ApJS, 67, 249.
\bibitem[Davis et~al.(1983)]{Davisetal83} Davis, M., Huchra, J., \& Latham, D., in {Early evolution of universe and its present structure}, eds. Abell, G.O., \& Chincarni, G., 1983, IAUS, 104, p. 167, Reidel, Dordrecht.
\bibitem[Dultzin-Hacyan et~al.(1999)]{DHetal99} Dultzin-Hacyan, D., Krongold, Y., Fuentes-Guridi, I., \& Marziani, P., 1999, ApJ, 513, L111.
\bibitem[Edelson(1987)]{Edelson87} Edelson, R.A. 1987, ApJ, 313, 651.
\bibitem[Evans et~al.(1991A)]{Evansetal91A} Evans, I.N., Ford, H.C.,Kinney, A.L., Antonucci, A.J., Armus, L., \& Caganoff, S. 1991A, ApJL, 369, L27.
\bibitem[Evans et~al.(1991B)]{Evansetal91B} Evans, I.N., Ford, H.C., Kinney, A.L., Antonucci, R.R.J., Armus, L., \& Caganoff, S. 1991B, ApJS, 76, 985.
\bibitem[Evans et~al.(1993)]{Evansetal93} Evans, I.N., Tsvetanov, Z., Kriss, G.A., Ford, H.C., Caganoff, S., \& Koratkar, A.P., 1993, ApJ, 417, 82.
\bibitem[Evans et~al.(1994)]{Evansetal94} Evans, I.N., Ford, H.C., Kriss, G.A., \& Tsvetanov, Z., in {First Stromlo Symposium: The physics of active galaxies}, eds Bicknell, G.V., Dopita, M.A., \& Quinn, P.J., 1994, ASP Conf. Ser., 54, p.~3.
\bibitem[Falcke, Wilson, \& Simpson(1998)]{FWS98} Falcke, H., Wilson, A.S., \& Simpson, C. 1998, ApJ, 502, 199.
\bibitem[Falcke et~al.(2000)]{Falckeetal00} Falcke, H., Nagar, N.M., Wilson, A.S., \& Ulvestad, J.S. 2000, ApJ, 542, 197.
\bibitem[Gallimore et~al.(2010)]{Gallimoreetal10} Gallimore, J.F., Yzaguirre, A., Jakoboski, J., Stevenosky, M.J., Axon, D.J., Baum, S.A., Buchanan, C.L., Elitzur, M., Elvis, M., O'Dea, C.P., \& Robinson, A., 2010, ApJS, 187, 172.
\bibitem[Gonz\'alez-Delgado et~al.(1998)]{GDetal98} Gonz\'alez-Delgado, Rosa M., Heckman, T., Leitherer, C., Meurer, G., Krolik, J., Wilson, A.S., Kinney, A., Koratkar, A., 1998, ApJ, 505, 174. 
\bibitem[Gonz\'alez-Delgado et~al.(2001)]{GDetal01} Gonz\'alez-Delgado, R.M., Heckman, T., \& Leitherer, C., 2001, ApJ, 546, 845. 
\bibitem[Goodrich(1989)]{Goodrich89} Goodrich, R.W., 1989, ApJ, {\bf 340}, 190.
\bibitem[Heckman(1990A)]{Heckman90A} Heckman, T.M., in {Paired and Interacting Galaxies}, eds. Sulentic, J.W., Keel, W.C., \& Telesco, C.M. 1990A, IAUC, 124, p.~359, NASA.
\bibitem[Heckman(1990B)]{Heckman90B} Heckman, T.M., in {Massive stars in starbursts}, eds. Walborn, N., \& Leitherer, C., 1990B, Proc. of the ST~ScI symposium. 
\bibitem[Heckman et~al.(1997)]{Heckmanetal97} Heckman, T.M., Gonzalez-Delgado, R., Leitherer, C., Meurer, G. R., Krolik, J., Wilson, A. S., Koratkar, A., \& Kinney, A., 1997, ApJ, 482, 114.
\bibitem[Heckman et~al.(2001)]{Hetal2001} Heckman, T.M., Sembach, K.R., Meurer, G.R., Leitherer, C., Calzetti, D., \& Martin, C.L., 2001, ApJ, 558, 56.
\bibitem[Ho, Filippenko, \& Sargent(1996)]{HFS96} Ho, L.C., Filippenko, A.V., \& Sargent, W.L.W., 1996, ApJ, 462, 183.
\bibitem[Ho et~al.(1997)]{Hoetal97} Ho, L.C., Filippenko, A.V., Sargent, W.L.W., Peng, C.Y., 1997, ApJS, 112, 391.
\bibitem[Ho \& Ulvestad(2001)]{HoU01} Ho, L.C., \& Ulvestad, J.S., 2001, ApJS, 133, 77.
\bibitem[Huchra et~al.(1983)]{Huchraetal83} Huchra, J., Davis, M., Latham, D., \& Torny, J., 1983, ApJS, 52, 89.
\bibitem[Huchra \& Burg(1992)]{HB92} Huchra, J., \& Burg, R., 1992, ApJ, 393, 90.
\bibitem[Hunt \& Malkan(1999)]{HuntMalkan99} Hunt, L.K., \& Malkan, M.A., 1999, ApJ, 516, 660.
\bibitem[Johnston et~al.(1982)]{Johnstonetal82} Johnston, K.J., Elvis, M., Kjer, D., \& Shen, B.S.P. 1982, ApJ, 262, 61.
\bibitem[Kapahi \& Saikia(1982)]{KS82} Kapahi, V.K., \& Saikia, D.J., 1982, JApA, 3, 465.
\bibitem[Kauffmann et~al.(2003A)]{Ketal03A} Kauffmann, G. et~al. 2003A, MNRAS, 341, 33.
\bibitem[Kauffmann et~al.(2003B)]{Ketal03B} Kauffmann, G. et~al. 2003B, MNRAS, 341, 54.
\bibitem[Kellermann et~al.(1989)]{Kellermannetal89} Kellermann, K.I., Sramek, R., Schmidt, M., Shaffer, D.B., \& Green, R. 1989, AJ, 98, 1195.
\bibitem[Keel(1983)]{Keel83} Keel, W.C., 1983, ApJS, 52, 229.
\bibitem[Khachikian \& Weedman(1974)]{KW74} Khachikian, E.Y., \& Weedman, D.W. 1974 ApJ, 192, 581.
\bibitem[Kinney et~al.(1991)]{Kinneyetal91} Kinney, A.L., Antonucci, R.R.J., Ward, M.J., Wilson, A.S., \& Whittle, M., 1991, ApJ, 377, 100.
\bibitem[Kim et~al.(1998)]{KVS98} Kim, D.-C., Veilleux, S., \& Sanders, D.B., 1998, ApJ, 508, 627.
\bibitem[Kriss et~al.(1994)]{Krissetal94} Kriss, G.A., Tsvetanov, Z., \& Davidsen, A.F., in {First Stromlo Symposium: The phy of active galaxies}, eds. Bicknell, G.V., Dopita, M.A., \& Quinn, P.J., 1994, ASP Conf. Ser., 54, p.~281.
\bibitem[Krongold et~al.(2002)]{Krongoldetal02} Krongold, Y., Dultzin-Hacyan, D., \& Marziani, P., 2002, ApJ, 572, 169.
\bibitem[Krolik(1999)]{Krolik99} Krolik, J.H. {Active Galactic Nuclei}, 1999, Princeton University Press.
\bibitem[Kukula et~al.(1995)]{Kukulaetal95} Kukula, M.J., Pedlar, A., Baum, S.A., \& O'Dea, C.P. 1995, MNRAS, 276, 1262.
\bibitem[Lal, Shastri, \& Gabuzda(2004)]{LSG04} Lal, D.V., Shastri, P., \& Gabuzda, D.C., 2004, A\&A, 425, 99.
\bibitem[Lal(2002)]{Lal02} Lal, D.V. 2002, PhD Thesis, Seyfert Galaxies: Nuclear Radio Structure and Unification.
\bibitem[Lawrence \& Elvis(1982)]{LE82} Lawrence, A., \& Elvis, M., 1982, ApJ, 256, 706.
\bibitem[Lawrence(1987)]{Lawrence87} Lawrence, A., 1987, PASP, 99, 309.
\bibitem[Lawrence \& Elvis(2010)]{LE10} Lawrence, A., \& Elvis, M., 2010, ApJ, 714, 561.
\bibitem[Levenson et~al.(2001)]{Levensonetal01} Levenson, N.A., Weaver, K.A., \& Heckman, T.M. 2001, ApJ, 550, 230.
\bibitem[Lipovetsky et~al.(1988)]{Lipovetskyetal88} Lipovetsky, V.A., Neizvestny, S.I., \& Neizvestnay, O.M. 1988, Soobshch. Spets. Astrofiz. Obs., 55, 5.
\bibitem[Maiolino et~al.(1997)]{Maiolinoetal97} Maiolino, R., Ruiz, M., Rieke, G.H., \& Papadopoulos, P., 1997, ApJ, 485, 552.
\bibitem[Mulchaey, Wilson, \& Tsvetanov(1996)]{MWT96} Mulchaey, J.S., Wilson, A.S., \& Tsvetanov, Z. 1996, ApJ, 467, 197.
\bibitem[Malkan, Gorjian, \& Tam(1998)]{MGT98} Malkan, M.A., Gorjian, V., \& Tam, R. 1998, ApJS, 117, 25.
\bibitem[Mu\~noz~Mar\'in et~al.(2007)]{Munozetal07} Mu\~noz~Mar\'in V.M., Gonz\'alez Delgado R.M., Schmitt H.R., Cid Fernandes R., P\'erez E., Storchi-Bergmann T., Heckman T., \& Leitherer C., 2007, AJ, 134, 648.
\bibitem[Markarian(1967)]{Markarian67} Markarian, B.E., Astrophysics, 1967, 3, 24.
\bibitem[Markarian et~al.(1986)]{Markarianetal86} Markarian, B.E., Stepanian, J.A., \& Erastova, L.K., 1986, Astronomy, 25, 345. 
\bibitem[Mas-Hesse et~al.(1994)]{MHetal94} Mas-Hesse, J.M., Rodriguez-Pascual, P.M., de~Cordoba, L.S.F., Mirabel, I.F., 1994, ApJS, 92, 599.
\bibitem[Meurs \& Wilson(1984)]{MW84} Meurs, E.J.A., \& Wilson, A.S., 1984, A\&A, 136, 206.
\bibitem[Miller \& Goodrich(1990)]{MG90} Miller, J.S., \& Goodrich, R.W., 1990, ApJ, 355, 456.
\bibitem[Morganti et~al.(1999)]{Morgantietal99} Morganti, R., Tsvetanov, Z.I., Gallimore, J., \& Allen, M.G., 1999, A\&AS, 137, 457.
\bibitem[Mundell et~al.(2000)]{Mundelletal00} Mundell, C.G., Wilson, A.S., Ulvestad, J.S., \& Roy, A.L. 2000, ApJ, 529, 816.
\bibitem[Mundell et~al.(2009)]{Mundelletal09} Mundell, C.G., Ferruit, P., Nagar, N. Wilson, A.S., 2009, ApJ, 703, 802.
\bibitem[Muxlow et~al.(1994)]{Muxlowetal94} Muxlow, T.W.B., T.W.B., Pedlar, A., Wilkinson, P.M., Axon, D.J., Sanders, E.M., \& de~Bruyn, A.G., 1994 MNRAS, 266, 455.
\bibitem[Nagar \& Wilson(1999)]{NW99} Nagar, N.M., \& Wilson, A.S. 1999, ApJ, 516, 79.
\bibitem[Nagar et~al.(1999)]{Nagaretal99} Nagar, N.M., Wilson, A.S., Mulchaey, J.S., \& Gallimore, J.F. 1999, ApJS, 120, 209.
\bibitem[Nagar et~al.(2000)]{Nagaretal00} Nagar, N.M., Falcke, H., Wilson, A.S., \& Ho, L.C. 2000, ApJ, 542, 186.
\bibitem[Nelson \& Whittle(1995)]{NW95} Nelson, C.H., \& Whittle, M. 1995, ApJS, 99, 67.
\bibitem[Nelson \& Whittle(1996)]{NW96} Nelson, C.H., \& Whittle, M., 1996, ApJ, 465, 96.
\bibitem[Norris et~al.(1988)]{Norrisetal88B} Norris, R.P., Kesteven, M.J., Wellington, K.J., \& Batti, M.J., 1988, ApJS, 67, 85.
\bibitem[Norris et~al.(1992A)]{Norrisetal92A} Norris, R.P., Roy, A.L., Allen, D.A., Kestevan, M.J., Troup, E.R., \& Reynolds, J.E., in {Relationships between active galactic nuclei and starburst galaxies}, ed. Filippenko, A.V. 1992A, ASP Conf Series, 31, p.~71.
\bibitem[Norris et~al.(1992B)]{Norrisetal92B} Norris, R.P., Kesteven, M.J., \& Calabretta, M.R., 1992B, Jl. Elect. of Elec. and Engg., Austr., 12, 205.
\bibitem[Osterbrock(1981)]{Osterbrock81} Osterbrock, D.E., 1981, ApJ, 249, 462.
\bibitem[Osterbrock \& Pogge(1985)]{OP85} Osterbrock, D.E., \& Pogge, R.W., 1985, ApJ, 297, 166.
\bibitem[Pedlar et~al.(1993)]{Pedlaretal93} Pedlar, A., Kukula, M.J., Longley, D.P.T., Muxlow, T.W.B., Axon, D.J., Baum, S.A., O'Dea, C.P., \& Unger, S.W., 1993, MNRAS, 263, 471.
\bibitem[Panessa \& Bassani(2002)]{PB02} Penessa, F., Bassani, L., 2002, A\&A, 394, 435.
\bibitem[Parra et~al.(2007)]{Parraetal07} Parra, R., Conway, J.E., Diamond, P.J., Thrall, H., Lonsdale, C.J., Lonsdale, C.J., \& Smith, H.E., 2007, ApJ, 659, 314
\bibitem[Peterson(1997)]{Peterson97} Peterson, B.M. {An introduction to the Active Galactic Nuclei}, 1997, Cambridge University Press. 
\bibitem[Pogge(1989)]{Pogge89} Pogge, R.W. 1989, ApJS, 71, 433.
\bibitem[Pringle et~al.(1999)]{Pringleetal99} Pringle, J.E., Antonucci, R.R.J., Clarke, C.J., Kinney, A.L., Schmitt, H.R., \& Ulvestad, J.S. 1999, ApJL, 526, L9.
\bibitem[Raimann et~al.(2003)]{Raimannetal03} Raimann, D., Storchi-Bergmann, T., Gonz\'alez Delgado, R.M., Cid Fernandes, R., Heckman, T., Leitherer, C., Schmitt, H.R., 2003, MNRAS, 339, 772.
\bibitem[Roy et~al.(1994)]{Royetal94} Roy, A.L., Norris, R.P., Kesteven, M.J., Troup, E.R., \& Reynolds, J.E. 1994, ApJ, 432, 496.
\bibitem[Rupen et~al.(1987)`]{Rupenetal87} Rupen, M.P., van~Gorkom, J.H., Knapp, G.R., Gunn, J.E., \& Schneider, D.P., 1987, AJ, 94, 61.
\bibitem[Sandage(1975)]{Sandage75} Sandage, A.R., in {Galaxies and the Universe}, eds. Sandage, A.R., Sandage. M., Kristian, J., {Stars and Stellar Systems}, 1975, 9, p.1, Univ. Chicago.
\bibitem[Schmidt \& Green(1983)]{SG83} Schmidt, M., \& Green, R.F. 1983, ApJ, 269, 352.
\bibitem[Schmitt et~al.(1999)]{Schmittetal99} Schmitt, H.R., Storchi-Bergmann, T., Cid~Fernandes, R., 1999, MNRAS, 303, 173.
\bibitem[Schmitt et~al.(2001)]{Schmittetal01} Schmitt, H.R., Ulvestad, J.S., Antonucci, R.R.J., \& Kinney, A.L. 2001, ApJS, 132, 199.
\bibitem[Shuder(1981)]{Shuder81} Shuder, J.M. 1981, ApJ, 244, 12.
\bibitem[Siegel \& Castellan(1981)]{SC81} Siegel, S., \& Castellan Jr., N.J. 1981, Nonparametric Statistics for the Behavioral Sciences.
\bibitem[Smith et~al.(1998)]{Smithetal98A} Smith, H.E., Lonsdale, C.J., Lonsdale, C.J., Diamond, P.J., 1998, ApJL, 493, L17.
\bibitem[Smith, Lonsdale, \& Lonsdale(1998)]{SLL98B} Smith, H.E., Lonsdale, C.J., \& Lonsdale, C.J. 1998, ApJ, 492, 137.
\bibitem[Storchi-Bergmann et~al.(2000)]{SB00}  Storchi-Bergmann T., Raimann D., Bica E. L.D., \& Fraquelli H.A., 2000, ApJ, 544, 747.
\bibitem[Storchi-Bergmann et~al.(2001)]{SB01}  Storchi-Bergmann T., Gonz\'alez Delgado, R.M., Schmitt, H.R., Cid~Fernandes, R., \& Heckman, T., 2001, ApJ, 559, 147.
\bibitem[Su et~al.(1996)]{Suetal96} Su, B.M., Muxlow, T.W.B., Pedlar, A., Holloway, A.J., Steffen, W., Kukula, M.J., \& Mutel, R.L. 1996, MNRAS, 279, 1111.
\bibitem[Terlevich(1990A)]{Terlevich90A} Terlevich, R., in {Windows on galaxies}, eds. Fabbiano, G., Gallagher, J. \& Renzini, A., 1990A, ASSL, p. 87, Kluwer: Dordrecht. 
\bibitem[Terlevich(1990B)]{Terlevich90B} Terlevich, R., in {Structure and dynamics of the inter-stellar medium}, eds. Tenorio-Tagle, G., Moles, M., \& Melnick, J., 1990B, p. 343, Springer-Verlag: Berlin. 
\bibitem[Terlevich \& Boyle(1993)]{TB93} Terlevich, R. \& Boyle, B.J., 1993, MNRAS, 262, 491.
\bibitem[Thean et~al.(2001)]{Theanetal01} Thean, A., Pedlar, A., Kukula, M.J., Baum, S.A., \& O'Dea, C.P. 2001, MNRAS, 325, 737.
\bibitem[Tran(2001)]{Tran01} Tran, H.D., 2001, ApJ, 554, L19.
\bibitem[Tran(2003)]{Tran03} Tran, H.D., 2003, ApJ, 583, 632.
\bibitem[Ulvestad \& Wilson(1981)]{UWS81} Ulvestad, J.S., Wilson, A.S., \& Sramek, R.A., 1981, ApJ, 247, 419.
\bibitem[Ulvestad(1982)]{Ulvestad1982} Ulvestad, J.S., 1982, ApJ, 259, 96.
\bibitem[Ulvestad \& Wilson(1984A)]{UW84A} Ulvestad, J.S., \& Wilson, A.S. 1984A, ApJ, 278, 544.
\bibitem[Ulvestad \& Wilson(1984B)]{UW84B} Ulvestad, J.S., \& Wilson, A.S. 1984B, ApJ, 285, 439.
\bibitem[Ulvestad(1986)]{Ulvestad1986} Ulvestad, J.S. 1986, ApJ, 310, 136.
\bibitem[Ulvestad \& Wilson(1986)]{UW86} Ulvestad, J.S., \& Wilson, A.S. 1986, MNRAS, 218, 711.
\bibitem[Ulvestad \& Wilson(1989)]{UW89} Ulvestad, J.S., \& Wilson, A.S. 1989, ApJ, 343, 659.
\bibitem[Ulvestad, Antonucci, \& Goodrich(1995)]{UAG95} Ulvestad, J.S., \& Antonucci, R.R.J., Goodrich, R.W. 1995, AJ, 109, 81.
\bibitem[Ulvestad et~al.(1998)]{Ulvestadetal98} Ulvestad, J.S., Roy, A.L, Colbert, J.M., \& Wilson, A.S. 1998, ApJ, 496, 196.
\bibitem[Ulvestad et~al.(1999A)]{Ulvestadetal99A} Ulvestad, J.S., Wrobel, J.M., \& Carilli, C.L. 1999A, ApJ, 516, 127.
\bibitem[Ulvestad et~al.(1999B)]{Ulvestadetal99B} Ulvestad, J.S., Wrobel, J.M., Roy, A.l., Wilson, A.S., Falcke, H., \& Krichbaum, T.P. 1999B, ApJL, 517, L81.
\bibitem[Ulvestad(1999)]{Ulvestad99} Ulvestad, J.S., 1999, VLBA operations memo No.~34, NRAO. 
\bibitem[Ulvestad et~al.(2000)]{Ulvestad00} Ulvestad, J.S., 2000, VLBA scientific memo No.~25, NRAO. 
\bibitem[Ulvestad \& Ho(2001)]{UHo01} Ulvestad, J.S., \& Ho, L.C. 2001, ApJ, 558, 561.
\bibitem[Unger et~al.(1984)]{Ungeretal84} Unger, S.W., Pedlar, A., Neff, S.G., \& de~Bruyn, A.G. 1984, MNRAS, 209, 15P.
\bibitem[Unger et~al.(1986)]{Ungeretal86} Unger, S.W., Pedlar, A., Booler, R.V., \& Harrison, B.A. 1986, MNRAS, 219, 387.
\bibitem[Urry \& Padovani(1995)]{UP95} Urry, C.M., \& Padovani, P. 1995, PASP, 107, 803.
\bibitem[de~Vaucouleurs et~al.(1991)]{deVetal91} de~Vaucouleurs, G., de~Vaucouleurs, A., Corwin (Jr.) H.G., Buta, R.J., Paturel, G., \& Fouque, P. {Third reference catalogue of bright galaxies (RC3), version~3.9}, 1991, Springer-Verlag, New York. 
\bibitem[Veron(1986)]{Veron86} Veron, P., in {Structure and evolution of active galactic nuclei}, eds. Giuricin, G., Mardirossian, F., Mezzetti, M., \& Ramella, M. 1986, ASSL, p .253, D. Reidel Publishing Co. 
\bibitem[Weedman(1977)]{Weedman77} Weedman, D.W. 1977, ARA\&A, 15, 69.
\bibitem[Whittle et~al.(1988)]{Whittleetal88} Whittle, M., Pedlar, A., Meurs, E.J.A., Unger, S.W., Axon, D.J., Ward, M.J., 1988, ApJ, 326, 125.
\bibitem[Whittle(1992A)]{Whittle92A} Whittle, M. 1992A, ApJS, 79, 49.
\bibitem[Whittle(1992B)]{Whittle92B} Whittle, M., 1992B, ApJ, 387, 109.
\bibitem[Whittle(1992C)]{Whittle92C} Whittle,M., 1992C, ApJ, 387, 121.
\bibitem[Wilson \& Ulvestad(1982)]{WU82} Wilson, A.S. \& Ulvestad, J.S., 1982, ApJ, 263, 576.
\bibitem[Wilson \& Ulvestad(1983)]{WU83} Wilson, A.S. \& Ulvestad, J.S., 1983, ApJ, 275, 8.
\bibitem[Wilson et~al.(1986)]{Wilsonetal86} Wilson, A.S., Baldwin, J.A., Sun, Sze-Dung, \& Wright, A.E., 1986, ApJ, 310, 121.
\bibitem[Wilson \& Keel(1989)]{WK89} Wilson, A.S. \& Keel, W.C. 1989, AJ, 98, 1581.
\bibitem[Wilson et~al.(1991)]{Wilsonetal91} Wilson, A.S., Helfer, T.T., Haniff, C.A., \& Ward, M.J. 1991, 381, 79.
\bibitem[Wilson(1991)]{Wilson91} Wilson, A.S., in {The interpretation of modern synthesis observations of spiral galaxies},eds. Duric, N., \& Crane, P.C. 1991, ASP, 18, p.~227.
\bibitem[Wrobel(1995)]{Wrobel95} Wrobel, J.M. 1995, ``VLBI observing strategies'', in {Very Long Baseline Interferometry and the VLBA}, eds. Zensus, J.A., Diamond,P.J., \& Napier, P.J. 1995, ASP Conf Series, 82, p.~411.
\bibitem[Wrobel(2000)]{Wrobel00} Wrobel, J.M. {VLBA observational status summary}, 2000A, NRAO. 
\bibitem[Yee(1980)]{Yee80} Yee, H.K.C. 1980, ApJ, 241, 894.

\end{thebibliography}
